\documentclass[fleqn,usenatbib]{mnras}

\usepackage{newtxtext,newtxmath}
\usepackage{tikz}
\usepackage[version=4]{mhchem}
\usetikzlibrary{decorations.pathmorphing}
\usepackage{longtable}
\usepackage{tabularx}
\usepackage{xltabular}
\usepackage{supertabular}
\usepackage{float}
\usepackage{placeins}
\usepackage{hyperref}

\usepackage[T1]{fontenc}

\DeclareRobustCommand{\VAN}[3]{#2}
\let\VANthebibliography\thebibliography
\def\thebibliography{\DeclareRobustCommand{\VAN}[3]{##3}\VANthebibliography}

\usepackage{graphicx}	
\usepackage{amsmath}	

\title[Detection of YMCs' ongoing self-enrichment with $^{26}$Al]{Using $^{26}$Al to detect ongoing self-enrichment in young massive star clusters}

\author[K. Nowak et al.]{Katarzyna Nowak,$^{1}$\thanks{E-mail: k.nowak@herts.ac.uk}, Martin G. H. Krause,$^{1}$ Thomas Siegert,$^{2}$ Jan Forbrich,$^{1}$ 
\newauthor{Robert M. Yates,$^{1}$, Laura Ram\'irez-Galeano,$^{3}$ Corinne Charbonnel$^{3,4}$ and Mark Gieles$^{5,6}$}
\\
$^{1}$Centre for Astrophysics Research, Department of Physics, Astronomy and Mathematics, University of Hertfordshire, College Lane, Hatfield AL10 9AB, UK \\
$^{2}$ Julius-Maximilians-Universit\"{a}t W\"{u}rzburg, Fakult\"{a}t f\"{u}r Physik und Astronomie, Institut f\"{u}r Theoretische Physik und Astrophysik,\\ Lehrstuhl f\"{u}r Astronomie, Emil-Fischer-Str. 31, D-97074 W\"{u}rzburg, Germany\\
$^{3}$ Department of Astronomy, University of Geneva, Chemin des Maillettes 51, CH-1290 Versoix, Switzerland\\
$^{4}$ IRAP, UMR 5277, CNRS and Universit\'e de Toulouse, 14, avenue \'Edouard Belin, F-31400 Toulouse, France\\
$^{5}$ ICREA, Pg. Llu\'is Companys 23, E-08010 Barcelona, Spain\\
$^{6}$ Institut de Ci\'encies del Cosmos (ICCUB), Universitat de Barcelona (IEEC-UB), Mart\'i i Franqu\'es 1, E-08028 Barcelona, Spain
}

\date{Accepted 2024 September 24. Received 2024 September 23; in original form 2024 August 21}

\pubyear{2024}

\begin{document}
\label{firstpage}
\pagerange{\pageref{firstpage}--\pageref{lastpage}}
\maketitle

\begin{abstract}
Self-enrichment is one of the leading explanations for chemical anomalies in globular clusters. In this scenario, various candidate polluter stars have been proposed to eject gas with altered chemical composition during the self-enrichment process. Most of the proposed polluters will also eject radioactive $^{26}$Al into the surroundings. Hence, any detection of $^{26}$Al in young massive star clusters (YMCs) would support the self-enrichment scenario if YMCs were indeed the progenitors of globular clusters. Observations of gamma-ray data from COMPTEL and INTEGRAL, as well as detections of $^{26}$AlF molecules by the Atacama Large Millimeter-submillimeter Array (ALMA), indicate the maturing of $^{26}$Al detection methods. Detection possibilities will be enhanced in the short- to mid-term by the upcoming launch of the Compton Spectrometer and Imager (COSI). The Square Kilometer Array (SKA) could in principle also detect radio recombination lines of the positronium formed from the decay products of $^{26}$Al. Here, we show for a sample of YMCs in the nearby Universe, where self-enrichment could plausibly take place. For some nearby galaxies, this could enhance $^{26}$Al by an order of one magnitude. Detecting $^{26}$AlF with ALMA appears feasible for many candidate self-enrichment clusters, although significant challenges remain with other detection methods. The Large Magellanic Cloud, with its YMC R136, stands out as the most promising candidate. Detecting a 1.8~MeV radioactive decay line of $^{26}$Al here would require at least 15 months of targeted observation with COSI, assuming ongoing self-enrichment in R136.

\end{abstract}




\section{Introduction}

Variations in abundances of C, N, O, Na, Mg, and Al have been extensively observed in globular clusters (GCs), highlighting the significant inhomogeneities in lighter elements within these stellar systems \citep{Denisenkov_1990, Langer_1993, Ventura_2001, Prantzos_2007, Gratton_2012, Charbonnel_2016, Prantzos_2017, Milone_2022}. Among these elements, the Na-O anticorrelation stands out as a prominent feature observed in the majority of GCs \citep{Carretta_2010}. Additionally, most Galactic and extra-galactic GCs display evidence of multiple sequences in the colour–magnitude diagram, indicating a spread in helium abundance \citep{Norris_2004, D'Antona_2005, Charbonnel_2016, Chantereau_2016}, which is supported by spectroscopic measurements conducted over the last 20 years \citep{Anderson_2002, Bedin_2004, Piotto_2007, Piotto_2015, Milone_2012, Milone_2013, Martocchia_2018}. These studies have provided crucial insights into the existence of multiple stellar populations within the majority of GCs \citep{Bastian_2018, Milone_2018}.

A leading explanation for the observed chemical peculiarities in GCs is self-enrichment, where certain stars pollute the surrounding gas with their yields \citep{Decressin_2007, DErcole_2008, Krause_2013, Wunsch_2017, Gieles_2018, Winter_2023}. It is assumed that yields from those polluters can be transported via stellar winds into their surroundings, mix with the pristine gas, and eventually accrete onto proto-stars \citep{Krause_2020}. Commonly proposed polluters include asymptotic giant branch stars \citep[AGBs;][]{Ventura_2001}, fast-rotating massive stars \citep[FRMSs;][]{Decressin_2007, Krause_2013}, and supermassive stars \citep[SMSs, M > 10$^4$ M$_{\odot}$;][]{Denissenkov_2014, Gieles_2018, Nowak_2022}. However, some of these propositions have shortcomings \citep{Bastian_2015, Prantzos_2017}. For example, AGB stars show an O-Na correlation \citep{Forestini_1997, Denissenkov_2003, Karakas_2007, Siess_2010, Ventura_2013, Doherty_2014, Renzini_2022} and He-burning products that are not found in the majority of GCs \citep{Karakas_2006, Decressin_2009, Yong_2014}. FRMSs are not capable of producing the Mg-Al anticorrelation without a strong He enrichment, which is not observed in GCs \citep{Decressin_2007, Martins_2021}. The situation is more advantageous for SMSs, where, according to \cite{Prantzos_2017}, the required temperature to activate the MgAl chain is reached at the very beginning of their evolution, when the He levels are still low, thereby accounting for the various abundances displayed in GCs \citep{Denissenkov_2014, Denissenkov_2015}.

The identity of the polluter in the self-enrichment scenario remains unclear. However, according to \cite{Gratton_2012} and \cite{Prantzos_2017}, all polluted material is produced from hydrogen burning at high temperatures via the CNO-cycle ($\gtrsim$ 20 MK), NeNa ($\gtrsim$ 45 MK), and MgAl ($\gtrsim$ 70 MK) chains. These conditions are highly conducive to producing the radioactive isotope $^{26}$Al \citep{Prantzos_1986}, with a half-life of 0.717~Myr \citep{Norris_1983, Basunia_2016}. $^{26}$Al is specifically produced in MgAl chains through the reaction $^{25}$Mg($p, \gamma$)$^{26}$Al in environments with temperatures between 40 and 90 MK \citep{Diehl_2021}. According to \cite{Prantzos_1996}, short-lived massive stars ($M$ > 40 M$_{\odot}$) play a crucial role in the production of $^{26}$Al. This conclusion is supported by the strong correlation between the intensity map of $^{26}$Al radioactive decay line at 1.8~MeV and 53 GHz microwave free-free emission, which indicates the presence of ionized regions \citep{Knodlseder_1999}. The production of $^{26}$Al in massive stars is to a significant degree facilitated through stellar wind ejection \citep{Prantzos_1986, Meynet_1997, Palacios_2005, Voss_2009}. Contrary to the stable isotope $^{27}$Al, the decay timescale of $^{26}$Al aligns with the timescales of pollution via winds of massive stars and star formation. As a result, $^{26}$Al emerges as a potential tracer of ongoing self-enrichment, distinguishing it from other radioactive isotopes such as $^{44}$Ti, $^7$Be, and $^{22}$Na, which primarily trace supernovae or novae events \citep{Diehl_1998}.   

In this study, our focus is on detecting ongoing nucleosynthesis in YMCs, which are considered plausible progenitors of GCs \citep{Bastian_2014, Cabrera_Ziri_2014, Cabrera_Ziri_2015}. If self-enrichment occurs in proto-GCs, then we should also expect the presence of radioactive nuclei e.g. $^{26}$Al. Thus, detecting $\gamma$-ray decay lines from a YMC would directly link to ongoing self-enrichment. 
The ejecta that pollute the clusters must be present at a very early age if the polluters are very massive stars (VMSs, $M \simeq~10^{2-3}$~M$_{\odot}$) or SMSs, due to their short lifetime. There is a debate in the literature \citep{Mucciarelli_2014, Krause_2016, Milone_2022} as to which properties YMCs must have to express multiple populations, or whether this occurs at all in the nearby Universe. We argue in Section \ref{YMCS} that the compactness of YMCs should allow the formation of multiple population clusters also today. 

There is a well-known "mass-budget" problem in the formation of GCs \citep{Bastian_2018}. For a normal IMF one expects not enough ejecta to explain the observed number of 2nd populations of stars. This is an important point for consideration of the mass of $^{26}$Al present. If YMCs are progenitors of multiple populations GCs, then we should expect the presence of sufficient amounts of ejecta to form the 2nd population. At this stage one can be agnostic about precise mechanism how the required ejecta mass is produced. Suggested mechanisms include top-heavy IMF \citep{Prantzos_2006} or a conveyor-belt recycling of gas through a SMS \citep{Gieles_2018}. As a result of that one must have a much higher abundance of $^{26}$Al in such a YMC of a given mass compared to YMCs that do not form multiple populations, which we estimate in this study. This is therefore a critical test for assessing if self-enrichment in YMCs actually proceeds in a way conducive to the expression of multiple populations as observed in old GCs. 

Here, we specifically concentrate on $^{26}$Al with its radioactive decay line at 1.8~MeV, the recombination and annihilation of the emitted decay positrons and millimetre-wave rotational lines of the molecule $^{26}$AlF, recently observed for the first time by \cite{Kaminski_2018} as the only molecule containing $^{26}$Al. At present no extragalactic source has ever been detected in $\gamma$-ray decay lines due to sensitivity limitations of current $\gamma$-ray instruments \citep{Bouchet_2015}. We investigate here the best possible case, namely the assumption that all the processed material is ejected quickly during the formation period of YMCs when they are still embedded. At this point, we expect that approximately half of the $^{26}$Al will remain mixed within the cold gas. This maximises also the chances that the positron produced in the beta decay of $^{26}$Al forms a Ps atom by capturing an electron from neutral gas. The latter would then produce radio recombination lines and eventually decay and emit the 511~keV line.

The paper is organised as follows: in Section \ref{Al26}, we provide a summary of the formation process and possible detection of $^{26}$Al, positronium (Ps) radio recombination lines as well as the $^{26}$AlF molecule. We outline the selection criteria for YMCs in Section \ref{YMCS}, followed by a description of the methods used to estimate the flux for $^{26}$Al and the 511~keV line, the background emission from massive stars that do not contribute to self-enrichment for both lines, and the Ps radio recombination lines in these YMCs in Section \ref{flux}. In Section \ref{observatories} we explore the capabilities of future observatories for detecting $\gamma$-ray lines and Ps radio recombination lines and the results are discussed in Section \ref{results}. Section \ref{discussion} focuses on our findings, and we argue that, given the proposed and planned missions, we currently lack the ability to detect most of the aforementioned $^{26}$Al tracers within our selected sample of YMCs. However, the R136 cluster in the LMC could potentially be detected for the 1.8 MeV and 511 keV lines using targeted observations with COSI instrument spanning from 15 months. Additionally, the rotational line of $^{26}$AlF remains detectable across all our selected YMCs. Finally, we summarize our work in Section \ref{conclusions}.

\section{Aluminium 26}\label{Al26}

\begin{figure*}
\vspace{0.5cm}
\begin{tikzpicture}
  \path (0,-2) rectangle (4,2);
  \pgfmathdeclarerandomlist{color}{{red}{white}}
  \pgfmathsetseed{1}

  \foreach \A/\R in {8/0.2, 5/0.13, 1/0}{
    \pgfmathsetmacro{\S}{360/\A}
    \foreach \B in {0,\S,...,360}{
      \pgfmathrandomitem{\C}{color}
      \pgfmathsetmacro{\Radius}{\R*3pt}
      \shade[ball color=\C] (\B+\A:\Radius) circle (\Radius);
    }
  }
  \node at (-1.3,1.3) {\ce{^{26}Al}};

  \begin{scope}[shift={(5,0)}]
\foreach \A/\R in {8/0.2, 5/0.13, 1/0}{
    \pgfmathsetmacro{\S}{360/\A}
    \foreach \B in {0,\S,...,360}{
      \pgfmathrandomitem{\C}{color}
      \pgfmathsetmacro{\Radius}{\R*3pt}
      \shade[ball color=\C] (\B+\A:\Radius) circle (\Radius);
      }
    }
    \node at (-1.3,1.3) {\ce{^{26}Mg^*}};
  \end{scope}

  \draw[->, line width=1pt] (1.5,0) -- (3.5,0);
  \node at (2.5,0.25)  {$\beta+$ decay};
  \draw[->, line width=1pt] (6.5,0) -- (8.5,0);
  \node at (7.5,0.25)  {de-excitation};
  \draw[->, line width=2pt, decorate, decoration={snake, amplitude=.4mm, segment length=2mm, post length=1mm}] (7.5, 1.0) --  (7.5, 2.5) node[above] {1.8 MeV};
  \draw[->, line width=1pt] (5.0, 1.5) -- (5.0,2.5) node[above] {$\nu_e$};
  \draw[->, line width=1pt] (5.0, -1.5) -- (5.0, -2.7); 

  \begin{scope}[shift={(10,0)}]
\foreach \A/\R in {8/0.2, 5/0.13, 1/0}{
    \pgfmathsetmacro{\S}{360/\A}
    \foreach \B in {0,\S,...,360}{
      \pgfmathrandomitem{\C}{color}
      \pgfmathsetmacro{\Radius}{\R*3pt}
      \shade[ball color=\C] (\B+\A:\Radius) circle (\Radius);
      }
    }
    \node at (-1.3,1.3) {\ce{^{26}Mg}};
  \end{scope}

\draw (5,-4.3) circle (1.2cm and 1.2cm);
\draw[dashed] (5,-4.3) circle (0.75cm and 0.75cm);

\shade[ball color=red!20!white,opacity=1.50] (5, -3.1) circle (0.2);
\node[align=center](moon) at (5,-3.1) {e$^+$};
\shade[ball color=red!20!white,opacity=1.5] (5.5, -3.8) circle (0.2);
\node[align=center](moon) at (5.5,-3.8) {e$^+$};
\draw[->, line width=1pt, color=red] (5.15, -3.3) -- (5.35,-3.65);

\shade[ball color=blue!20!white,opacity=1.50] (5, -5.5) circle (0.2);
\node[align=center](moon) at (5,-5.5) {e$^-$};
\shade[ball color=blue!20!white,opacity=1.5] (4.5, -4.8) circle (0.2);
\node[align=center](moon) at (4.5,-4.8) {e$^-$};
\draw[->, line width=1pt, color=blue] (4.85,-5.3) -- (4.6,-5.);

\draw[->, line width=2pt, decorate, decoration={snake, amplitude=.4mm, segment length=2mm, post length=1mm}] (3.5, -4.3) --  (2.0, -4.3) node[left] {$\nu_\mathrm{rec}$};
\node[align=center](moon) at (5., -6.1) {Ps radio recombination};

\shade[ball color=red!20!white,opacity=1.50] (9.5, -3.3) circle (0.2);
\node[align=center](moon) at (9.5, -3.3) {e$^+$};

\shade[ball color=blue!20!white,opacity=1.50] (9.5, -3.7) circle (0.2);
\node[align=center](moon) at (9.5, -3.7) {e$^-$};
\draw[->, line width=1pt] (6.4,-3.5) -- (8.6,-3.5);

\draw[->, line width=2pt, decorate, decoration={snake, amplitude=.4mm, segment length=2mm, post length=1mm}] (10.0, -3.3) --  (11.5, -3.1) node[right] {511 keV};
\draw[->, line width=2pt, decorate, decoration={snake, amplitude=.4mm, segment length=2mm, post length=1mm}] (10.0, -3.7) --  (11.5, -3.9) node[right] {511 keV};
\node[align=center](moon) at (7.4, -3.3) {decay of para-Ps};

\shade[ball color=red!20!white,opacity=1.50] (9.5, -5.0) circle (0.2);
\node[align=center](moon) at (9.5, -5.0) {e$^+$};

\shade[ball color=blue!20!white,opacity=1.50] (9.5, -5.4) circle (0.2);
\node[align=center](moon) at (9.5, -5.4) {e$^-$};
\draw[->, line width=1pt] (6.4,-5.2) -- (8.6,-5.2);

\draw[->, line width=2pt, decorate, decoration={snake, amplitude=.4mm, segment length=2mm, post length=1mm}] (10.0, -5.0) --  (11.5, -5.0);
\draw[->, line width=2pt, decorate, decoration={snake, amplitude=.4mm, segment length=2mm, post length=1mm}] (10.0, -5.2) --  (11.5, -5.2) node[right] {\{total = 1022 keV\}};
\draw[->, line width=2pt, decorate, decoration={snake, amplitude=.4mm, segment length=2mm, post length=1mm}] (10.0, -5.4) --  (11.5, -5.4);
\node[align=center](moon) at (7.4, -5.0) {decay of ortho-Ps};
\end{tikzpicture}
\vspace{0.5cm}
\caption{The schematic depicts the decay process of $^{26}$Al. From the leftmost, $^{26}$Al undergoes $\beta^+$ decay, resulting in the formation of an excited state of $^{26}$Mg, along with an electron neutrino ($\nu_\mathrm{e}$) and a positron (e$^+$). The excited state subsequently de-excites to the ground state of $^{26}$Mg, emitting a 1.8~MeV $\gamma$-ray photon. The positron from the $\beta^+$ decay process usually loses energy during the in-flight phase and forms a short-lived state called positronium (Ps) when thermalised with the surrounding medium, by binding with an electron (e$^-$). This schematic assumes all positrons produced from the decay process will form a Ps state via radiative recombination. In this process, Ps undergoes radiative cascade by de-exciting to the ground state, emitting subsequent radio recombination lines ($\nu_\mathrm{rec}$). Finally, depending on the spin state, Ps will decay into two 511 keV $\gamma$-ray photons (para-Ps) or into three $\gamma$-ray photons with the total energy of 1022 keV (ortho-Ps).}
\label{Al26 decay}
\end{figure*}
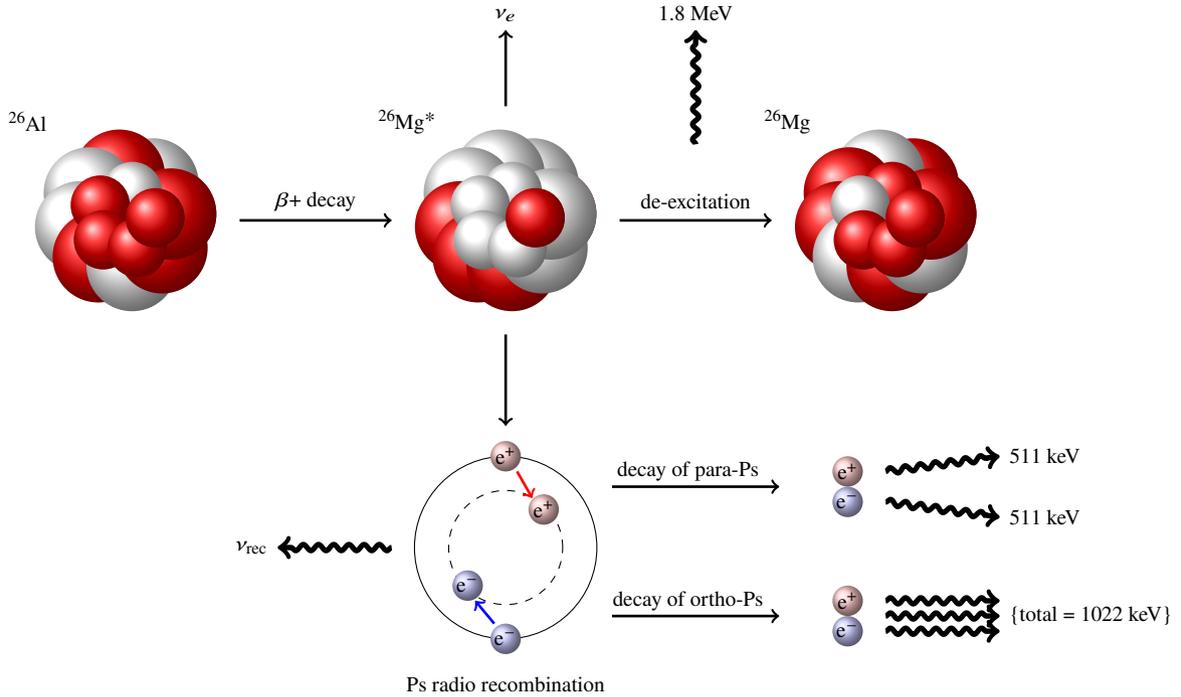

As illustrated in Figure \ref{Al26 decay}, $^{26}$Al undergoes $\beta^+$ decay from the ground state of $^{26}$Al to an excited state of $^{26}$Mg, accompanied by the emission of a positron (e$^+$) and an electron neutrino ($\nu_e$). During the subsequent de-excitation process to the ground state of $^{26}$Mg, it emits a characteristic $\gamma$-ray at 1.809~MeV \citep{Diehl_2021}. The emitted positron thermalises by interaction with the surrounding medium \citep{Ellis_2009} and either rapidly annihilates with an electron, resulting in the emission of two 511 keV $\gamma$-ray photons \citep{Jean_2009, Prantzos_2011} or forms a short-lived bound state, Ps, with an electron \citep{Mohorovicic_1934, Deutsch_1951}. In the latter scenario, para-Ps is formed 1/4 of the time, while ortho-Ps is formed 3/4 of the time \citep{Ore_1949}. In the Milky Way, annihilation seems to occur primarily in gas of temperature $T~\approx$~10$^4$~K \citep{Churazov_2005, Jean_2006, Churazov_2011, Siegert_2016}. This results in a fraction of positrons forming positronium before annihilation of approximately 95-98 per cent \citep{Guessoum_2005}. When Ps is formed in an excited state, as illustrated in Figure \ref{Al26 decay}, it undergoes radiative cascade by transitioning to the ground state, followed by annihilation and the emission of two 511~keV $\gamma$-ray photons for para-Ps or three $\gamma$-ray photons for ortho-Ps resulting in a continuum up to 511 keV \citep{Ore_1949}. Transitions between excited states give rise to the production of positronium radio recombination lines, allowing for the detection of Ps \citep{Mohorovicic_1934, Anantharamaiah_1989, Puxley_1996, Ellis_2009, StaveleySmith_2022}. 

Regardless of the specific polluter model, it is likely that we obtain $^{26}$Al ejecta in the stellar winds \citep{Prantzos_1986, Prantzos_1996}, including massive stars and VMSs \citep{Martinet_2022, Higgins_2023}, close binaries \citep{Brinkman_2019}, and possibly supernova explosions \citep{Limongi_2006, Limongi_2018}. Other sources such as SMSs up to a few times 10$^5$~M$_{\odot}$ \citep{Hillebrandt_1987}, novae \citep{Bennett_2013, Canete_2023}, and AGB stars \citep{Wasserburg_2006, Mowlavi_2005, Wasserburg_2017, Canete_2023} have also been investigated. AGB stars are believed to contribute between 0.1 and 0.6 M$_{\odot}$ to the total Galactic $^{26}$Al yield, with their main production occurring during hydrogen shell burning \citep{Mowlavi_2005, Canete_2023}. In contrast, massive stars (< 100 M$_{\odot}$) produce $^{26}$Al primarily during hydrogen core burning \citep{Diehl_2021}. For a 60~M${\odot}$ star, the maximum surface abundance is established around 3.6 Myr, but rotational effects enhance transport processes, causing the maximum abundance to occur in less than 2 Myr \citep{Martinet_2022}. Binaries also play a crucial role for stars <~40~M$_\odot$, significantly increasing $^{26}$Al yields \citep{Brinkman_2019}.

\cite{Martinet_2022} conducted an investigation on VMSs within the mass range of 120~M$_{\odot}$ to 300~M$_{\odot}$, which produce $^{26}$Al during hydrogen core burning and release it to the surface rapidly, around 0.1 Myr. Their findings suggest that these stars, characterised by metallicities between Z~=~0.006 and 0.020, could have a significant impact on the $^{26}$Al budget in the interstellar medium. This is due to their substantial rates of mass loss via stellar winds, potentially leading to a significant increase of up to 150 per cent in the total abundance of $^{26}$Al within a galaxy compared to stellar populations within the mass range of 8 to 120~M$_{\odot}$. $^{26}$Al might be particularly interesting in the context of the conveyor-belt model for SMSs \citep{Gieles_2018}. Here, the SMS is fed by lower mass stars and sheds processed material via a wind, during which it is expected that all the ejected material would be enriched in $^{26}$Al. This could increase the typical yield per star substantially. For instance, while a 100 (500)M$_{\odot}$ star is expected to yield on the order of 10$^{-3}$ (10$^{-2}$)~M$_{\odot}$ \citep{Higgins_2023}, the yield from an SMS could exceed 1 M$_{\odot}$ (compare Section \ref{SMS}, below) comparable to the estimated $^{26}$Al mass of the entire Milky Way, ranging between 1.7 and 3.5~M$_{\odot}$ \citep{Knodlseder_1999, Diehl_2006, Wang_2009, Pleintinger_2019, Pleintinger_2023, Siegert_2023}.

Furthermore, $^{26}$Al has also been detected in molecular form as $^{26}$AlF. \cite{Kaminski_2018} detected the molecule in four rotational transitions towards CK Vul, marking the first astronomical detection of $^{26}$AlF in any astronomical object. Additionally, \cite{Pavlenko_2022} conducted a search for $^{26}$Al$^1$H lines in the Proxima Centauri spectrum, but no noticeable features were found. It should be stressed that the formation of $^{26}$AlF depends strongly on the availability of fluorine (F), not only $^{26}$Al. Within a forming star cluster, regions of relatively cool gas are expected to be present, with $ T \sim 10^2$ K \citep{Conroy_2011}, and high column densities reaching $N_\mathrm{H_2}$~=~10$^{17}$~cm$^{-2}$ \citep{Krause_2013}. In such environments, the photodissociation of H$_2$ molecules can be suppressed, allowing for the production of molecules such as AlF which is even harder to dissociate than H$_2$ \citep{Wang_2024}.

\section{Selection of star clusters}\label{YMCS}

In this section, we focus on the selection criteria for identifying YMCs capable of hosting multiple populations. We specifically examine three key parameters: age, mass, and compactness.

We initially selected around 4000 YMCs within the Local Group from the review paper by \cite{PortegiesZwart_2010} and the Legacy ExtraGalactic UV Survey (LEGUS, see \cite{Calzetti_2015}) through their publicly released catalog\footnote{Publicly available at \href{https://gillenbrown.com/LEGUS-sizes/}{https://gillenbrown.com/LEGUS-sizes/}.} \citep{Brown_2021}. The ages were determined from optical observations as reported in the relevant literature. We do not strictly require the ejecta to still be present inside the YMC at the time of observation. Generally, we consider here scenarios in which the pollution event takes place early in the evolution of a star cluster. The half-life of $^{26}$Al of 0.7~Myr then implies that we should restrict our attention to not much after the pollution phase. YMCs older than 30 Myr are generally gas-free \citep{Longmore_2015}. Even at a few Myr, YMCs may be devoid of gas and hence past the time when any polluted stars may form \citep{Bastian_2014, Hollyhead_2015}. However, more recently, \cite{Calzetti_2023} have found a population of dust-buried cluster candidates with ages as high as 5-6~Myr. Also, models by \cite{Gieles_2018} indicate that an SMS, even at around 6~Myr, still has a high mass and can produce and eject $^{26}$Al via its winds (even, if the cluster would be gas-free at that point, the SMS wind could still produce $^{26}$Al). Therefore, we set our upper age limit to 10 Myr. To ensure the inclusion of clusters with potential for long-term survival and the hosting of multiple populations, we also adopted a mass criterion of > 10$^4$ M$_\odot$ \citep{Carretta_2010}. According to \cite{PortegiesZwart_2010}), clusters with lower masses tend not to survive more than 1 Gyr.

\cite{Milone_2022} argued that detecting the presence of multiple populations in star clusters younger than $\sim$~2~Gyr is challenging from an observational standpoint, due to the colour and magnitudes of main sequence stars being influenced by stellar rotation. Through the use of photometric data and high-resolution spectroscopy, they deduced that these young clusters exhibit chemical homogeneity. The key object in this argument is an intermediate-age (1-3 Gyr) GC, NGC~1806, analysed by \cite{Mucciarelli_2014}. Their study indicated no variations in C and N within this cluster, implying the absence of multiple populations. However, a recent detection of  multiple stellar populations on the main sequence in the colour-magnitude diagram has been reported for an intermediate-age ($\sim$ 1.5~Gyr) star cluster,  NGC 1783 in the LMC \citep{Cadelano_2022}. This is the first observation of light-element variations in a star cluster younger than 2~Gyr and suggests that multiple populations could potentially form in clusters at any cosmic age.

\cite{Krause_2016} suggested compactness rather than age as secondary parameter for the expression of multiple populations in a cluster. They calculated a compactness index denoted as C$_5$ in their Table B.1 for GCs displaying multiple populations. This index is formulated as C$_5$~$\equiv$~($M_*$/10$^5$~M$_{\odot}$)/($r_\mathrm{h}$/pc), with $M_*$ representing the stellar mass of the cluster and $r_\mathrm{h}$ denoting its half-mass radius. \cite{Krause_2016} find a possibly metallicity-dependent threshold of C$_5$ $\approx$ 0.1, where multiple populations would occur above this value. The physical explanation could be that gas expulsion during the formation epoch becomes much more difficult above this value, so that the longer-remaining gas could break down ejecta from the polluter \citep[see also Section 4.1 in][]{Gieles_2018}. In the sample of \cite{Krause_2016}, NGC~288 is the multiple population cluster with the lowest value of C$_5$, which is 0.05. This aligns with the general trend of clusters expanding with age, suggesting a higher C$_5$ during its YMC phase. NGC~1806 \citep{Baumgardt_2018}, the homogeneous cluster mentioned above, also has a C$_5$ value of 0.05, but is much younger ($\sim$~2~Gyr vs. $\sim$~12~Gyr). On the other hand, Rup 106, a massive star cluster without multiple populations, has a C$_5$~=~0.06. We also checked the C$_5$ index for the intermediate-age globular clusters that do not exhibit multiple populations \citep{Baumgardt_2018}: NGC~1755 \citep{Milone_2016} with C$_5$~=~0.04, NGC~419 \citep{Martocchia_2017} with C$_5$~=~0.09 and NGC~1806 \citep{Mucciarelli_2014} with C$_5$~=~0.05. Based on this criterion, we selected only YMCs with C$_5$ > 0.10, narrowing down the initial set of approximately 4000 candidates to 101. The complete list of the selected star clusters and their properties can be found in Table~\ref{tab:YMC_table}.

\begin{table*}
\renewcommand{\arraystretch}{1.16} 
\centering
\caption{Properties of selected young massive star clusters in the Local Group. \\
References: 1, \protect\cite{PortegiesZwart_2010} and references therein; 2, \protect\cite{Crowther_2016}; 3, \protect\cite{Pietrzynski_2013}; 4, \protect\cite{DeCia_2024}; 5, \protect\cite{Hunter_1995}; 6, \protect\cite{Andersen_2009}, 7, \protect\cite{Leroy_2018}; 8, \protect\cite{Mills_2021}; 9, \protect\cite{RicoVillas_2020}; 10, \protect\cite{Rekola_2005}; 11, \protect\cite{Beck_2022}; 12, \protect\cite{Levy_2021}; 13, \protect\cite{Ostlin_2007}; 14, \protect\cite{Bik_2018}; 15, \protect\cite{McQuinn_2016}; 16, \protect\cite{Bastian_2008}; 17, \protect\cite{Kaleida_2010}; 18, \protect\cite{Bastian_2006}; 19, \protect\cite{McCrady_2003}; 20, \protect\cite{Freedman_1994}; 21, \protect\cite{Schreiber_2001}; 22, \protect\cite{Smith_2006} ; 23, \protect\cite{Moll_2007}; 24, \protect\cite{Mengel_2003} ; 25, \protect\cite{Mengel_2008}; 26,  \protect\cite{deGrijs_2005}; 27, \protect\cite{MaizApellaniz_2001}; 28, \protect\cite{Israel_1988}; 29, \protect\cite{Hunter_2000}; 30, \protect\cite{Schweizer_2008}; 31, \protect\cite{Lardo_2015}; 32, \protect\cite{Mengel_2002}; 33, \protect\cite{Sabbi_2018}; 34, \protect\cite{Engelbracht_2008}; 35, \protect\cite{Adamo_2017}; 36, \protect\cite{Calzetti_2015}; 37, \protect\cite{Cook_2019}; 38, \protect\cite{Freedman_1988}; 39, \protect\cite{Bresolin_2022}; 40, \protect\cite{Dopita_2010}; 41, \protect\cite{Kobulnicky_1996}; 42, \protect\cite{Kobulnicky_1999}.\\
Uncertainties estimated based on: $^*$ minimum and maximum values found in the literature, $^\mathrm{a}$ NGC 253 distance values, $^\mathrm{b}$ NGC 4038 metallicity values, $^\mathrm{c}$ M82 mass values, $^\mathrm{d}$ ESO338-IG age values.}
\label{tab:YMC_table}
\begin{tabular}{lcllllccr}
\hline
Galaxy & Name / Legus ID & \multicolumn{1}{c}{Age} & \multicolumn{1}{c}{log Mass} & \multicolumn{1}{c}{Distance} & \multicolumn{1}{c}{Metallicity} & $r_\mathrm{h}$ & $C_5$ & References  \\
&  &\multicolumn{1}{c}{[Myr]}& \multicolumn{1}{c}{[M$_{\odot}$]} & \multicolumn{1}{c}{[Mpc]} & \multicolumn{1}{c}{[Z$_{\odot}$]} & [pc] \\
\hline
LMC & R136 & 1.50$\begin{smallmatrix}+0.30 \\-0.70 \end{smallmatrix}$  & 4.78$\begin{smallmatrix}+0.22 \\-0.44 \end{smallmatrix}^*$  & 0.04997$\begin{smallmatrix}+0.00019 \\-0.00019 \end{smallmatrix}$ & 0.47$\begin{smallmatrix}+0.46 \\-0.24 \end{smallmatrix}$  & 2.27 & 0.27 &  1, 2, 3, 4, 5, 6\\
NGC 253 & 4 & 1.50$\begin{smallmatrix}+1.50 \\-1.40 \end{smallmatrix}^*$ & 5.00$\begin{smallmatrix}+0.2 \\-0.0 \end{smallmatrix}^*$ & 3.34$\begin{smallmatrix}+0.26 \\-0.38 \end{smallmatrix}$ & 1.00$\begin{smallmatrix}+0.50 \\-0.50 \end{smallmatrix}^*$ & 0.85 & 1.18 & 7, 8, 9, 10, 11, 12\\
NGC 253 & 5 & 1.50$\begin{smallmatrix}+1.50 \\-1.40 \end{smallmatrix}^*$ & 5.40$\begin{smallmatrix}+0.0 \\-0.2 \end{smallmatrix}^*$ & 3.34$\begin{smallmatrix}+0.26 \\-0.38 \end{smallmatrix}$ & 1.00$\begin{smallmatrix}+0.50 \\-0.50 \end{smallmatrix}^*$ & 1.30 & 1.93 & 7, 8, 9, 10, 11, 12\\
NGC 253 & 6 & 1.60$\begin{smallmatrix}+1.40 \\-1.50 \end{smallmatrix}^*$ & 5.30$\begin{smallmatrix}+0.0 \\-0.5 \end{smallmatrix}^*$ & 3.34$\begin{smallmatrix}+0.26 \\-0.38 \end{smallmatrix}$ & 1.00$\begin{smallmatrix}+0.50 \\-0.50 \end{smallmatrix}^*$ & - & - & 7, 8, 9, 10, 11\\
NGC 253 & 9 & 1.60$\begin{smallmatrix}+1.40 \\-1.50 \end{smallmatrix}^*$ & 5.50$\begin{smallmatrix}+0.0 \\-0.3 \end{smallmatrix}^*$ & 3.34$\begin{smallmatrix}+0.26 \\-0.38 \end{smallmatrix}$ & 1.00$\begin{smallmatrix}+0.50 \\-0.50 \end{smallmatrix}^*$ & 0.78 & 4.05 & 7, 8, 9, 10, 11, 12\\
NGC 253 & 10 & 1.60$\begin{smallmatrix}+1.40 \\-1.50 \end{smallmatrix}^*$ & 5.30$\begin{smallmatrix}+0.0 \\-0.7 \end{smallmatrix}^*$ & 3.34$\begin{smallmatrix}+0.26 \\-0.38 \end{smallmatrix}$ & 1.00$\begin{smallmatrix}+0.50 \\-0.50 \end{smallmatrix}^*$ & 2.24 & 0.89 & 7, 8, 9, 10, 11, 12\\
NGC 253 & 11 & 1.50$\begin{smallmatrix}+1.50 \\-1.40 \end{smallmatrix}^*$ & 5.60$\begin{smallmatrix}+0.0 \\-0.5 \end{smallmatrix}^*$ & 3.34$\begin{smallmatrix}+0.26 \\-0.38 \end{smallmatrix}$ & 1.00$\begin{smallmatrix}+0.50 \\-0.50 \end{smallmatrix}^*$ & 0.22 & 18.10 & 7, 8, 9, 10, 11, 12\\
NGC 253 & 12 & 1.60$\begin{smallmatrix}+1.40 \\-1.50 \end{smallmatrix}^*$ & 6.00$\begin{smallmatrix}+0.0 \\-1.1 \end{smallmatrix}^*$ & 3.34$\begin{smallmatrix}+0.26 \\-0.38 \end{smallmatrix}$ & 1.00$\begin{smallmatrix}+0.50 \\-0.50 \end{smallmatrix}^*$ & 2.19 & 4.57 & 7, 8, 9, 10, 11, 12\\
NGC 253 & 14 & 1.50$\begin{smallmatrix}+1.50 \\-1.40 \end{smallmatrix}^*$ & 5.50$\begin{smallmatrix}+0.0 \\-0.2 \end{smallmatrix}^*$ & 3.34$\begin{smallmatrix}+0.26 \\-0.38 \end{smallmatrix}$ & 1.00$\begin{smallmatrix}+0.50 \\-0.50 \end{smallmatrix}^*$ & 0.90 & 3.51 & 7, 8, 9, 10, 11, 12\\
ESO338-IG & 23 & 6.00$\begin{smallmatrix}+4.00 \\-2.00 \end{smallmatrix}$ & 6.70$\begin{smallmatrix}+1.1 \\-0.1 \end{smallmatrix}^*$ & 37.50$\begin{smallmatrix}+2.92\\-4.27 \end{smallmatrix}^\mathrm{a}$ & 0.12$\begin{smallmatrix}+0.01 \\-0.01 \end{smallmatrix}^\mathrm{b}$ & 8.84 & 5.67 &  1, 13, 14\\
M51 & 3c1-b & 5.00$\begin{smallmatrix}+8.90 \\-4.00 \end{smallmatrix}$ & 5.91$\begin{smallmatrix}+0.09 \\-0.12 \end{smallmatrix}^\mathrm{c}$ & 8.58$\begin{smallmatrix}+0.10 \\-0.10 \end{smallmatrix}$ & 2.50$\begin{smallmatrix}+0.50 \\-0.50 \end{smallmatrix}^*$ & 3.91 & 2.08 & 1, 15, 16, 17\\
M51 & a1 & 5.00$\begin{smallmatrix}+6.00 \\-4.00 \end{smallmatrix}$ & 5.47$\begin{smallmatrix}+0.08 \\-0.11 \end{smallmatrix}^\mathrm{c}$ & 8.58$\begin{smallmatrix}+0.10 \\-0.10 \end{smallmatrix}$ & 2.50$\begin{smallmatrix}+0.50 \\-0.50 \end{smallmatrix}^*$ & 7.14 & 0.41 &  1, 15, 16, 17\\
M82 & MGG 9 & 7.94$\begin{smallmatrix}+3.28 \\-2.32 \end{smallmatrix}$ & 6.18$\begin{smallmatrix}+0.08 \\-0.10 \end{smallmatrix}$ & 3.63$\begin{smallmatrix}+0.34 \\-0.34 \end{smallmatrix}$ & 1.00$\begin{smallmatrix}+0.08 \\-0.06 \end{smallmatrix}^\mathrm{b}$ & 4.42 & 3.42 &  1, 18, 19, 20, 21\\
M82 & A1 & 6.40$\begin{smallmatrix}+0.50 \\-0.50 \end{smallmatrix}$ & 6.00$\begin{smallmatrix}+0.11 \\-0.15 \end{smallmatrix}$ & 3.63$\begin{smallmatrix}+0.34 \\-0.34 \end{smallmatrix}$ & 1.00$\begin{smallmatrix}+0.08 \\-0.06 \end{smallmatrix}^\mathrm{b}$ & 5.10 & 1.96 &  1, 20, 21, 22\\
NGC 1140 & 1 & 5.00$\begin{smallmatrix}+1.00 \\-1.00 \end{smallmatrix}$ & 6.04$\begin{smallmatrix}+0.11 \\-0.14 \end{smallmatrix}$ & 20.00$\begin{smallmatrix}+1.56\\-2.28 \end{smallmatrix}^\mathrm{a}$ & 0.57$\begin{smallmatrix}+0.04 \\-0.03  \end{smallmatrix}^\mathrm{b}$ & 13.60 & 0.81 &  1, 23\\
NGC 1487 & 2 & 8.50$\begin{smallmatrix}+0.50 \\-0.50 \end{smallmatrix}$ & 5.20$\begin{smallmatrix}+0.12 \\-0.05 \end{smallmatrix}$ & 9.30$\begin{smallmatrix}+0.72\\-1.06 \end{smallmatrix}^\mathrm{a}$ & 0.25$\begin{smallmatrix}+0.15 \\-0.10 \end{smallmatrix}^*$ & 2.04 & 0.78 & 1, 24, 25, 26\\
NGC 1487 & 1 & 8.10$\begin{smallmatrix}+0.50 \\-0.50 \end{smallmatrix}$ & 5.18$\begin{smallmatrix}+0.12 \\-0.07 \end{smallmatrix}$ & 9.30$\begin{smallmatrix}+0.72\\-1.06 \end{smallmatrix}^\mathrm{a}$ & 0.25$\begin{smallmatrix}+0.15 \\-0.10 \end{smallmatrix}^*$ & 3.91 & 0.39 & 1, 24, 25, 26\\
NGC 1487 & 3 & 7.90$\begin{smallmatrix}+0.50 \\-0.50 \end{smallmatrix}$ & 4.88$\begin{smallmatrix}+0.18 \\-0.22 \end{smallmatrix}$ & 9.30$\begin{smallmatrix}+0.72\\-1.06 \end{smallmatrix}^\mathrm{a}$ & 0.25$\begin{smallmatrix}+0.15 \\-0.10 \end{smallmatrix}^*$ & 3.57 & 0.21 & 1, 24, 25, 26\\
NGC 1569 & C & 3.00$\begin{smallmatrix}+2.00 \\-2.00 \end{smallmatrix}$ & 5.16$\begin{smallmatrix}+0.08 \\-0.11 \end{smallmatrix}^\mathrm{c}$ & 2.20$\begin{smallmatrix}+0.60 \\-0.60 \end{smallmatrix}$ & 0.29$\begin{smallmatrix}+0.02 \\-0.02 \end{smallmatrix}^\mathrm{b}$ & 4.93 & 0.29 & 1, 27, 28, 29\\
NGC 4038 & S2\textunderscore1 & 9.00$\begin{smallmatrix}+0.30 \\-0.30 \end{smallmatrix}$ & 5.47$\begin{smallmatrix}+0.15 \\-0.09 \end{smallmatrix}$ & 22.00$\begin{smallmatrix}+3.00 \\-3.00 \end{smallmatrix}$ & 1.17$\begin{smallmatrix}+0.09 \\-0.07 \end{smallmatrix}$ & 6.29 & 0.47 & 1, 25, 30, 31\\
NGC 4038 & W99-1 & 8.10$\begin{smallmatrix}+0.50 \\-0.50 \end{smallmatrix}$ & 5.86$\begin{smallmatrix}+0.09 \\-0.12  \end{smallmatrix}^\mathrm{c}$ & 22.00$\begin{smallmatrix}+3.00 \\-3.00 \end{smallmatrix}$ & 1.17$\begin{smallmatrix}+0.09 \\-0.07 \end{smallmatrix}$ & 6.12 & 1.18 & 1, 30, 31, 32\\
NGC 4038 & W99-16 & 10.00$\begin{smallmatrix}+2.00 \\-2.00 \end{smallmatrix}$ & 5.46$\begin{smallmatrix}+0.08 \\-0.11  \end{smallmatrix}^\mathrm{c}$ & 22.00$\begin{smallmatrix}+3.00 \\-3.00 \end{smallmatrix}$ & 1.17$\begin{smallmatrix}+0.09 \\-0.07 \end{smallmatrix}$ & 10.20 & 0.28 & 1, 30, 31, 32\\
NGC 4038 & W99-2 & 6.60$\begin{smallmatrix}+0.30 \\-0.30 \end{smallmatrix}$ & 6.43$\begin{smallmatrix}+0.11 \\-0.09 \end{smallmatrix}$ & 22.00$\begin{smallmatrix}+3.00 \\-3.00 \end{smallmatrix}$ & 1.17$\begin{smallmatrix}+0.09 \\-0.07 \end{smallmatrix}$ & 13.60 & 1.98 & 1, 25, 30, 31\\
NGC 4038 & W99-15 & 8.70$\begin{smallmatrix}+0.30 \\-0.30 \end{smallmatrix}$ & 5.70$\begin{smallmatrix}+0.08 \\-0.10 \end{smallmatrix}$ & 22.00$\begin{smallmatrix}+3.00 \\-3.00 \end{smallmatrix}$ & 1.17$\begin{smallmatrix}+0.09 \\-0.07 \end{smallmatrix}$ & 2.38 & 2.11 & 1, 25, 30, 31\\
NGC 4038 & S1\textunderscore1 & 8.00$\begin{smallmatrix}+0.30 \\-0.30 \end{smallmatrix}$ & 5.85$\begin{smallmatrix}+0.10 \\-0.07 \end{smallmatrix}$ & 22.00$\begin{smallmatrix}+3.00 \\-3.00 \end{smallmatrix}$ & 1.17$\begin{smallmatrix}+0.09 \\-0.07 \end{smallmatrix}$ & 6.12 & 1.16 & 1, 25, 30, 31\\
NGC 4038 & S1\textunderscore2 & 8.30$\begin{smallmatrix}+0.30 \\-0.30 \end{smallmatrix}$ & 5.70$\begin{smallmatrix}+0.20 \\-0.10 \end{smallmatrix}$ & 22.00$\begin{smallmatrix}+3.00 \\-3.00 \end{smallmatrix}$ & 1.17$\begin{smallmatrix}+0.09 \\-0.07 \end{smallmatrix}$ & 6.12 & 0.82 & 1, 25, 30, 31\\
NGC 4038 & S1\textunderscore5 & 8.50$\begin{smallmatrix}+0.30 \\-0.30 \end{smallmatrix}$ & 5.48$\begin{smallmatrix}+0.05 \\-0.08 \end{smallmatrix}$ & 22.00$\begin{smallmatrix}+3.00 \\-3.00 \end{smallmatrix}$ & 1.17$\begin{smallmatrix}+0.09 \\-0.07 \end{smallmatrix}$ & 1.53 & 1.97 & 1, 25, 30, 31\\
NGC 4038 & 2000\textunderscore1 & 8.50$\begin{smallmatrix}+0.30 \\-0.30 \end{smallmatrix}$ & 6.23$\begin{smallmatrix}+0.18 \\-0.19 \end{smallmatrix}$ & 22.00$\begin{smallmatrix}+3.00 \\-3.00 \end{smallmatrix}$ & 1.17$\begin{smallmatrix}+0.09 \\-0.07 \end{smallmatrix}$ & 6.12 & 2.77 & 1, 25, 30, 31\\
NGC 4038 & S2\textunderscore2 & 9.00$\begin{smallmatrix}+0.30 \\-0.30 \end{smallmatrix}$ & 5.60$\begin{smallmatrix}+0.08 \\-0.03 \end{smallmatrix}$ & 22.00$\begin{smallmatrix}+3.00 \\-3.00 \end{smallmatrix}$ & 1.17$\begin{smallmatrix}+0.09 \\-0.07 \end{smallmatrix}$ & 4.25 & 0.94 & 1, 25, 30, 31\\
NGC 4038 & S2\textunderscore3 & 9.00$\begin{smallmatrix}+0.30 \\-0.30 \end{smallmatrix}$ & 5.38$\begin{smallmatrix}+0.11 \\-0.06 \end{smallmatrix}$ & 22.00$\begin{smallmatrix}+3.00 \\-3.00 \end{smallmatrix}$ & 1.17$\begin{smallmatrix}+0.09 \\-0.07 \end{smallmatrix}$ & 5.10 & 0.47 & 1, 25, 30, 31\\
NGC 4449 & N-2 & 3.00$\begin{smallmatrix}+2.00 \\-2.00 \end{smallmatrix}$ & 5.00$\begin{smallmatrix}+0.08 \\-0.10  \end{smallmatrix}^\mathrm{c}$ & 4.01$\begin{smallmatrix}+0.30 \\-0.30 \end{smallmatrix}$ & 0.33$\begin{smallmatrix}+0.01 \\-0.01 \end{smallmatrix}$ & 9.86 & 0.10 & 1, 27, 33, 34\\
NGC 4449 & 76 & 5.00$\begin{smallmatrix}+3.33 \\-1.67  \end{smallmatrix}^\mathrm{d}$ & 5.24$\begin{smallmatrix}+0.61 \\-0.09 \end{smallmatrix}^*$ & 4.01$\begin{smallmatrix}+0.30 \\-0.30 \end{smallmatrix}$ & 0.33$\begin{smallmatrix}+0.01 \\-0.01 \end{smallmatrix}$ & 3.82 & 0.45 & 33, 34, 35, 36, 37\\
NGC 4449 & 132 & 5.00$\begin{smallmatrix}+3.33 \\-1.67  \end{smallmatrix}^\mathrm{d}$ & 5.38$\begin{smallmatrix}+0.62 \\-0.09 \end{smallmatrix}^*$ & 4.01$\begin{smallmatrix}+0.30 \\-0.30 \end{smallmatrix}$ & 0.33$\begin{smallmatrix}+0.01 \\-0.01 \end{smallmatrix}$ & 3.77 & 0.64 & 33, 34, 35, 36, 37\\
NGC 4449 & 181 & 5.00$\begin{smallmatrix}+3.33 \\-1.67  \end{smallmatrix}^\mathrm{d}$ & 4.74$\begin{smallmatrix}+0.99 \\-0.11 \end{smallmatrix}^*$ & 4.01$\begin{smallmatrix}+0.30 \\-0.30 \end{smallmatrix}$ & 0.33$\begin{smallmatrix}+0.01 \\-0.01 \end{smallmatrix}$ & 3.38 & 0.16 & 33, 34, 35, 36, 37\\
NGC 4449 & 186 & 5.00$\begin{smallmatrix}+1.00 \\-0.00  \end{smallmatrix}^*$ & 4.68$\begin{smallmatrix}+0.03 \\-0.11 \end{smallmatrix}^*$ & 4.01$\begin{smallmatrix}+0.30 \\-0.30 \end{smallmatrix}$ & 0.33$\begin{smallmatrix}+0.01 \\-0.01 \end{smallmatrix}$ & 2.95 & 0.16 & 33, 34, 35, 36, 37\\
NGC 4449 & 1497 & 10.00$\begin{smallmatrix}+6.67 \\-3.33  \end{smallmatrix}^\mathrm{d}$ & 4.81$\begin{smallmatrix}+0.03 \\-0.02 \end{smallmatrix}^*$ & 4.01$\begin{smallmatrix}+0.30 \\-0.30 \end{smallmatrix}$ & 0.33$\begin{smallmatrix}+0.01 \\-0.01 \end{smallmatrix}$ & 2.79 & 0.23 & 33, 34, 35, 36, 37\\
NGC 4449 & 1673 & 10.00$\begin{smallmatrix}+6.67 \\-3.33  \end{smallmatrix}^\mathrm{d}$ & 4.93$\begin{smallmatrix}+0.02 \\-0.03 \end{smallmatrix}^*$ & 4.01$\begin{smallmatrix}+0.30 \\-0.30 \end{smallmatrix}$ & 0.33$\begin{smallmatrix}+0.01 \\-0.01 \end{smallmatrix}$ & 6.18 & 0.14 & 33, 34, 35, 36, 37\\
NGC 4449 & 1828 & 10.00$\begin{smallmatrix}+6.67 \\-3.33  \end{smallmatrix}^\mathrm{d}$ & 5.40$\begin{smallmatrix}+0.03 \\-0.03 \end{smallmatrix}^*$ & 4.01$\begin{smallmatrix}+0.30 \\-0.30 \end{smallmatrix}$ & 0.33$\begin{smallmatrix}+0.01 \\-0.01 \end{smallmatrix}$ & 6.15 & 0.41 & 33, 34, 35, 36, 37\\
NGC 2403 & II & 4.50$\begin{smallmatrix}+2.50 \\-2.50 \end{smallmatrix}$ & 5.35$\begin{smallmatrix}+0.08 \\-0.11  \end{smallmatrix}^\mathrm{c}$ & 3.20$\begin{smallmatrix}+0.40 \\-0.40 \end{smallmatrix}$ & 0.56$\begin{smallmatrix}+0.04 \\-0.03  \end{smallmatrix}^\mathrm{b}$ & 20.06 & 0.11 & 1, 27, 38, 39\\
NGC 4214 & I-A & 3.50$\begin{smallmatrix}+0.50 \\-0.50 \end{smallmatrix}$ & 5.44$\begin{smallmatrix}+0.08 \\-0.11  \end{smallmatrix}^\mathrm{c}$ & 2.98$\begin{smallmatrix}+0.13 \\-0.13 \end{smallmatrix}$ & 0.25$\begin{smallmatrix}+0.02 \\-0.01  \end{smallmatrix}^\mathrm{b}$ & 28.05 & 0.10 & 1, 27, 40, 41\\
NGC 1313 & 640 & 1.00$\begin{smallmatrix}+1.00 \\-0.00 \end{smallmatrix}^*$ & 4.70$\begin{smallmatrix}+0.03 \\-0.08 \end{smallmatrix}^*$ & 4.30$\begin{smallmatrix}+0.24 \\-0.24 \end{smallmatrix}$ & 0.57$\begin{smallmatrix}+0.04 \\-0.03  \end{smallmatrix}^\mathrm{b}$ & 3.88 & 0.13 & 33, 35, 36, 37\\
NGC 1313 & 2419 & 8.00$\begin{smallmatrix}+5.33 \\-2.67  \end{smallmatrix}^\mathrm{d}$ & 4.58$\begin{smallmatrix}+0.07 \\-0.09  \end{smallmatrix}^\mathrm{c}$ & 4.30$\begin{smallmatrix}+0.24 \\-0.24 \end{smallmatrix}$ & 0.57$\begin{smallmatrix}+0.04 \\-0.03  \end{smallmatrix}^\mathrm{b}$ & 1.13 & 0.33 & 33, 35, 36, 37\\
NGC 1433 & 155 & 2.00$\begin{smallmatrix}+1.33 \\-0.67  \end{smallmatrix}^\mathrm{d}$ & 4.72$\begin{smallmatrix}+0.07 \\-0.10  \end{smallmatrix}^\mathrm{c}$ & 9.10$\begin{smallmatrix}+1.00 \\-1.00 \end{smallmatrix}$ & 1.43$\begin{smallmatrix}+0.11 \\-0.09  \end{smallmatrix}^\mathrm{b}$ & 3.66 & 0.15 & 33, 35, 36, 37\\
NGC 1433 & 194 & 1.00$\begin{smallmatrix}+1.00 \\-0.00 \end{smallmatrix}^*$ & 4.64$\begin{smallmatrix}+0.07 \\-0.10  \end{smallmatrix}^\mathrm{c}$ & 9.10$\begin{smallmatrix}+1.00 \\-1.00 \end{smallmatrix}$ & 1.43$\begin{smallmatrix}+0.11 \\-0.09  \end{smallmatrix}^\mathrm{b}$ & 1.64 & 0.27 & 33, 35, 36, 37\\
NGC 1433 & 213 & 2.00$\begin{smallmatrix}+1.00 \\-0.00 \end{smallmatrix}^*$ & 4.51$\begin{smallmatrix}+0.07 \\-0.09  \end{smallmatrix}^\mathrm{c}$ & 9.10$\begin{smallmatrix}+1.00 \\-1.00 \end{smallmatrix}$ & 1.43$\begin{smallmatrix}+0.11 \\-0.09  \end{smallmatrix}^\mathrm{b}$ & 3.10 & 0.10 & 33, 35, 36, 37\\
NGC 1433 & 1005 & 2.00$\begin{smallmatrix}+1.00 \\-0.00 \end{smallmatrix}^*$ & 4.87$\begin{smallmatrix}+0.07 \\-0.10  \end{smallmatrix}^\mathrm{c}$ & 9.10$\begin{smallmatrix}+1.00 \\-1.00 \end{smallmatrix}$ & 1.43$\begin{smallmatrix}+0.11 \\-0.09  \end{smallmatrix}^\mathrm{b}$ & 4.28 & 0.17 & 33, 35, 36, 37\\
\hline
\end{tabular}
\end{table*}

\begin{table*}
\renewcommand{\arraystretch}{1.16} 
\centering
\contcaption{}
\begin{tabular}{lcllllccr}
\hline
Galaxy & Name / Legus ID & \multicolumn{1}{c}{Age} & \multicolumn{1}{c}{log Mass} & \multicolumn{1}{c}{Distance} & \multicolumn{1}{c}{Metallicity} & $r_\mathrm{h}$ & $C_5$ & References  \\
&  &\multicolumn{1}{c}{[Myr]}& \multicolumn{1}{c}{[M$_{\odot}$]} & \multicolumn{1}{c}{[Mpc]} & \multicolumn{1}{c}{[Z$_{\odot}$]} & [pc] \\
\hline
NGC 1566 & 171 & 1.00$\begin{smallmatrix}+0.67 \\-0.33  \end{smallmatrix}^\mathrm{d}$ & 4.88$\begin{smallmatrix}+0.10 \\-0.66 \end{smallmatrix}^*$ & 15.60$\begin{smallmatrix}+0.60 \\-0.60 \end{smallmatrix}$ & 1.43$\begin{smallmatrix}+0.11 \\-0.09  \end{smallmatrix}^\mathrm{b}$ & 6.59 & 0.12 & 33, 35, 36, 37\\
NGC 1566 & 790 & 9.00$\begin{smallmatrix}+6.00 \\-3.00  \end{smallmatrix}^\mathrm{d}$ & 4.88$\begin{smallmatrix}+0.03 \\-0.02 \end{smallmatrix}^*$ & 15.60$\begin{smallmatrix}+0.60 \\-0.60 \end{smallmatrix}$ & 1.43$\begin{smallmatrix}+0.11 \\-0.09  \end{smallmatrix}^\mathrm{b}$ & 7.37 & 0.10 & 33, 35, 36, 37\\
NGC 1566 & 792 & 1.00$\begin{smallmatrix}+0.67 \\-0.33  \end{smallmatrix}^\mathrm{d}$ & 5.21$\begin{smallmatrix}+0.05 \\-0.05 \end{smallmatrix}^*$ & 15.60$\begin{smallmatrix}+0.60 \\-0.60 \end{smallmatrix}$ & 1.43$\begin{smallmatrix}+0.11 \\-0.09  \end{smallmatrix}^\mathrm{b}$ & 13.36 & 0.12 & 33, 35, 36, 37\\
NGC 1566 & 869 & 9.00$\begin{smallmatrix}+6.00 \\-3.00  \end{smallmatrix}^\mathrm{d}$ & 5.15$\begin{smallmatrix}+0.03 \\-0.05 \end{smallmatrix}^*$ & 15.60$\begin{smallmatrix}+0.60 \\-0.60 \end{smallmatrix}$ & 1.43$\begin{smallmatrix}+0.11 \\-0.09  \end{smallmatrix}^\mathrm{b}$ & 14.66 & 0.10 & 33, 35, 36, 37\\
NGC 1566 & 1094 & 1.00$\begin{smallmatrix}+0.67 \\-0.33  \end{smallmatrix}^\mathrm{d}$ & 4.93$\begin{smallmatrix}+0.08 \\-0.09 \end{smallmatrix}^*$ & 15.60$\begin{smallmatrix}+0.60 \\-0.60 \end{smallmatrix}$ & 1.43$\begin{smallmatrix}+0.11 \\-0.09  \end{smallmatrix}^\mathrm{b}$ & 4.52 & 0.19 & 33, 35, 36, 37\\
NGC 1566 & 1565 & 9.00$\begin{smallmatrix}+6.00 \\-0.00 \end{smallmatrix}^*$ & 4.99$\begin{smallmatrix}+0.26 \\-0.00 \end{smallmatrix}^*$ & 15.60$\begin{smallmatrix}+0.60 \\-0.60 \end{smallmatrix}$ & 1.43$\begin{smallmatrix}+0.11 \\-0.09  \end{smallmatrix}^\mathrm{b}$ & 4.09 & 0.24 & 33, 35, 36, 37\\
NGC 1566 & 1579 & 3.00$\begin{smallmatrix}+2.00 \\-1.00  \end{smallmatrix}^\mathrm{d}$ & 4.89$\begin{smallmatrix}+0.29 \\-0.59 \end{smallmatrix}^*$ & 15.60$\begin{smallmatrix}+0.60 \\-0.60 \end{smallmatrix}$ & 1.43$\begin{smallmatrix}+0.11 \\-0.09  \end{smallmatrix}^\mathrm{b}$ & 5.08 & 0.15 & 33, 35, 36, 37\\
NGC 1566 & 1626 & 5.00$\begin{smallmatrix}+3.33 \\-1.67  \end{smallmatrix}^\mathrm{d}$ & 4.67$\begin{smallmatrix}+0.54 \\-0.60 \end{smallmatrix}^*$ & 15.60$\begin{smallmatrix}+0.60 \\-0.60 \end{smallmatrix}$ & 1.43$\begin{smallmatrix}+0.11 \\-0.09  \end{smallmatrix}^\mathrm{b}$ & 2.58 & 0.18 & 33, 35, 36, 37\\
NGC 1566 & 1627 & 1.00$\begin{smallmatrix}+2.00 \\-0.00 \end{smallmatrix}^*$ & 4.89$\begin{smallmatrix}+0.05 \\-0.18 \end{smallmatrix}^*$ & 15.60$\begin{smallmatrix}+0.60 \\-0.60 \end{smallmatrix}$ & 1.43$\begin{smallmatrix}+0.11 \\-0.09  \end{smallmatrix}^\mathrm{b}$ & 3.54 & 0.22 & 33, 35, 36, 37\\
NGC 1566 & 1916 & 9.00$\begin{smallmatrix}+6.00 \\-3.00  \end{smallmatrix}^\mathrm{d}$ & 4.91$\begin{smallmatrix}+0.29 \\-0.03 \end{smallmatrix}^*$ & 15.60$\begin{smallmatrix}+0.60 \\-0.60 \end{smallmatrix}$ & 1.43$\begin{smallmatrix}+0.11 \\-0.09  \end{smallmatrix}^\mathrm{b}$ & 8.37 & 0.10 & 33, 35, 36, 37\\
NGC 1566 & 2039 & 1.00$\begin{smallmatrix}+0.67 \\-0.33  \end{smallmatrix}^\mathrm{d}$ & 4.97$\begin{smallmatrix}+0.05 \\-0.04 \end{smallmatrix}^*$ & 15.60$\begin{smallmatrix}+0.60 \\-0.60 \end{smallmatrix}$ & 1.43$\begin{smallmatrix}+0.11 \\-0.09  \end{smallmatrix}^\mathrm{b}$ & 5.66 & 0.17 & 33, 35, 36, 37\\
NGC 1566 & 2151 & 1.00$\begin{smallmatrix}+0.67 \\-0.33  \end{smallmatrix}^\mathrm{d}$ & 5.26$\begin{smallmatrix}+0.07 \\-0.63 \end{smallmatrix}^*$ & 15.60$\begin{smallmatrix}+0.60 \\-0.60 \end{smallmatrix}$ & 1.43$\begin{smallmatrix}+0.11 \\-0.09  \end{smallmatrix}^\mathrm{b}$ & 2.90 & 0.62 & 33, 35, 36, 37\\
NGC 3351 & 1372 & 4.00$\begin{smallmatrix}+2.67 \\-1.33  \end{smallmatrix}^\mathrm{d}$ & 5.28$\begin{smallmatrix}+0.03 \\-0.03 \end{smallmatrix}^*$ & 9.30$\begin{smallmatrix}+0.90 \\-0.90 \end{smallmatrix}$ & 1.43$\begin{smallmatrix}+0.11 \\-0.09  \end{smallmatrix}^\mathrm{b}$ & 2.43 & 0.79 & 33, 35, 36, 37\\
NGC 3351 & 1377 & 4.00$\begin{smallmatrix}+2.67 \\-1.33  \end{smallmatrix}^\mathrm{d}$ & 5.33$\begin{smallmatrix}+0.03 \\-0.03 \end{smallmatrix}^*$ & 9.30$\begin{smallmatrix}+0.90 \\-0.90 \end{smallmatrix}$ & 1.43$\begin{smallmatrix}+0.11 \\-0.09  \end{smallmatrix}^\mathrm{b}$ & 9.37 & 0.23 & 33, 35, 36, 37\\
NGC 3351 & 1380 & 6.00$\begin{smallmatrix}+0.00 \\-4.00 \end{smallmatrix}^*$ & 4.75$\begin{smallmatrix}+0.41 \\-0.03 \end{smallmatrix}^*$ & 9.30$\begin{smallmatrix}+0.90 \\-0.90 \end{smallmatrix}$ & 1.43$\begin{smallmatrix}+0.11 \\-0.09  \end{smallmatrix}^\mathrm{b}$ & 4.34 & 0.13 & 33, 35, 36, 37\\
NGC 3351 & 1382 & 5.00$\begin{smallmatrix}+1.00 \\-3.00 \end{smallmatrix}^*$ & 5.17$\begin{smallmatrix}+0.11 \\-0.29 \end{smallmatrix}^*$ & 9.30$\begin{smallmatrix}+0.90 \\-0.90 \end{smallmatrix}$ & 1.43$\begin{smallmatrix}+0.11 \\-0.09  \end{smallmatrix}^\mathrm{b}$ & 1.85 & 0.80 & 33, 35, 36, 37\\
NGC 3351 & 1383 & 2.00$\begin{smallmatrix}+1.00 \\-1.00 \end{smallmatrix}^*$ & 4.70$\begin{smallmatrix}+0.07 \\-0.13 \end{smallmatrix}^*$ & 9.30$\begin{smallmatrix}+0.90 \\-0.90 \end{smallmatrix}$ & 1.43$\begin{smallmatrix}+0.11 \\-0.09  \end{smallmatrix}^\mathrm{b}$ & 2.11 & 0.23 & 33, 35, 36, 37\\
NGC 4395 & 588 & 2.00$\begin{smallmatrix}+1.00 \\-1.00 \end{smallmatrix}^*$ & 4.68$\begin{smallmatrix}+0.09 \\-0.08 \end{smallmatrix}^*$ & 4.54$\begin{smallmatrix}+0.18 \\-0.18 \end{smallmatrix}$ & 0.29$\begin{smallmatrix}+0.02 \\-0.02  \end{smallmatrix}^\mathrm{b}$ & 0.60 & 0.80 & 33, 35, 36, 37\\
NGC 4656 & 1208 & 10.00$\begin{smallmatrix}+6.67 \\-3.33  \end{smallmatrix}^\mathrm{d}$ & 5.22$\begin{smallmatrix}+0.02 \\-0.03 \end{smallmatrix}^*$ & 7.90$\begin{smallmatrix}+0.70 \\-0.70 \end{smallmatrix}$ & 0.29$\begin{smallmatrix}+0.02 \\-0.02  \end{smallmatrix}^\mathrm{b}$ & 4.53 & 0.37 & 33, 35, 36, 37\\
NGC 4656 & 1233 & 5.00$\begin{smallmatrix}+3.33 \\-1.67  \end{smallmatrix}^\mathrm{d}$ & 5.12$\begin{smallmatrix}+0.02 \\-0.03 \end{smallmatrix}^*$ & 7.90$\begin{smallmatrix}+0.70 \\-0.70 \end{smallmatrix}$ & 0.29$\begin{smallmatrix}+0.02 \\-0.02  \end{smallmatrix}^\mathrm{b}$ & 2.40 & 0.55 & 33, 35, 36, 37\\
NGC 5194 & 4204 & 4.00$\begin{smallmatrix}+1.00 \\-0.00 \end{smallmatrix}^*$ & 4.81$\begin{smallmatrix}+0.03 \\-0.04 \end{smallmatrix}^*$ & 7.40$\begin{smallmatrix}+0.42 \\-0.42 \end{smallmatrix}$ & 1.43$\begin{smallmatrix}+0.11 \\-0.09  \end{smallmatrix}^\mathrm{b}$ & 4.20 & 0.15 & 33, 35, 36, 37\\
NGC 5194 & 7509 & 1.00$\begin{smallmatrix}+0.67 \\-0.33  \end{smallmatrix}^\mathrm{d}$ & 4.60$\begin{smallmatrix}+0.01 \\-0.02\end{smallmatrix}^*$ & 7.40$\begin{smallmatrix}+0.42 \\-0.42 \end{smallmatrix}$ & 1.43$\begin{smallmatrix}+0.11 \\-0.09  \end{smallmatrix}^\mathrm{b}$ & 2.10 & 0.19 & 33, 35, 36, 37\\
NGC 5194 & 8849 & 7.00$\begin{smallmatrix}+4.67 \\-2.33  \end{smallmatrix}^\mathrm{d}$ & 4.59$\begin{smallmatrix}+0.03 \\-0.02\end{smallmatrix}^*$ & 7.40$\begin{smallmatrix}+0.42 \\-0.42 \end{smallmatrix}$ & 1.43$\begin{smallmatrix}+0.11 \\-0.09  \end{smallmatrix}^\mathrm{b}$ & 1.03 & 0.38 & 33, 35, 36, 37\\
NGC 5194 & 11596 & 4.00$\begin{smallmatrix}+2.67 \\-1.33  \end{smallmatrix}^\mathrm{d}$ & 4.73$\begin{smallmatrix}+0.02 \\-0.03\end{smallmatrix}^*$ & 7.40$\begin{smallmatrix}+0.42 \\-0.42 \end{smallmatrix}$ & 1.43$\begin{smallmatrix}+0.11 \\-0.09  \end{smallmatrix}^\mathrm{b}$ & 1.53 & 0.35 & 33, 35, 36, 37\\
NGC 5194 & 13404 & 1.00$\begin{smallmatrix}+0.67 \\-0.33  \end{smallmatrix}^\mathrm{d}$ & 4.52$\begin{smallmatrix}+0.02 \\-0.02 \end{smallmatrix}^*$ & 7.40$\begin{smallmatrix}+0.42 \\-0.42 \end{smallmatrix}$ & 1.43$\begin{smallmatrix}+0.11 \\-0.09  \end{smallmatrix}^\mathrm{b}$ & 3.47 & 0.10 & 33, 35, 36, 37\\
NGC 5194 & 13671 & 2.00$\begin{smallmatrix}+4.00 \\-0.00 \end{smallmatrix}^*$ & 4.53$\begin{smallmatrix}+0.01 \\-0.40 \end{smallmatrix}^*$ & 7.40$\begin{smallmatrix}+0.42 \\-0.42 \end{smallmatrix}$ & 1.43$\begin{smallmatrix}+0.11 \\-0.09  \end{smallmatrix}^\mathrm{b}$ & 1.28 & 0.26 & 33, 35, 36, 37\\
NGC 5194 & 14243 & 4.00$\begin{smallmatrix}+2.67 \\-1.33  \end{smallmatrix}^\mathrm{d}$ & 4.54$\begin{smallmatrix}+0.01 \\-0.02 \end{smallmatrix}^*$ & 7.40$\begin{smallmatrix}+0.42 \\-0.42 \end{smallmatrix}$ & 1.43$\begin{smallmatrix}+0.11 \\-0.09  \end{smallmatrix}^\mathrm{b}$ & 2.13 & 0.16 & 33, 35, 36, 37\\
NGC 5194 & 14849 & 1.00$\begin{smallmatrix}+1.00 \\-0.00 \end{smallmatrix}^*$ & 5.54$\begin{smallmatrix}+0.01 \\-0.10 \end{smallmatrix}^*$ & 7.40$\begin{smallmatrix}+0.42 \\-0.42 \end{smallmatrix}$ & 1.43$\begin{smallmatrix}+0.11 \\-0.09  \end{smallmatrix}^\mathrm{b}$ & 2.03 & 1.70 & 33, 35, 36, 37\\
NGC 5194 & 15696 & 2.00$\begin{smallmatrix}+0.00 \\-1.00 \end{smallmatrix}^*$ & 4.58$\begin{smallmatrix}+0.08 \\-0.02 \end{smallmatrix}^*$ & 7.40$\begin{smallmatrix}+0.42 \\-0.42 \end{smallmatrix}$ & 1.43$\begin{smallmatrix}+0.11 \\-0.09  \end{smallmatrix}^\mathrm{b}$ & 3.60 & 0.11 & 33, 35, 36, 37\\
NGC 5194 & 16045 & 3.00$\begin{smallmatrix}+2.00 \\-1.00  \end{smallmatrix}^\mathrm{d}$ & 4.89$\begin{smallmatrix}+0.01 \\-0.02 \end{smallmatrix}^*$ & 7.40$\begin{smallmatrix}+0.42 \\-0.42 \end{smallmatrix}$ & 1.43$\begin{smallmatrix}+0.11 \\-0.09  \end{smallmatrix}^\mathrm{b}$ & 1.30 & 0.60 & 33, 35, 36, 37\\
NGC 5194 & 16216 & 4.00$\begin{smallmatrix}+2.67 \\-1.33  \end{smallmatrix}^\mathrm{d}$ & 5.01$\begin{smallmatrix}+0.01 \\-0.00 \end{smallmatrix}^*$ & 7.40$\begin{smallmatrix}+0.42 \\-0.42 \end{smallmatrix}$ & 1.43$\begin{smallmatrix}+0.11 \\-0.09  \end{smallmatrix}^\mathrm{b}$ & 1.84 & 0.55 & 33, 35, 36, 37\\
NGC 5194 & 16866 & 4.00$\begin{smallmatrix}+2.67 \\-1.33  \end{smallmatrix}^\mathrm{d}$ & 5.16$\begin{smallmatrix}+0.00 \\-0.02 \end{smallmatrix}^*$ & 7.40$\begin{smallmatrix}+0.42 \\-0.42 \end{smallmatrix}$ & 1.43$\begin{smallmatrix}+0.11 \\-0.09  \end{smallmatrix}^\mathrm{b}$ & 1.12 & 1.28 & 33, 35, 36, 37\\
NGC 5194 & 16883 & 4.00$\begin{smallmatrix}+0.00 \\-1.00 \end{smallmatrix}^*$ & 5.17$\begin{smallmatrix}+0.02 \\-0.02 \end{smallmatrix}^*$ & 7.40$\begin{smallmatrix}+0.42 \\-0.42 \end{smallmatrix}$ & 1.43$\begin{smallmatrix}+0.11 \\-0.09  \end{smallmatrix}^\mathrm{b}$ & 2.61 & 0.57 & 33, 35, 36, 37\\
NGC 5194 & 17204 & 4.00$\begin{smallmatrix}+2.67 \\-1.33  \end{smallmatrix}^\mathrm{d}$ & 4.77$\begin{smallmatrix}+0.01 \\-0.02 \end{smallmatrix}^*$ & 7.40$\begin{smallmatrix}+0.42 \\-0.42 \end{smallmatrix}$ & 1.43$\begin{smallmatrix}+0.11 \\-0.09  \end{smallmatrix}^\mathrm{b}$ & 0.83 & 0.71 & 33, 35, 36, 37\\
NGC 5194 & 17462 & 3.00$\begin{smallmatrix}+2.00 \\-1.00  \end{smallmatrix}^\mathrm{d}$ & 5.26$\begin{smallmatrix}+0.02 \\-0.01 \end{smallmatrix}^*$ & 7.40$\begin{smallmatrix}+0.42 \\-0.42 \end{smallmatrix}$ & 1.43$\begin{smallmatrix}+0.11 \\-0.09  \end{smallmatrix}^\mathrm{b}$ & 7.08 & 0.26 & 33, 35, 36, 37\\
NGC 5194 & 17512 & 4.00$\begin{smallmatrix}+2.67 \\-1.33  \end{smallmatrix}^\mathrm{d}$ & 4.84$\begin{smallmatrix}+0.01 \\-0.02 \end{smallmatrix}^*$ & 7.40$\begin{smallmatrix}+0.42 \\-0.42 \end{smallmatrix}$ & 1.43$\begin{smallmatrix}+0.11 \\-0.09  \end{smallmatrix}^\mathrm{b}$ & 1.11 & 0.62 & 33, 35, 36, 37\\
NGC 5194 & 17988 & 2.00$\begin{smallmatrix}+1.33 \\-0.67  \end{smallmatrix}^\mathrm{d}$ & 4.56$\begin{smallmatrix}+0.02 \\-0.01 \end{smallmatrix}^*$ & 7.40$\begin{smallmatrix}+0.42 \\-0.42 \end{smallmatrix}$ & 1.43$\begin{smallmatrix}+0.11 \\-0.09  \end{smallmatrix}^\mathrm{b}$ & 1.42 & 0.26 & 33, 35, 36, 37\\
NGC 5194 & 18184 & 2.00$\begin{smallmatrix}+0.00 \\-1.00 \end{smallmatrix}^*$ & 4.62$\begin{smallmatrix}+0.07 \\-0.02 \end{smallmatrix}^*$ & 7.40$\begin{smallmatrix}+0.42 \\-0.42 \end{smallmatrix}$ & 1.43$\begin{smallmatrix}+0.11 \\-0.09  \end{smallmatrix}^\mathrm{b}$ & 0.94 & 0.44 & 33, 35, 36, 37\\
NGC 5194 & 19033 & 5.00$\begin{smallmatrix}+3.33 \\-1.67  \end{smallmatrix}^\mathrm{d}$ & 5.43$\begin{smallmatrix}+0.01 \\-0.02 \end{smallmatrix}^*$ & 7.40$\begin{smallmatrix}+0.42 \\-0.42 \end{smallmatrix}$ & 1.43$\begin{smallmatrix}+0.11 \\-0.09  \end{smallmatrix}^\mathrm{b}$ & 5.37 & 0.50 & 33, 35, 36, 37\\
NGC 5194 & 19361 & 2.00$\begin{smallmatrix}+1.33 \\-0.67  \end{smallmatrix}^\mathrm{d}$ & 4.71$\begin{smallmatrix}+0.15 \\-0.02 \end{smallmatrix}^*$ & 7.40$\begin{smallmatrix}+0.42 \\-0.42 \end{smallmatrix}$ & 1.43$\begin{smallmatrix}+0.11 \\-0.09  \end{smallmatrix}^\mathrm{b}$ & 2.99 & 0.17 & 33, 35, 36, 37\\
NGC 5194 & 19653 & 4.00$\begin{smallmatrix}+2.67 \\-1.33  \end{smallmatrix}^\mathrm{d}$ & 5.44$\begin{smallmatrix}+0.01 \\-0.02 \end{smallmatrix}^*$ & 7.40$\begin{smallmatrix}+0.42 \\-0.42 \end{smallmatrix}$ & 1.43$\begin{smallmatrix}+0.11 \\-0.09  \end{smallmatrix}^\mathrm{b}$ & 3.38 & 0.81 & 33, 35, 36, 37\\
NGC 5194 & 20267 & 5.00$\begin{smallmatrix}+3.33 \\-1.67  \end{smallmatrix}^\mathrm{d}$ & 5.06$\begin{smallmatrix}+0.03 \\-0.02 \end{smallmatrix}^*$ & 7.40$\begin{smallmatrix}+0.42 \\-0.42 \end{smallmatrix}$ & 1.43$\begin{smallmatrix}+0.11 \\-0.09  \end{smallmatrix}^\mathrm{b}$ & 2.24 & 0.51 & 33, 35, 36, 37\\
NGC 5194 & 20504 & 3.00$\begin{smallmatrix}+3.00 \\-0.00 \end{smallmatrix}^*$ & 5.08$\begin{smallmatrix}+0.02 \\-0.34 \end{smallmatrix}^*$ & 7.40$\begin{smallmatrix}+0.42 \\-0.42 \end{smallmatrix}$ & 1.43$\begin{smallmatrix}+0.11 \\-0.09  \end{smallmatrix}^\mathrm{b}$ & 9.38 & 0.13 & 33, 35, 36, 37\\
NGC 5194 & 20888 & 5.00$\begin{smallmatrix}+3.33 \\-1.67  \end{smallmatrix}^\mathrm{d}$ & 4.85$\begin{smallmatrix}+0.01 \\-0.02 \end{smallmatrix}^*$ & 7.40$\begin{smallmatrix}+0.42 \\-0.42 \end{smallmatrix}$ & 1.43$\begin{smallmatrix}+0.11 \\-0.09  \end{smallmatrix}^\mathrm{b}$ & 1.55 & 0.46 & 33, 35, 36, 37\\
NGC 5194 & 21179 & 3.00$\begin{smallmatrix}+2.00 \\-1.00  \end{smallmatrix}^\mathrm{d}$ & 4.53$\begin{smallmatrix}+0.02 \\-0.01 \end{smallmatrix}^*$ & 7.40$\begin{smallmatrix}+0.42 \\-0.42 \end{smallmatrix}$ & 1.43$\begin{smallmatrix}+0.11 \\-0.09  \end{smallmatrix}^\mathrm{b}$ & 2.81 & 0.12 & 33, 35, 36, 37\\
NGC 5194 & 21725 & 2.00$\begin{smallmatrix}+1.33 \\-0.67  \end{smallmatrix}^\mathrm{d}$ & 4.74$\begin{smallmatrix}+0.02 \\-0.01 \end{smallmatrix}^*$ & 7.40$\begin{smallmatrix}+0.42 \\-0.42 \end{smallmatrix}$ & 1.43$\begin{smallmatrix}+0.11 \\-0.09  \end{smallmatrix}^\mathrm{b}$ & 2.92 & 0.19 & 33, 35, 36, 37\\
NGC 5194 & 22291 & 3.00$\begin{smallmatrix}+1.00 \\-0.00 \end{smallmatrix}^*$ & 4.54$\begin{smallmatrix}+0.02 \\-0.02 \end{smallmatrix}^*$ & 7.40$\begin{smallmatrix}+0.42 \\-0.42 \end{smallmatrix}$ & 1.43$\begin{smallmatrix}+0.11 \\-0.09  \end{smallmatrix}^\mathrm{b}$ & 1.25 & 0.27 & 33, 35, 36, 37\\
NGC 5194 & 26974 & 5.00$\begin{smallmatrix}+0.00 \\-1.00 \end{smallmatrix}^*$ & 4.64$\begin{smallmatrix}+0.06 \\-0.02 \end{smallmatrix}^*$ & 7.40$\begin{smallmatrix}+0.42 \\-0.42 \end{smallmatrix}$ & 1.43$\begin{smallmatrix}+0.11 \\-0.09  \end{smallmatrix}^\mathrm{b}$ & 1.83 & 0.24 & 33, 35, 36, 37\\
NGC 5194 & 28599 & 2.00$\begin{smallmatrix}+0.00 \\-1.00 \end{smallmatrix}^*$ & 4.95$\begin{smallmatrix}+0.08 \\-0.01 \end{smallmatrix}^*$ & 7.40$\begin{smallmatrix}+0.42 \\-0.42 \end{smallmatrix}$ & 1.43$\begin{smallmatrix}+0.11 \\-0.09  \end{smallmatrix}^\mathrm{b}$ & 0.29 & 3.11 & 33, 35, 36, 37\\
NGC 5194 & 29178 & 2.00$\begin{smallmatrix}+1.33 \\-0.67  \end{smallmatrix}^\mathrm{d}$ & 4.61$\begin{smallmatrix}+0.02 \\-0.01 \end{smallmatrix}^*$ & 7.40$\begin{smallmatrix}+0.42 \\-0.42 \end{smallmatrix}$ & 1.43$\begin{smallmatrix}+0.11 \\-0.09  \end{smallmatrix}^\mathrm{b}$ & 2.38 & 0.17 & 33, 35, 36, 37\\
NGC 5253 & 1 & 1.00$\begin{smallmatrix}+0.67 \\-0.33  \end{smallmatrix}^\mathrm{d}$ & 4.87$\begin{smallmatrix}+0.04 \\-0.05 \end{smallmatrix}^*$ & 3.32$\begin{smallmatrix}+0.25 \\-0.25 \end{smallmatrix}$ & 0.25$\begin{smallmatrix}+0.05 \\-0.05 \end{smallmatrix}^*$ & 4.28 & 0.17 & 33, 35, 36, 37, 42\\
NGC 5253 & 595 & 10.00$\begin{smallmatrix}+6.67 \\-3.33  \end{smallmatrix}^\mathrm{d}$ & 4.60$\begin{smallmatrix}+0.05 \\-0.00 \end{smallmatrix}^*$ & 3.32$\begin{smallmatrix}+0.25 \\-0.25 \end{smallmatrix}$ & 0.25$\begin{smallmatrix}+0.05 \\-0.05 \end{smallmatrix}^*$ & 3.73 & 0.11 & 33, 35, 36, 37, 42\\
NGC 5253 & 606 & 10.00$\begin{smallmatrix}+1.00 \\-0.00 \end{smallmatrix}^*$ & 4.69$\begin{smallmatrix}+0.06 \\-0.05 \end{smallmatrix}^*$ & 3.32$\begin{smallmatrix}+0.25 \\-0.25 \end{smallmatrix}$ & 0.25$\begin{smallmatrix}+0.05 \\-0.05 \end{smallmatrix}^*$ & 2.96 & 0.16 & 33, 35, 36, 37, 42\\
NGC 5253 & 615 & 1.00$\begin{smallmatrix}+0.67 \\-0.33  \end{smallmatrix}^\mathrm{d}$ & 4.77$\begin{smallmatrix}+0.05 \\-0.05 \end{smallmatrix}^*$ & 3.32$\begin{smallmatrix}+0.25 \\-0.25 \end{smallmatrix}$ & 0.25$\begin{smallmatrix}+0.05 \\-0.05 \end{smallmatrix}^*$ & 5.24 & 0.11 & 33, 35, 36, 37, 42\\
NGC 628 & 579 & 2.00$\begin{smallmatrix}+6.00 \\-1.00 \end{smallmatrix}^*$ & 4.71$\begin{smallmatrix}+0.05 \\-0.60 \end{smallmatrix}^*$ & 8.80$\begin{smallmatrix}+0.70 \\-0.70 \end{smallmatrix}$ & 1.43$\begin{smallmatrix}+0.11 \\-0.09  \end{smallmatrix}^\mathrm{b}$ & 2.19 & 0.23 & 33, 35, 36, 37\\
\hline
\end{tabular}
\end{table*}

\section{Flux estimations}\label{flux}

In this study we assume that each YMC in our sample is undergoing self-enrichment, regardless of the polluter. Consequently, a fraction of its total mass will be in the polluted ejecta. Based on this assumption, we calculate the mass of $^{26}$Al and the flux of its tracers. Here, we outline our methodology for estimating the mass of $^{26}$Al and computing its fluxes, while addressing their associated uncertainties. For the 1.8 MeV and 511 keV lines, we consider two scenarios: an optimistic one, assuming simultaneous ejection of all polluted material, and a realistic scenario where the decay of $^{26}$Al is factored in over the cluster's age.

\subsection{\texorpdfstring{$^{26}$Al mass}{26Al mass}}
Future instruments capable of detecting $^{26}$Al may not have the resolution to distinguish individual YMCs within a galaxy. Therefore, we calculate the total mass of star clusters by summing the masses of YMCs in a given galaxy, apart from R136 in the LMC. We assume that approximately half of this total mass corresponds to wind ejecta. This fiducial value comes from the assumption that YMCs form multiple populations in a similar way as observed in GCs, which we would like to test here. We further assume that a comparable amount of ejecta mass condensed in stars is detectable in the interstellar medium. This value is consistent with the wide range of observed pollution fractions in Galactic GCs, which varies from less than 40 per cent to more than 90 per cent \citep{Milone_2017, Milone_2020, Dondoglio_2021}. In any self-enrichment scenario, this mass must have been ejected by stellar winds. However, it seems unlikely that all of the ejecta would end up in the second population stars.

To determine the fraction of $^{26}$Al present in the wind ejecta, we rely on the stellar evolution models developed by \cite{Martinet_2022} and \cite{Higgins_2023}. \cite{Martinet_2022} find that 60 per cent of the initial mass for their 250~M$_{\odot}$ model is ejected during the H-burning phase. Taking the $^{26}$Al yields from their Table A.1 for the rotating model at solar metallicity, as 2.63~$\times$~10$^{-3}$ M$_{\odot}$, we obtain a corresponding mass fraction for $^{26}$Al in the wind ejecta of 1.8~$\times$~10$^{-5}$. We use this fiducial value for the $^{26}$Al abundance in freshly ejected processed gas. \cite{Martinet_2022} found that the $^{26}$Al mass in stellar wind ejecta increases with metallicity, following the relation M$_{\mathrm{Al26}} \propto$ Z$^{1.45}$. Therefore, we scale the mass of $^{26}$Al accordingly for each galaxy based on its metallicity.  Alternatively, \cite{Higgins_2023} reports higher $^{26}$Al yields in stellar ejecta, but highlights a larger percentage of ejecta from the initial mass. For example, their 300 M$_{\odot}$ star loses almost 90 per cent of its mass during the H-burning phase. Hence, their $^{26}$Al mass fraction for the ejected gas is 4.9~$\times$~10$^{-5}$. 

\subsection{\texorpdfstring{Optimistic case for $^{26}$Al $\gamma$-ray lines}{26Al gamma-ray lines}}
The flux for $^{26}$Al, measured in units of ph~cm$^{-2}$~s$^{-1}$, is calculated using the following formula:
\begin{equation}
    F_{\mathrm{Al} 26} = \frac{M_{\mathrm{Al26}}}{m_{\mathrm{nuc, Al}26}} \frac{1}{4\pi D^2\tau}.
    \label{flux_Al26}
\end{equation}
In the above equation, M$_{\mathrm{Al26}}$ represents the total mass of $^{26}$Al in the galaxy, considering only contributions from individual chosen YMCs. $m_{\mathrm{nuc, Al}26}$ denotes the nuclear mass of $^{26}$Al, $\tau$~=~1.0~Myr refers to the exponential decay time of $^{26}$Al and $D$ represents the distance to the galaxy. The formula assumes that all the $^{26}$Al was ejected significantly less than $\tau$ before the time of observation. We then take into consideration that only 99.7 per cent of $^{26}$Al atoms will emit 1.8~MeV line and only 82 per cent will produce a positron \citep{Diehl_2021}. Hence, we scale Eq.\ref{flux_Al26} to obtain the flux for 1.8 MeV emission line as:
\begin{equation}
    F_{1.8\mathrm{MeV}} = 0.997F_{\mathrm{Al} 26}.
    \label{1.8MeV}
\end{equation}

We also use this fraction to calculate the positron flux as follows:
\begin{equation}
    F_{\mathrm{e^+}} = 0.82 \, F_{\mathrm{Al}26}.
\end{equation} 

In our calculations we assume a Ps fraction of 100 per cent, hence all positrons produced from the decay process of $^{26}$Al will form Ps state. According to \cite{Brown_1987} only 25 per cent of Ps will decay via para-Ps emitting two 511 keV photons, hence 511 keV emission line flux is calculated using the below conversion:

\begin{equation}
    F_{511\mathrm{keV}} = 0.5F_{\mathrm{e^+}} = 0.41F_{\mathrm{Al} 26}.
    \label{511keV}
\end{equation}

\subsection{\texorpdfstring{Realistic case for $^{26}$Al $\gamma$-ray lines}{Realistic case for 26Al gamma-ray lines}}
However, the calculations above assume that the mass has been ejected all at once. To account for a more realistic scenario, we assume that $M_{\mathrm{Al26}}$ will decrease as $^{26}$Al decays over time. Thus, we calculate the mass of the decayed $^{26}$Al for each selected YMC using the expression:
\begin{equation}
    M_{\mathrm{Al} 26, \mathrm{decay}} = M_{\mathrm{Al26}}\exp\left({\frac{-t_{\mathrm{age}}}{\tau}}\right).
    \label{Al26mass_decay}
\end{equation}
Here, $t_{\mathrm{age}}$ is the age of each YMC in our sample. Following a similar procedure as in the optimistic case, we sum $M_{\mathrm{Al} 26, \mathrm{decay}}$ for each YMC in a given galaxy. Subsequently, we utilize these mass calculations to compute fluxes for $^{26}$Al $\gamma$-ray lines using Equations \ref{flux_Al26}, \ref{1.8MeV}, and \ref{511keV}.

\subsection{Background emission for 1.8 MeV and 511 keV fluxes for each galaxy}
Here, we estimate the background emission for the 1.8 MeV and 511 keV fluxes, assuming that the galaxy does not host YMCs undergoing self-enrichment, so the contribution to the production of $^{26}$Al comes primarily from massive stars that are not polluters. Since none of the YMCs in the Milky Way meet our criteria, we infer that the Galaxy similarly lacks such clusters. Using the total $^{26}$Al mass in the Milky Way as a reference, we estimate the background emission of $^{26}$Al produced by massive stars not involved in self-enrichment. To obtain the total mass of $^{26}$Al (M$_{\mathrm{back,Al}{26}}$) for each galaxy, we scaled the best current estimate value of 2 M$_\odot$ \citep{Diehl_2021} for the Milky Way with the star formation rate (SFR), as $^{26}$Al is produced during various stages of the evolution of massive stars. 

We scaled the Milky Way's 511~keV flux from \cite{Siegert_2016} with the total stellar mass of any given galaxy to obtain galaxy's overall positron production rate. We note that this might be an underestimation of the total production rate since only the annihilation rate is measured. In addition, the stellar mass may be only one factor that scales the production rate, and other factors, such as the dark matter halo and SFR may impact the production rate. The background emission for $^{26}$Al and positron production rate for each galaxy is calculated in Table \ref{tab:background}.

\begin{table*}
\centering
\caption{Properties of galaxies hosting selected YMCs and their estimated background emission at 1.8~MeV and positron production, calculated assuming their YMCs were not undergoing self-enrichment. All properties refer to the host galaxies except the once with "YMCs" in their name. The latter are the signal from the YMCs detection which we consider against the background of the host galaxy. References: 1, \protect\cite{McMillan_2017}; 2,  \protect\cite{Diehl_2021} and references therein; 3, \protect\cite{Siegert_2016}; 4, \protect\cite{Wik_2014}; 5, \protect\cite{Bailin_2011}; 6, \protect\citep{Ostlin_2001}; 7, \protect\cite{Pineda_2018}; 8, \protect\cite{Wei_2021} ; 9, \protect\cite{Yoast-Hull_2013}; 10, \protect\cite{Chiang_2011}; 11, \protect\cite{Moll_2007}; 12, \protect\cite{Fumagalli_2010}; 13, \protect\cite{Buzzo_2021}; 14, \protect\citep{Fahrion_2022}; 15, \protect\cite{Greggio_1998}; 16, \protect\cite{Johnson_2012}; 17, \protect\cite{Lardo_2015}; 18, \protect\cite{Seille_2022}; 19, \protect\cite{Kennicutt_2003}; 20, \protect\cite{Leroy_2008}; 21, \protect\cite{Hartwell_2004}; 22, \protect\cite{Karachentsev_2004}; 23, \protect\cite{Calzetti_2015}. \\
$^1$ production rate = annihilation rate if a steady state is assumed, and if all positrons annihilate inside the galaxy \\
$^2$ we do not take into account instrumental effects}
\label{tab:background}
\resizebox{\textwidth}{!}{
\begin{tabular}{lcccccccccr}
\hline
Galaxy & SFR & \multicolumn{1}{c}{log Stellar mass} & $M_{\mathrm{back,Al26}}$ & $M_{\mathrm{YMCs/back,Al26}}$ & \multicolumn{1}{c}{$F_{\mathrm{Al} 26}$} & \multicolumn{1}{c}{$F_{\mathrm{Al} 26}$} & e$^{+}$ & \multicolumn{1}{c}{$F_{\mathrm{e^+}}$} & \multicolumn{1}{c}{$F_{\mathrm{e^+}}$} &  Reference  \\
& & &  &  &\multicolumn{1}{c}{background$^2$} & \multicolumn{1}{c}{YMCs} &production rate$^1$ & \multicolumn{1}{c}{background$^2$} & \multicolumn{1}{c}{YMCs} \\
& [M$_{\odot}$ yr$^{-1}$] & \multicolumn{1}{c}{[M$_{\odot}$]} & [M$_{\odot}$] &  & \multicolumn{2}{c}{[10$^{-9}$ ph cm$^{-2}$ s$^{-1}$] } & [10$^{42}$ e$^{+}$ s$^{-1}$] & \multicolumn{2}{c}{[10$^{-9}$ ph cm$^{-2}$ s$^{-1}$] } & \\
\hline
Milky Way & 2.00 & 10.73 & 2.00 & - & - & - & 50.0 & - & - & 1, 2, 3\\
NGC 253 & 5.00 & 10.64 & 5.00 & 13.63 & 5.300 & 72.00 & 41.0 & 31.00 & 30.00 & 4, 5\\
ESO338-IG & 3.20 & 9.60 & 3.20 & 1.77 & 0.026 & 0.047 & 3.70 & 0.022 & 0.019 &  6\\
M51 & 4.80 &  10.48 & 4.80 & 21.35 & 0.770 & 16.00  & 28.0 & 3.200 & 6.700 &7, 8\\
M82 & 10.0 & 10.80  & 10.0 & 6.16 & 9.000 & 55.00 & 58.0 & 37.00 & 23.00 & 9, 10\\
NGC 1140 & 0.70 & 9.58  & 0.70 & 16.99 &  0.021 & 0.350 & 3.50 & 0.073 & 0.140 &11, 12\\
NGC 1487 & 0.37 & 8.95  & 0.37 & 3.43 & 0.050  & 0.170 & 0.82 & 0.079 & 0.071 &13, 14\\
NGC 1569 & 0.50 & 8.45  & 0.50 & 1.18 & 1.200  & 1.400 & 0.26 & 0.450 & 0.590 &15, 16\\
NGC 4038 & 20.0 & 10.65  & 20.0 & 12.84 & 0.490 & 6.200 & 42.0 & 0.730 & 2.600 &17, 18\\
NGC 2403 & 1.30 & 9.70 & 1.30 & 1.82 & 1.500 & 2.700 & 4.60  & 5.500 & 1.100 &19, 20\\
NGC 4214 & 0.04 & 9.18 & 0.04 & 22.5 & 0.053 & 1.200 & 1.40  & 1.300 & 0.490 &21, 22\\
NGC 1313 & 1.15 & 9.41  & 1.15 & 0.08 & 0.730 & 0.610 & 2.40  & 1.100 & 0.250 &23\\
NGC 1433 & 0.27 & 10.23  & 0.27 & 30.89 & 0.038 & 1.200 & 16.0 & 1.600 & 0.490 &23\\
NGC 1566 & 5.67 & 10.43  & 5.67 & 8.68 & 0.270 & 2.400 & 25.0 & 0.860 & 0.980 &23\\
NGC 3351 & 1.57 & 10.32  & 1.57 & 17.26 & 0.210 & 3.700 & 19.0 & 1.800 & 1.500 &23\\
NGC 4395 & 0.34 & 8.78  & 0.34 & 0.56 & 0.190 & 0.110 & 0.56 & 0.230 & 0.046 &23\\
NGC 4449 & 0.94 & 9.04  & 0.94 & 5.31 & 0.690 & 3.700 & 1.00 & 0.520 & 1.500 &23\\
NGC 4656 & 0.50 & 8.60  & 0.50 & 2.42 & 0.095 & 0.230 & 0.37 & 0.050 & 0.094 &23\\
NGC 5194 & 6.88 & 10.38  & 6.88 & 16.41 & 1.500 & 24.00 & 22.0  & 3.400 & 10.00 &23\\
NGC 5253 & 0.10 & 8.34  & 0.10 & 7.30 & 0.110 & 0.780 & 0.20 & 0.150 & 0.320 &23\\
NGC 628 & 3.67 & 10.04  & 3.67 & 0.57 & 0.560 & 0.320 & 10.0 & 1.100 & 0.130 &23\\
\hline
\end{tabular}}
\end{table*}

\subsection{\texorpdfstring{$^{26}$Al $\gamma$-ray lines for an SMS scenario}{26Al gamma-ray lines for an SMS scenario}}\label{SMS}
For comparison, we calculate the amount of $^{26}$Al in a YMC at a distance of one Mpc undergoing self-enrichment, with an SMS as the polluter. We assume that the $^{26}$Al fraction in the star's winds remains approximately constant, similar to its core value, estimated at 10$^{-4.7}$ for an SMS with $M_\mathrm{SMS}$ = 2$\times$10$^{4}$~M$_\odot$, particularly in the case of a more metal-poor cluster (Ramírez-Galeano et al., in prep). Following Figure 3 in \cite{Gieles_2018}, the mass of the SMS winds with this specific mass is approximately 1$\times$10$^4$~M$\odot$. Using these estimates, we determine the mass of $^{26}$Al and calculate the fluxes for the $\gamma$-ray lines using Equations \ref{flux_Al26}, \ref{1.8MeV}, and \ref{511keV}, considering only the optimistic case.

\subsection{Ps radio recombination lines}

The flux of Ps radio recombination lines in units of $\mu$Jy is calculated following \cite{Anantharamaiah_1989}:
\begin{equation}
    F_{\mathrm{Ps}} = \frac{L_{\nu}}{4\pi D^2 \Delta\nu},
    \label{flux_Ps}
\end{equation}
where where $L_{\nu}$ is given by:
\begin{equation}
    L_{\nu} = N_{\mathrm{Ps}}\mathrm{h}\nu.
\end{equation}
Here, $\nu$ is the frequency of the line, h is Planck's constant and $N_{\mathrm{Ps}}$ defines the rate of production of positronium atoms. The width of the line, $\Delta\nu$ is calculated using the expression by \cite{Wallyn_1996}:
\begin{equation}
    \Delta\nu =  7.8\times10^{-4}\nu,
\end{equation}
for T~=~10$^4$~K, where most positrons form Ps. The line frequencies are calculated using the general Rydberg formula:
\begin{equation}
    \nu = cR_{\mathrm{Ps}}\left(\frac{1}{n^2_1} - \frac{1}{n^2_2}\right),
\end{equation}
where $c$ is the speed of light, $n_1$ is the principal quantum number of the lower energy level, and $n_2$ is the principal quantum number of the upper energy level. The Rydberg constant for Ps is $R_{\mathrm{Ps}}$~=~5.483069~$\times$~10$^6$~m$^{-1}$, approximately half of the value for hydrogen, $R_{\mathrm{H}}$ = 1.0966139 $\times$ 10$^7$ m$^{-1}$. Hence, the frequencies of Ps recombination lines for a specific level will be approximately half the frequency of the corresponding H lines \citep{StaveleySmith_2022}.

We calculated fluxes for the radio recombination lines Ps87$\alpha$, along with lines Ps61$\alpha$ to Ps208$\alpha$, which were then stacked across line frequencies ranging from 362 MHz to 14 GHz, same as frequency ranges for SKA-Mid (see Section \ref{observatories}). Specifically, we computed Ps87$\alpha$ at a line frequency of 9.8 GHz as an illustrative example, similar to the line investigated by \cite{Anantharamaiah_1989}. In all calculations, we assumed, similarly to \cite{Anantharamaiah_1989}, that every positron from the decay process of $^{26}$Al will form positronium in an excited state and cascade down to the ground state, resulting in annihilation.

\subsection{\texorpdfstring{$^{26}$AlF rotational line}{26AlF rotational line}}

The ALMA telescope can resolve individual star clusters, so estimations for the $^{26}$AlF rotational line are made for each YMC. To calculate the flux for the $^{26}$AlF rotational line, we use the observation of the evolved star CK Vul by \cite{Kaminski_2018} as a benchmark. This is currently the only empirical evidence of $^{26}$Al in molecular form, and these estimates, based on a single star, are presumably conservative. The flux for the rotational line (6$\rightarrow$5) of $^{26}$AlF is estimated using the value of 8 mJy obtained in \cite{Kaminski_2018} for CK Vul, scaled according to the distance to each galaxy hosting the selected YMCs, D, as well as the respective mass of $^{26}$Al calculated from the available fluorine content in each YMCs, denoted as $M_\mathrm{Al26, F}$:

\begin{equation}
    F_\mathrm{^{26}AlF} = 8 \mathrm{mJy} \left(\frac{D_{\mathrm{CK Vul}}}
    {D}\right)^{2}\left(\frac{M_{\mathrm{Al26, F}}}{M_{\mathrm{Al26, CKVul}}}\right),
\end{equation}

where the distance to CK~Vul, $D_{\mathrm{CK Vul}}$, is 3.5~kpc \citep{Kaminski_2021}  and $M_{\mathrm{Al26, CKVul}}$~=~1.71$\times$10$^{-9}$~M$_{\odot}$ \citep{Kaminski_2018}, its mass of $^{26}$Al. To calculate $M_\mathrm{Al26, F}$, we used the solar fluorine abundance, (3.63 $\pm$ 0.11)~$\times$~10$^{-8}$ \citep{Asplund_2009} and multiply it with the mass of each YMC. According to \cite{Franco_2021}, Wolf Rayet (WR) stars, which are very likely present in YMCs, play a crucial role in producing the fluorine abundance, potentially increasing $^{19}$F abundance to 10 to 70 times its initial value \citep{Meynet_2000}. Employing the conservative factor of 10, we first multiplied $M_\mathrm{Al26, F}$ and then further adjusted it by directly multiplying with the metallicity of each galaxy. We calculated the relation between $^{19}$F abundance and metallicity from the data provided in \cite{Limongi_2018}.

Not all $^{19}$F will contribute to the production of this molecule and the mass of $^{26}$Al that forms $^{26}$AlF is a small fraction of the available $^{26}$Al; therefore, we halved the derived value to obtain $M_\mathrm{Al26, F}$. 

\subsection{Error propagation}

The errors listed in Table \ref{tab:YMC_table} are taken from the literature and typically fall within 1-$\sigma$. When specific details were unavailable, we resorted to using maximum and minimum values. If such values were also unavailable, we estimated errors based on comparable sources, such as NGC 253 for distance, ESO338-IG for age, M82 for mass, and NGC 4038 for metallicity.

Regarding the fraction of $^{26}$Al yields in the stellar wind ejecta provided by \cite{Martinet_2022} and \cite{Higgins_2023}, no explicit errors were available.  Obtaining such errors would require rerunning the models with different cross-sections, considering various nuclear reactions, which is a resource-intensive process. Currently, models with such variations are unavailable to our knowledge. Therefore, we estimated the uncertainty by comparing the differences between the two stellar evolutionary models. It's worth noting that these models are 1D stellar models, inherently containing numerous uncertainties. However, recent advancements, such as the work by \cite{Rizzuti_2023}, who performed a set of 3D simulations of a convective neon-burning shell in a 20 M$_{\odot}$ star from its early development to fuel exhaustion, could hopefully provide new insights into $^{26}$Al yield fractions in the future.

\section{Future observatories}\label{observatories}
Here, we discuss the specifications of (potential) future $\gamma$-ray and radio telescopes that we believe would give us the best chance at detecting the chemical traces discussed above. 

The Compton Spectrometer and Imager \citep[COSI;][]{Tomsick_2019, Tomsick_2023} is a wide-field-of-view telescope, covering 25 per cent of the sky, designed to survey soft $\gamma$-ray emission in the 0.2-5~MeV energy range, often referred to as the "MeV Gap." COSI aims to understand the origins of positrons in the Galaxy by mapping the 511~keV emission in the sky with a 3$\sigma$ line sensitivity of 1.2~$\times$~10$^{-5}$~ph~cm$^{-2}$~s$^{-1}$ \citep{Tomsick_2023}. COSI will image $^{26}$Al with improved 3$\sigma$ sensitivity of 3.0~$\times$~10$^{-6}$~ph~cm$^{-2}$~s$^{-1}$ \citep{Tomsick_2023} compared to previous observations, which reached values on the order of 1.0~$\times$~10$^{-5}$~ph~cm$^{-2}$~s$^{-1}$ for specific regions by COMPTEL \citep{Knodlseder_1999comptel, Oberlack_2000} and INTEGRAL \citep{Diehl_2006, Siegert_2017, Krause_2018}. Those sensitives are for a two year all-sky survey. 

The proposed space mission, e-ASTROGAM \citep{DeAngelis_2017}, is expected to have a 3$\sigma$ line sensitivity for detecting $^{26}$Al, reaching 3.0~$\times$~10$^{-6}$~ph~cm$^{-2}$~s$^{-1}$ for a 1~Ms observation, and for a hypothetical two year all-sky survey it would be 1.0~$\times$~10$^{-6}$~ph~cm$^{-2}$~s$^{-1}$. Following \cite{DeAngelis_2017} we have extrapolated the sensitivity for 1~Ms and assumed field of view 2.5~sr to a two-year all sky survey. For this we used 42~Ms observing time, a typical value for the INTEGRAL observatory including downtime. With an improved 3$\sigma$ line sensitivity for 511~keV of 4.1~$\times$~10$^{-6}$~ph~cm$^{-2}$~s$^{-1}$ for a 1~Ms observation and 1.4~$\times$~10$^{-6}$~ph~cm$^{-2}$~s$^{-1}$ for a hypothetical two-year all sky survey, e-ASTROGAM would facilitate deep Galactic surveys of the positron annihilation radiation and the search for potential point-like sources. 

The Square Kilometre Array \citep[SKA,][]{Dewdney_2009} is currently under construction across sites in Australia and Africa and is set to become the world's largest radio astronomy observatory. It will operate across a wide range of frequencies, including SKA-Low (50-350~MHz) and SKA-Mid (350~MHz -  15.4~GHz). The first phase of SKA, expected to be completed in 2028, will provide approximately 10 per cent of the total collecting area. The completion of the full array, known as Phase 2 or SKA-2, is over the next decade. SKA-2 point source sensitivity limits for the chosen Ps recombination lines  are 1.45 and 0.46~$\mu$Jy for 100- and 1000-hour run respectively \citep{Braun_2019}. All the respective flux limits are used in Figs \ref{fig:AL26} - \ref{fig:Ps_stacked} below.

\section{Results}\label{results}

\begin{figure*}
        \includegraphics[width=\linewidth]{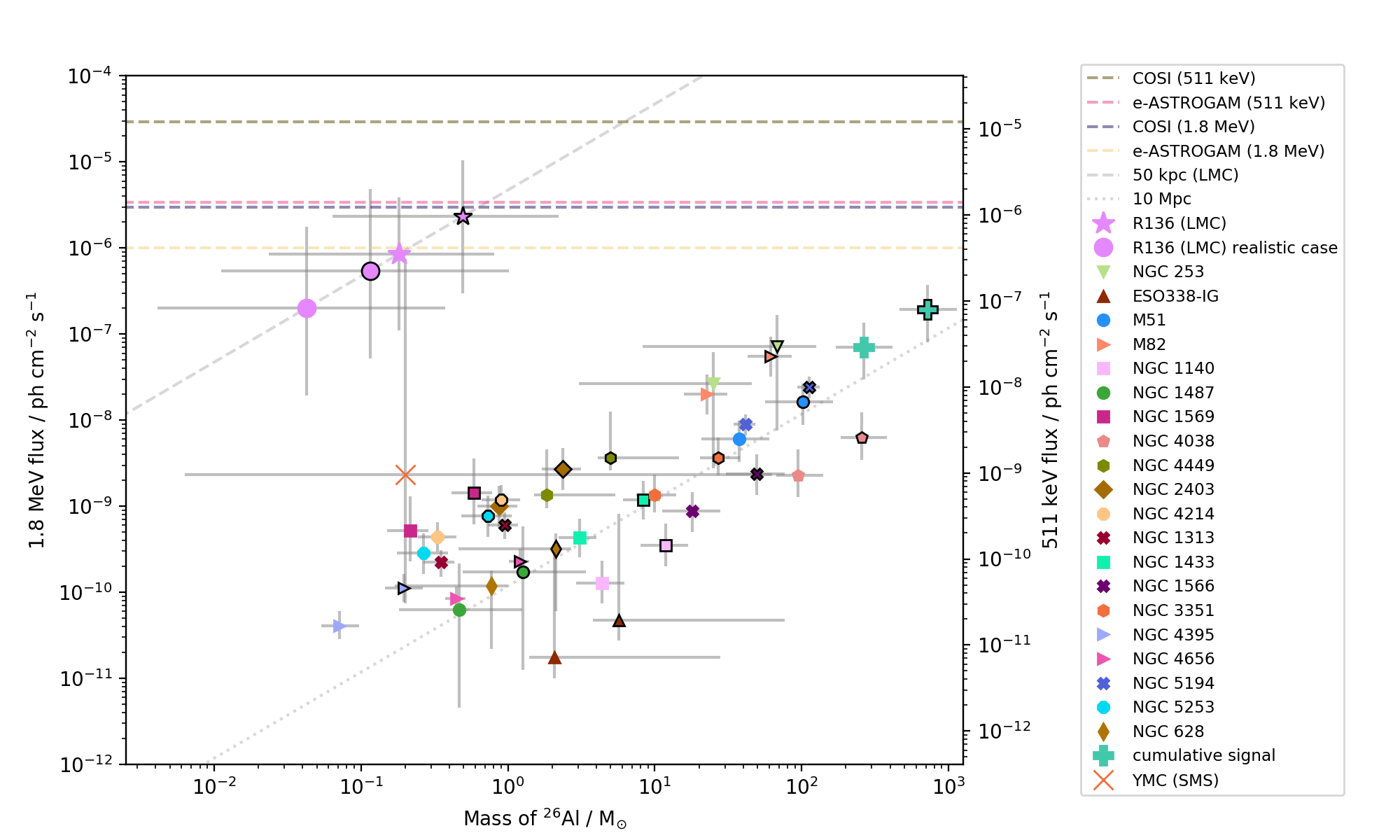}
        \caption{The figure shows the estimated $^{26}$Al flux (left y-axis) and the positron annihilation line at 511~keV (right y-axis) for the optimistic case i.e. $^{26}$Al ejection all at once. Different coloured markers represent galaxies hosting the chosen sample of YMCs used for this prediction. The markers with bold outlines represent the $^{26}$Al yield in the stellar ejecta from \protect\cite{Higgins_2023}, while the markers without bold outlines represent the yields from \protect\cite{Martinet_2022}. Dashed horizontal lines of different colors indicate the line sensitivities for the upcoming COSI and proposed e-ASTROGAM missions (3$\sigma$ for two-year all sky survey). Grey diagonal lines represent distances from the Milky Way used as a reference. R136 is not included in the cumulative signal, as it can be resolved.}
        \label{fig:AL26}
\end{figure*}

Figure \ref{fig:AL26} displays the predicted fluxes of the 1.8~MeV $\gamma$-ray line of $^{26}$Al (left y-axis) and $\gamma$-ray photons of 511~keV (right y-axis) for galaxies hosting YMCs from our sample. Bold outlined markers indicate results derived using the fraction of $^{26}$Al in stellar ejecta as proposed by \cite{Higgins_2023}, while markers without outlines represent fractions from the study of \cite{Martinet_2022}. The detection thresholds for e-ASTROGAM and COSI are also shown for comparison (dashed horizontal lines).

Our estimations suggest that, among the selected clusters in the shown galaxies, only R136 in the LMC is detectable with e-ASTROGAM within its nominal two-year mission time. This is based on the optimistic scenario with yield fractions from \cite{Higgins_2023}, which predicts a 1.8 MeV line flux of 2.33~$\times$~10$^{-6}$~ph~cm$^{-2}$ and a 511 keV line flux of 2.17~$\times$~10$^{-6}$~ph~cm$^{-2}$. The same cluster, based on the yield estimates from \cite{Martinet_2022}, could possibly be detected within its error bars. Similarly, the fluxes that consider the more realistic scenario, where $^{26}$Al decays over the life of the cluster, might also fall within detectable ranges. However, we find that COSI would not be able to detect any systems at 511 keV, and only potentially R136 at 1.8 MeV if we assume the optimistic scenario and/or \cite{Higgins_2023} yields. We also evaluated the cumulative signal for all clusters, except R136, to be 1.93~$\times$~10$^{-7}$~ph~cm$^{-2}$~s$^{-1}$ for the 1.8~MeV line assuming the optimistic scenario, represented by the large, bold outlined green plus. This signal remains below the detection threshold for both e-ASTROGAM and the upcoming COSI mission. To detect these signals, the surveys would need to be up to two orders of magnitude more sensitive. We have also taken into consideration the latitude constraint arising from the inability of the gamma-ray mission to differentiate the 1.8~MeV emission from the Galaxy from other sources. Hence, galaxies located in the Galactic plane, i.e., |b| < 20 deg, such as NGC 1140, NGC 4038, NGC 3351, and NGC 628, could be excluded from the cumulative signal, resulting in the 1.8~MeV line cumulative signal of 1.82~$\times$~10$^{-7}$~ph~cm$^{-2}$~s$^{-1}$. The fluxes in Figure \ref{fig:AL26} are, on average, one to two orders of magnitudes higher than in the more realistic case where $^{26}$Al ejection via stellar winds is spread out over each cluster’s age. Furthermore, the 1.8 MeV flux of 2.35~$\times$~10$^{-9}$~ph~cm$^{-2}$~s$^{-1}$ for a hypothetical cluster undergoing self-enrichment with an SMS as the polluter at a distance of one Mpc is also below the detection threshold for both instruments. It is worth noting that the uncertainties are significant due to varying estimates of wind mass loss for this star, as reported by \cite{Gieles_2018}. If that specific YMC were located at a distance of 50 kpc, like R136, its 1.8 MeV flux would be 9.39~$\times$~10$^{-7}$~ph~cm$^{-2}$~s$^{-1}$, still below the sensitivities of both COSI and e-ASTROGAM.

Table \ref{tab:background} shows the background emission for $^{26}$Al and the positron production rate for each galaxy, assuming no clusters are undergoing self-enrichment. When comparing these values with those derived from Figure \ref{fig:AL26} using the $^{26}$Al yield ratio from \cite{Higgins_2023}, some differences emerge. The scaled background emission is significantly lower for most galaxies, varying by one to two orders of magnitude, compared to values assuming contributions from polluters in YMCs. This becomes clearer when considering the ratio of the mass of $^{26}$Al in a galaxy with only background production to that with contributions from YMCs undergoing self-enrichment ($M_{\mathrm{YMCs/back,Al26}}$). Our calculations show that enrichment in some galaxies can boost $M_{\mathrm{Al26}}$ by up to 31 times, as in NGC 1433, or 21 times, as in M51. Galaxies with higher background emissions, like NGC 1313, NGC 4395, and NGC 628, typically exhibit lower contributions towards the mass of $^{26}$Al from YMCs than the estimated M$_{\mathrm{Al26}}$ for the entire galaxy. The expected background emission at 511 keV for galaxies is either of the same order of magnitude as the 511 keV emission from YMCs or differs by up to one order of magnitude. This suggests that there may be alternative sources of positrons besides the radioactive decay of $^{26}$Al (see Section \ref{discussion}). Hence, we conclude that the 511 keV line is not a reliable reference for determining whether there is ongoing self-enrichment in any given galaxy.

\begin{figure*}
        \includegraphics[width=\linewidth]{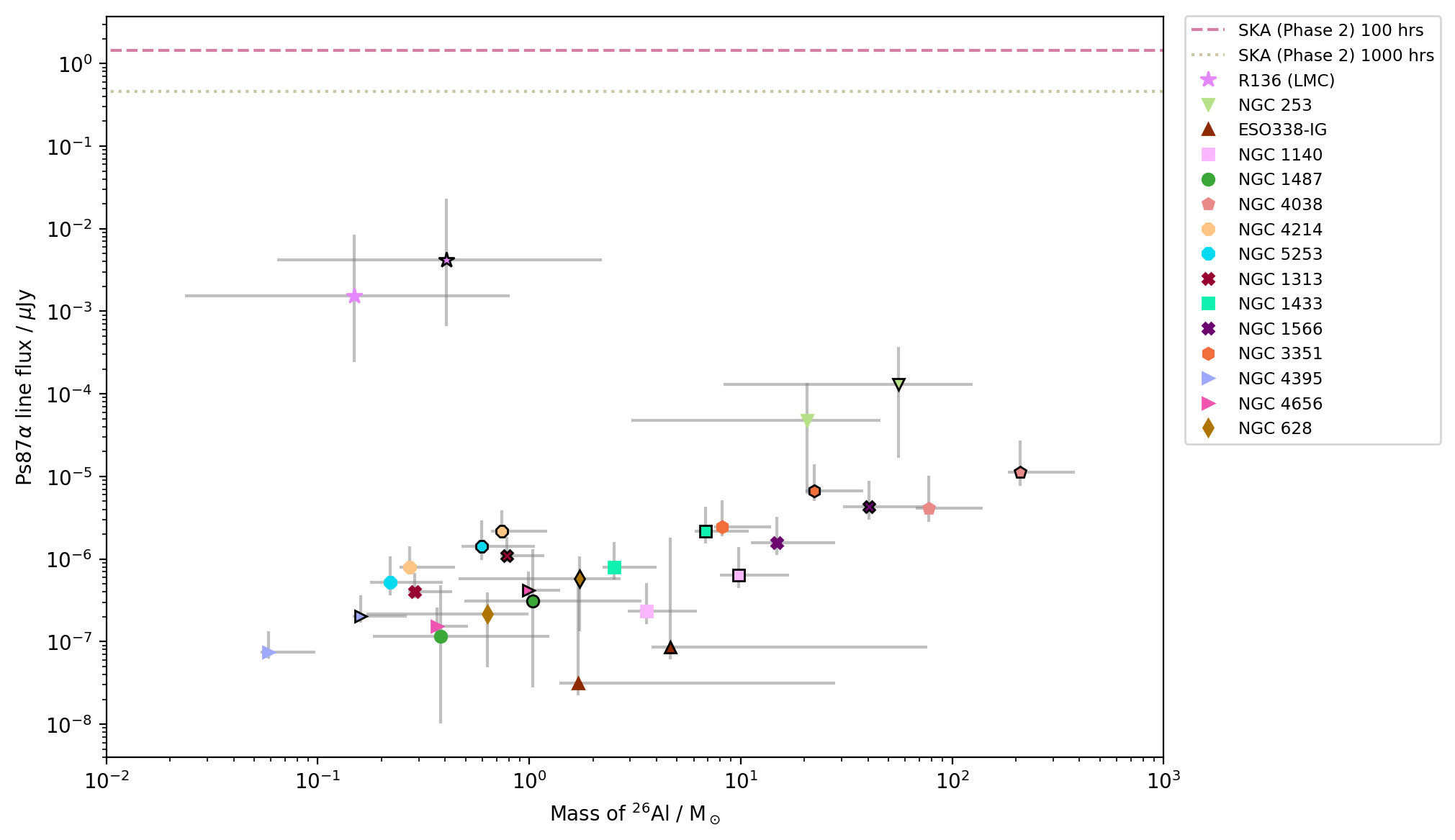}
        \caption{Flux estimation for positronium radio recombination Ps87$\alpha$ line for the optimistic case i.e. $^{26}$Al ejection all at once. Different coloured markers represent galaxies hosting the chosen sample of YMCs used for the estimation and visible using SKA telescope. The markers with bold outlines represent the $^{26}$Al yield in the stellar ejecta from \protect\cite{Higgins_2023}, while the markers without bold outlines represent the yields from \protect\cite{Martinet_2022}. The pink dashed line indicates the line sensitivity for SKA2 for the run of 100 hours, and yellow dotted line for 1000~hours.}
        \label{fig:Ps87alpha}
\end{figure*}

\begin{figure*}
        \includegraphics[width=\linewidth]{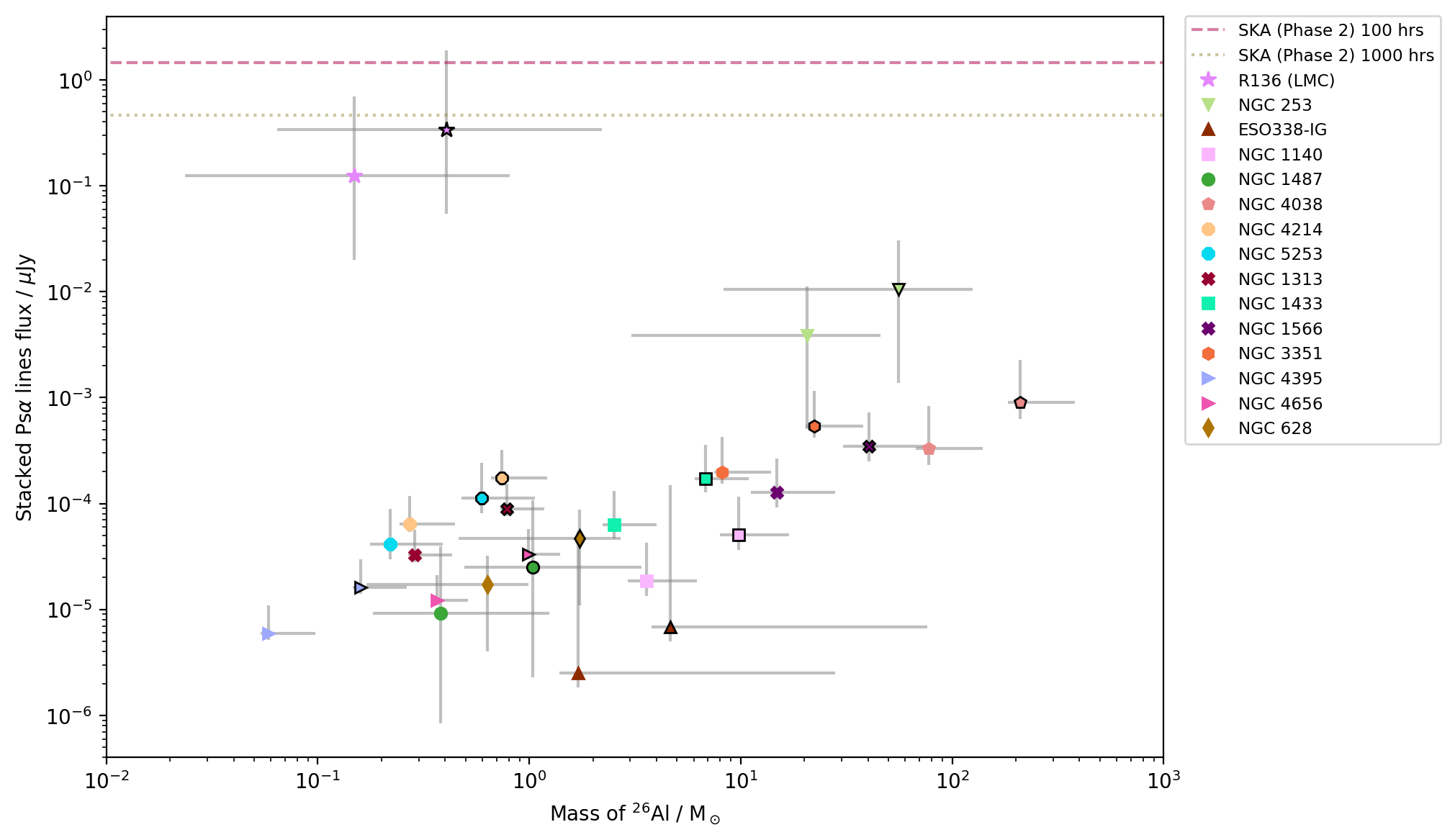}
        \caption{Flux estimations for stacked positronium radio recombination lines ranging from Ps61$\alpha$ to Ps208$\alpha$ for optimistic case i.e. $^{26}$Al ejection all at once. Different coloured markers represent galaxies hosting the chosen sample of YMCs used for the estimation and visible using SKA telescope. The markers with bold outlines represent the $^{26}$Al yield in the stellar ejecta from \protect\cite{Higgins_2023}, while the markers without bold outlines represent the yields from \protect\cite{Martinet_2022}. Additionally, the figure features the pink dashed line indicating the line sensitivity of SKA2 for a 100-hour run, and a yellow dotted line for a 1000-hour run.}
        \label{fig:Ps_stacked}
\end{figure*}

Figure \ref{fig:Ps87alpha} illustrates the estimated fluxes for Ps87$\alpha$ radio recombination line in the same galaxies. As with Figure \ref{fig:AL26}, we present both the $^{26}$Al yield in the stellar ejecta from \cite{Higgins_2023} and the yields from \cite{Martinet_2022}. We compare these estimated fluxes to the projected sensitivity of the SKA Phase 2, whether for a 100-hour observation or even a 1,000-hour observation. We also examined the fluxes for Ps radio recombination lines in the case of stacked lines for both $^{26}$Al mass fractions (Figure \ref{fig:Ps_stacked}). Similar to the individual Ps87$\alpha$ line, SKA Phase 2 would not be able to detect the signal in any galaxy, even with stacking across these extreme frequencies.

\begin{figure*}
        \includegraphics[width=\linewidth]{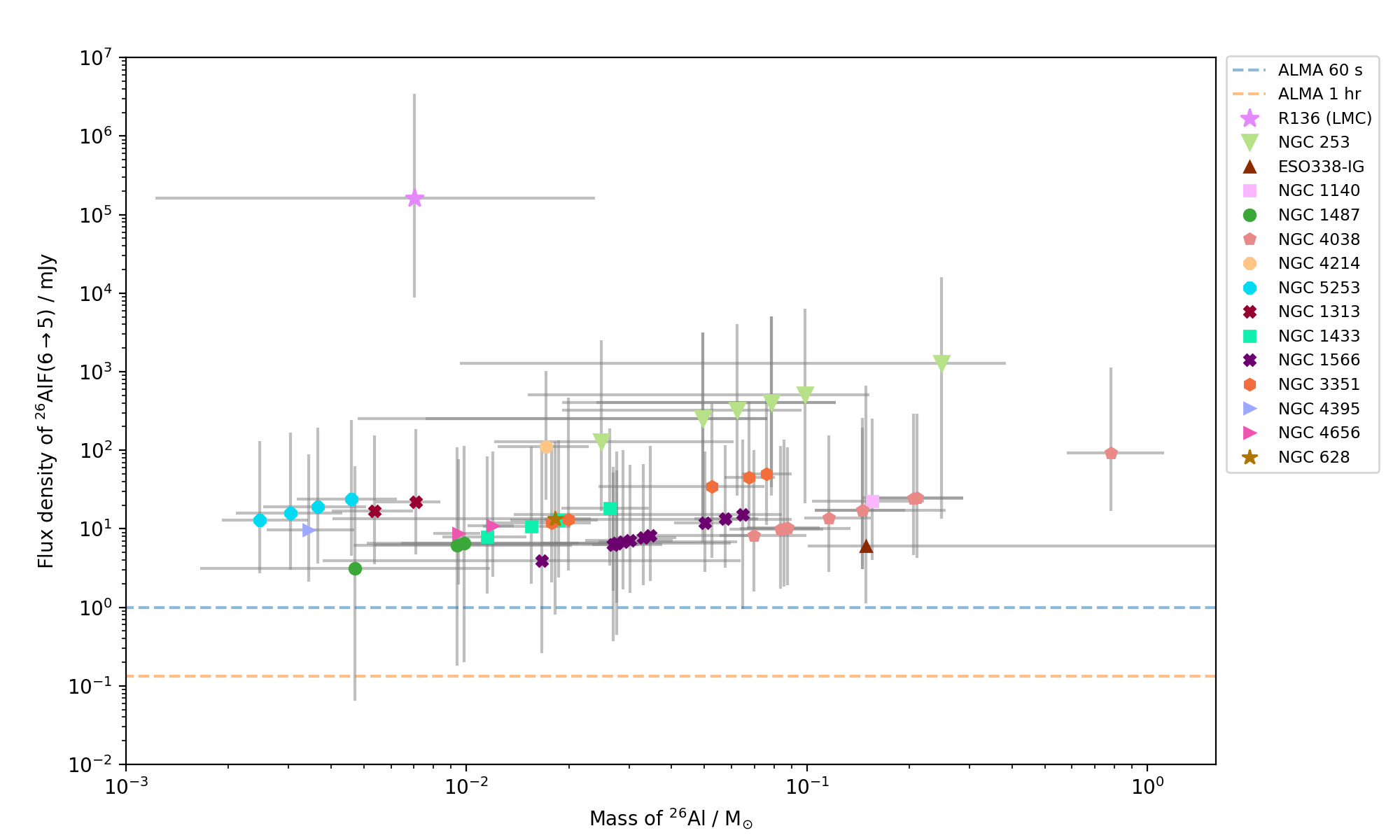}
        \caption{Flux estimations for rotational line (6$\rightarrow$5) for $^{26}$AlF molecule for each individual YMC in our sample. Galaxies hosting the selected YMCs are represented by different colored markers, observable using the ALMA telescope. Blue dashed line indicating the line sensitivity of ALMA with 43 antennas used for 60-seconds run, and an orange dashed one for a 1-hour run.}
        \label{fig:AlF_molecule}
\end{figure*}

The flux density for $^{26}$AlF rotational line (6$\rightarrow$5) for each individual YMC is estimated and grouped by host galaxy in Figure \ref{fig:AlF_molecule}. The highest flux is observed for R136 at 161.4 Jy, and 1.3 Jy for the cluster 12 in NGC 253. ALMA line sensitivity for exposure of 60~seconds and 1~hour is indicated with a blue dashed (orange) line. The exposure was calculated assuming 43 antennas, similarly to the observations performed by \cite{Kaminski_2018}. All the chosen star clusters are likely detectable using this ALMA setup with the exposure times of $\sim$60 seconds and $\sim$1 hour, giving us the best chances to measure $^{26}$Al in our sample (see Section \ref{discAlF} for further discussion).

\section{Discussion}\label{discussion}
\subsection{R136 cluster}
In this study, we have demonstrated that fluxes of the 1.8~MeV and the 511~keV lines for the chosen sample of YMCs fall beyond the detection capabilities of COSI or e-ASTROGAM for two year all-sky surveys with the published target sensitivities. However, R136, located in the 30 Doradus star-forming region, is an exception, as it is detectable within its error bars in some cases. This cluster harbors the most massive stars observed to date, with some reaching up to 300 M$_{\odot}$ \citep{Crowther_2010}. Notably, R136 contributes to approximately 30-50 per cent of the wind power in the 30 Doradus region \citep{Doran_2013}. Given its proximity to the Milky Way and its feedback contribution to the star formation region, R136 serves as an ideal laboratory for exploring the potential existence of multiple stellar populations. However, a targeted survey of the LMC with COSI would require a minimum of one year and three months to detect the predicted 1.8 MeV signal, even for the most optimistic scenario. In the more realistic scenario detecting the signal would require at least 22 years. If in the future more sensitive $\gamma$-ray instruments are developed and measure a signal for the 1.8 MeV line for the R136 cluster lower than $\approx$ 5~$\times$~10$^{-8}$~ph~cm$^{-2}$~s$^{-1}$, representing the lowest flux for the realistic case, then it would serve as evidence of no self-enrichment present in this YMC.

\subsection{NGC 253}
Furthermore, the starburst galaxy NGC~253, at a distance of 3.34~Mpc, exhibits the highest estimated flux of YMCs apart from the R136 for all the $^{26}$Al proxies for ongoing self-enrichment we studied. High-resolution observations with ALMA \citep{Leroy_2018, Levy_2021} confirm that the cluster population in NGC~253 may indeed be particularly suitable for our suggested test. In their study, \cite{Leroy_2018} found 14 candidates for proto-super star clusters (proto-SSCs). From this selection, we have chosen eight clusters that meet the criteria of hosting a massive polluter and being young enough to retain some $^{26}$Al. The authors argue that these proto-SSCs are still in the process of formation, as evidenced by their estimation of gas and stellar mass. Consequently, this environment provides an ideal setting for the detection of ejected $^{26}$Al through winds from massive stars. We note that cluster 11 which has the highest compactness index in our entire sample of YMCs also hosts a kilomaser, which \cite{Nowak_2022} suggested to interpret as an SMS accretion disc. As SMS are candidate polluters in the context of self-enrichment, this is a case in point for our compactness selection criterion.

\subsection{M82: background emission}
Another starburst galaxy, M82, located at the same distance as NGC~253, was chosen by \cite{Lacki_2014} to perform one-zone models of cosmic ray populations and investigate the MeV background for star-forming galaxies. Their model incorporated various parameters including the $^{26}$Al nuclear line, inverse Compton and bremsstrahlung, as well as positron annihilation lines and others. Their estimate for the 1.8~MeV line from normal star formation in M82 is 1.5~$\times$~10$^{-8}$~ph~cm$^{-2}$~s$^{-1}$, exceeding our estimations derived from Table \ref{tab:background} (9.0~$\times$~10$^{-9}$~ph~cm$^{-2}$~s$^{-1}$) by a less than a factor of two but still a factor of two below our estimate of the enhanced emission predicted on the basis of ongoing self-enrichment in M82's YMCs. The reference $^{26}$Al mass for the Milky Way is approximately twice in \cite{Lacki_2014} compared to what we used. The value for the 511~keV annihilation line flux reported here, 3.7~$\times$~10$^{-8}$~ph~cm$^{-2}$~s$^{-1}$, is around a factor of four larger than the estimate provided by \cite{Lacki_2014}, amounting to 1.0~$\times$~10$^{-8}$~ph~cm$^{-2}$~s$^{-1}$. Their M82 model takes into consideration additional positron sources, beyond the ones implied by our simple scaling of the Milky Way flux with mass. That the values actually agree within a factor of a few lends credibility to our approach. \cite{Siegert_2016} discuss various origins of positrons in the Galaxy, including contributions from supernovae, pair plasma ejection from accreting binary systems, and past AGN-jet activity of Sgr A$^*$. However, the origin of positrons is beyond the scope of this study (see \cite{Siegert_2023_positronreview} for a recent review). Since the uncertainties about the origins of positrons, apart from massive stars, are significant, we chose an empirical approach here, scaling the Milky Way flux at 511 keV with only the stellar mass of any given galaxy. 

\subsection{\texorpdfstring{$^{26}$AlF}{26AlF}}\label{discAlF}
The most promising prospect among all our estimations is the detection of $^{26}$AlF through its rotational line (6$\rightarrow$5) at a frequency of 200.95 GHz. With ALMA's capability to resolve individual YMCs, it should be able to detect individual signals and address any potential issues with beam dilution. Moreover, if the detection is weak, there is the potential to stack signals from each individual YMC within a specific galaxy.  However, this approach requires matching the signals both spatially and in velocity space. It is important to note that our assumptions include a 50 per cent binding of available $^{19}$F with $^{26}$Al to form this molecule. If this fraction were as low as one per cent, the flux for the rotational line from each YMC would decrease by up to two orders of magnitude, but most individual YMCs could still be detected with ALMA. As previously mentioned, the first observation of this species of molecule on a cosmic scale was conducted by \cite{Kaminski_2018}, and we are not aware of any other reported detections of this isotopologue of the aluminum monofluoride molecule. It is also worth mentioning that there are other molecules that carry aluminum, such as AlCl, AlO, AlOH, AlH, and AlCN. However, none of these molecules or their radioactive isotopologues containing $^{26}$Al were detected by \cite{Kaminski_2018}, but they might be observable in a YMC rather than an evolved star.

\subsection{Stellar evolution modelling}
In our calculations to estimate the fluxes for $^{26}$Al tracers, we employed data from two stellar evolution studies of VMSs: \cite{Martinet_2022}, who used the Geneva stellar evolution code, and \cite{Higgins_2023}, who relied on the MESA code. Both of those studies provided the fraction of $^{26}$Al in the stellar ejecta. The latter model predicts a higher $^{26}$Al fraction, with a larger percentage of ejecta from the initial mass at the end of the H-burning phase compared to \cite{Martinet_2022}. Their main difference is in the mass-loss prescriptions used. Specifically, \cite{Martinet_2022} employed the mass loss prescription outlined in \cite{Vink_2001}, while for Wolf-Rayet stars, they followed the prescription from \cite{Nugis_2000}. In contrast, \cite{Higgins_2023} utilized mass-loss rates that consider enhanced optically thick winds produced by VMS with high Eddington luminosities, as described in \cite{Vink_2011}. However, since the total loss cancels for the mass fractions, the results remain relatively similar, differing by only a factor of a few. Both of these studies computed the yields for VMS, which have been suggested as candidates for polluters in proto-globular clusters by \cite{Vink_2018}. In this scenario, VMS, with their inflated and cooler temperatures ($\simeq$~15~kK), are believed to have slower winds that do not exceed the escape velocity of GCs. This allows the yields from VMS to enrich the intracluster medium. \cite{Vink_2018} highlights the need for developing self-consistent and evolutionary models to test this hypothesis. 

Additionally, we considered the fraction of $^{26}$Al in the stellar winds of an SMS from detailed MESA model calculations from Ram\'irez-Galeano (in prep). Depending on the convection treatment in the stellar evolution code, the SMS can be fully convective or not. For the present study we used the low metallicity model since it is fully convective. If the SMS were not fully convective, we would not expect to find any $^{26}$Al in its stellar ejecta. Neither would the star be able to cause multiple stellar populations.

In any case, if we were to detect any of the lines mentioned in this paper, the efficiency of detection would be low. Most of the ejected material could be already lost from the intracluster medium or the cluster's environment at the time of the detection, either through accretion onto protostars or radioactive decay. As a result, we might only trace a small fraction of the material compared to the total amount ejected from potential polluters in YMCs within a given galaxy. 

\subsection{\texorpdfstring{Relation to stable $^{27}$Al}{Relation to stable 27Al}}
It is worth mentioning that a previous attempt to detect Al in YMCs, but in its stable form, was performed by \cite{Lard0_2017}. In their study, they tried to identify the presence of the Al variations in three super star clusters within the NGC~4038 galaxy, part of the Antennae systems. The clusters selected for analysis were older than our chosen range (10$^{6.8}$ - 10$^{7.6}$~yr), with the aim of exploring the potential existence of multiple populations in intermediate-age YMCs. Using J-band integrated spectra, they reported no observable Al variations in any of the chosen clusters, contrary to what is observed in most GCs. This lack of variation led them to conclude that multiple populations are not present in these YMCs. The unavailability of r$_{h}$ values for their sample of clusters prevents the calculation of the compactness index, a parameter that could potentially account for the absence of observable Al variations in their study.

\section{Conclusions}\label{conclusions}

We investigated possible tests of ongoing self-enrichment in YMCs that could lead to multiple populations, akin to those observed in old GCs. Specifically, we examined proxies for $^{26}$Al, which is likely produced in many polluter scenarios proposed to explain multiple populations. We selected YMCs younger than 10 Myr, massive and compact enough to trigger self-enrichment, and estimated fluxes for the 1.8 MeV and 511 keV lines in each galaxy. Our analysis revealed that these fluxes for almost all of our targets fall below the sensitivity thresholds of both the currently accepted $\gamma$-ray mission COSI and the proposed mission e-ASTROGAM for two-year all-sky surveys. 

R136 is the only YMC in our sample that is potentially detectable at 1.8 MeV and 511 keV by e-ASTROGRAM. A COSI targeted survey i.e. 25 per cent of the sky, lasting between one year and three months to 22 years, would be required to confirm our predicted signals for 1.8 MeV for the $^{26}$Al yield fractions from \cite{Higgins_2023} in both optimistic and realistic cases. Currently, such a survey is not viable.

We calculated the background emission for both lines in each galaxy by scaling values from the Milky Way, where none of the clusters currently meet the criteria for self-enrichment. Those resulted in 1.8 MeV fluxes of typically two orders of magnitude lower than the ones estimated using the ejecta in each YMC for most galaxies. Potentially detected fluxes of this line close to background levels would suggest a lack of self-enrichment in the observed YMCs, while values higher by a factor of a few to up to two orders of magnitude could indicate the presence of multiple populations in these young clusters. However, background emission estimates for the 511 keV line have proven to be an unreliable tool for this test due to large uncertainties in the background flux from any given galaxy. When comparing the mass of $^{26}$Al from the background contribution of massive stars to that produced by polluters in a self-enrichment scenario in YMCs, it is evident that some galaxies can have their $^{26}$Al mass increased by up to 31 times due to enrichment.

Similar to previous results, the planned SKA2 radio telescope would be unable to detect any single radio recombination line, such as Ps87$\alpha$, or the highly challenging stacked lines from Ps61$\alpha$ to Ps208$\alpha$ in most of the given galaxies. This limitation holds even with extended observation times of 100 or 1000 hours, due to its sensitivity limits. An exception might be R136, where the stacked line flux could potentially be detected within its error bars during a 1000-hour observation run, if such stacking is feasible.

Crucially, our study highlights the potential for detecting the rotational line (6$\rightarrow$5) of $^{26}$AlF using ALMA's 43 antennas, with an integration time as short as 60 seconds, emerging as a promising avenue for $^{26}$Al detection in our sample. Furthermore, ALMA capabilities would also allow us to pinpoint individual YMCs from our selection. Detection of this tracer in those objects would confirm the presence of $^{26}$Al in those forming stellar clusters, thereby further strengthening the scenario of self-enrichment as the cause of the chemical anomalies in GCs, as well as the assumption of YMCs being progenitors of GCs.

\section*{Acknowledgements}

The authors thank the reviewer, Chiaki Kobayashi and John A. Tomsick for their valuable comments and suggestions. This work was supported by Science and Technology Facilities Council (ST/V506709/1).

\section*{Data Availability}

The data presented in this article will be shared on reasonable request to the corresponding author.



\bibliographystyle{mnras}
\bibliography{bibliography} 

\begin{thebibliography}{}
\makeatletter
\relax
\def\mn@urlcharsother{\let\do\@makeother \do\$\do\&\do\#\do\^\do\_\do\%\do\~}
\def\mn@doi{\begingroup\mn@urlcharsother \@ifnextchar [ {\mn@doi@}
  {\mn@doi@[]}}
\def\mn@doi@[#1]#2{\def\@tempa{#1}\ifx\@tempa\@empty \href
  {http://dx.doi.org/#2} {doi:#2}\else \href {http://dx.doi.org/#2} {#1}\fi
  \endgroup}
\def\mn@eprint#1#2{\mn@eprint@#1:#2::\@nil}
\def\mn@eprint@arXiv#1{\href {http://arxiv.org/abs/#1} {{\tt arXiv:#1}}}
\def\mn@eprint@dblp#1{\href {http://dblp.uni-trier.de/rec/bibtex/#1.xml}
  {dblp:#1}}
\def\mn@eprint@#1:#2:#3:#4\@nil{\def\@tempa {#1}\def\@tempb {#2}\def\@tempc
  {#3}\ifx \@tempc \@empty \let \@tempc \@tempb \let \@tempb \@tempa \fi \ifx
  \@tempb \@empty \def\@tempb {arXiv}\fi \@ifundefined
  {mn@eprint@\@tempb}{\@tempb:\@tempc}{\expandafter \expandafter \csname
  mn@eprint@\@tempb\endcsname \expandafter{\@tempc}}}

\bibitem[\protect\citeauthoryear{{Adamo} et~al.,}{{Adamo}
  et~al.}{2017}]{Adamo_2017}
{Adamo} A.,  et~al., 2017, \mn@doi [\apj] {10.3847/1538-4357/aa7132}, \href
  {https://ui.adsabs.harvard.edu/abs/2017ApJ...841..131A} {841, 131}

\bibitem[\protect\citeauthoryear{{Anantharamaiah}, {Radhakrishnan}, {Morris},
  {Vivekanand}, {Downes}  \& {Shukre}}{{Anantharamaiah}
  et~al.}{1989}]{Anantharamaiah_1989}
{Anantharamaiah} K.~R.,  {Radhakrishnan} V.,  {Morris} D.,  {Vivekanand} M.,
  {Downes} D.,   {Shukre} C.~S.,  1989, in {Morris} M.,  ed.,  The 136th
  Symposium of the International Astronomical Union Vol. 136, The Center of the
  Galaxy. pp 607--616

\bibitem[\protect\citeauthoryear{{Andersen}, {Zinnecker}, {Moneti},
  {McCaughrean}, {Brandl}, {Brandner}, {Meylan}  \& {Hunter}}{{Andersen}
  et~al.}{2009}]{Andersen_2009}
{Andersen} M.,  {Zinnecker} H.,  {Moneti} A.,  {McCaughrean} M.~J.,  {Brandl}
  B.,  {Brandner} W.,  {Meylan} G.,   {Hunter} D.,  2009, \mn@doi [\apj]
  {10.1088/0004-637X/707/2/1347}, \href
  {https://ui.adsabs.harvard.edu/abs/2009ApJ...707.1347A} {707, 1347}

\bibitem[\protect\citeauthoryear{{Anderson}}{{Anderson}}{2002}]{Anderson_2002}
{Anderson} J.,  2002, in {van Leeuwen} F.,  {Hughes} J.~D.,   {Piotto} G.,
  eds,  Astronomical Society of the Pacific Conference Series Vol. 265, Omega
  Centauri, A Unique Window into Astrophysics. p.~87

\bibitem[\protect\citeauthoryear{{Asplund}, {Grevesse}, {Sauval}  \&
  {Scott}}{{Asplund} et~al.}{2009}]{Asplund_2009}
{Asplund} M.,  {Grevesse} N.,  {Sauval} A.~J.,   {Scott} P.,  2009, \mn@doi
  [\araa] {10.1146/annurev.astro.46.060407.145222}, \href
  {https://ui.adsabs.harvard.edu/abs/2009ARA&A..47..481A} {47, 481}

\bibitem[\protect\citeauthoryear{{Bailin}, {Bell}, {Chappell}, {Radburn-Smith}
  \& {de Jong}}{{Bailin} et~al.}{2011}]{Bailin_2011}
{Bailin} J.,  {Bell} E.~F.,  {Chappell} S.~N.,  {Radburn-Smith} D.~J.,   {de
  Jong} R.~S.,  2011, \mn@doi [\apj] {10.1088/0004-637X/736/1/24}, \href
  {https://ui.adsabs.harvard.edu/abs/2011ApJ...736...24B} {736, 24}

\bibitem[\protect\citeauthoryear{{Bastian} \& {Lardo}}{{Bastian} \&
  {Lardo}}{2018}]{Bastian_2018}
{Bastian} N.,  {Lardo} C.,  2018, \mn@doi [\araa]
  {10.1146/annurev-astro-081817-051839}, \href
  {https://ui.adsabs.harvard.edu/abs/2018ARA&A..56...83B} {56, 83}

\bibitem[\protect\citeauthoryear{{Bastian}, {Saglia}, {Goudfrooij},
  {Kissler-Patig}, {Maraston}, {Schweizer}  \& {Zoccali}}{{Bastian}
  et~al.}{2006}]{Bastian_2006}
{Bastian} N.,  {Saglia} R.~P.,  {Goudfrooij} P.,  {Kissler-Patig} M.,
  {Maraston} C.,  {Schweizer} F.,   {Zoccali} M.,  2006, \mn@doi [\aap]
  {10.1051/0004-6361:20054177}, \href
  {https://ui.adsabs.harvard.edu/abs/2006A&A...448..881B} {448, 881}

\bibitem[\protect\citeauthoryear{{Bastian}, {Gieles}, {Goodwin}, {Trancho},
  {Smith}, {Konstantopoulos}  \& {Efremov}}{{Bastian}
  et~al.}{2008}]{Bastian_2008}
{Bastian} N.,  {Gieles} M.,  {Goodwin} S.~P.,  {Trancho} G.,  {Smith} L.~J.,
  {Konstantopoulos} I.,   {Efremov} Y.,  2008, \mn@doi [\mnras]
  {10.1111/j.1365-2966.2008.13547.x}, \href
  {https://ui.adsabs.harvard.edu/abs/2008MNRAS.389..223B} {389, 223}

\bibitem[\protect\citeauthoryear{{Bastian}, {Hollyhead}  \&
  {Cabrera-Ziri}}{{Bastian} et~al.}{2014}]{Bastian_2014}
{Bastian} N.,  {Hollyhead} K.,   {Cabrera-Ziri} I.,  2014, \mn@doi [\mnras]
  {10.1093/mnras/stu1775}, \href
  {https://ui.adsabs.harvard.edu/abs/2014MNRAS.445..378B} {445, 378}

\bibitem[\protect\citeauthoryear{{Bastian}, {Cabrera-Ziri}  \&
  {Salaris}}{{Bastian} et~al.}{2015}]{Bastian_2015}
{Bastian} N.,  {Cabrera-Ziri} I.,   {Salaris} M.,  2015, \mn@doi [\mnras]
  {10.1093/mnras/stv543}, \href
  {https://ui.adsabs.harvard.edu/abs/2015MNRAS.449.3333B} {449, 3333}

\bibitem[\protect\citeauthoryear{{Basunia} \& {Hurst}}{{Basunia} \&
  {Hurst}}{2016}]{Basunia_2016}
{Basunia} M.~S.,  {Hurst} A.~M.,  2016, \mn@doi [Nuclear Data Sheets]
  {10.1016/j.nds.2016.04.001}, \href
  {https://ui.adsabs.harvard.edu/abs/2016NDS...134....1B} {134, 1}

\bibitem[\protect\citeauthoryear{{Baumgardt} \& {Hilker}}{{Baumgardt} \&
  {Hilker}}{2018}]{Baumgardt_2018}
{Baumgardt} H.,  {Hilker} M.,  2018, \mn@doi [\mnras] {10.1093/mnras/sty1057},
  \href {https://ui.adsabs.harvard.edu/abs/2018MNRAS.478.1520B} {478, 1520}

\bibitem[\protect\citeauthoryear{{Beck} et~al.,}{{Beck}
  et~al.}{2022}]{Beck_2022}
{Beck} A.,  et~al., 2022, \mn@doi [\aap] {10.1051/0004-6361/202243822}, \href
  {https://ui.adsabs.harvard.edu/abs/2022A&A...665A..85B} {665, A85}

\bibitem[\protect\citeauthoryear{{Bedin}, {Piotto}, {Anderson}, {Cassisi},
  {King}, {Momany}  \& {Carraro}}{{Bedin} et~al.}{2004}]{Bedin_2004}
{Bedin} L.~R.,  {Piotto} G.,  {Anderson} J.,  {Cassisi} S.,  {King} I.~R.,
  {Momany} Y.,   {Carraro} G.,  2004, \mn@doi [\apjl] {10.1086/420847}, \href
  {https://ui.adsabs.harvard.edu/abs/2004ApJ...605L.125B} {605, L125}

\bibitem[\protect\citeauthoryear{{Bennett} et~al.,}{{Bennett}
  et~al.}{2013}]{Bennett_2013}
{Bennett} M.~B.,  et~al., 2013, \mn@doi [\prl]
  {10.1103/PhysRevLett.111.232503}, \href
  {https://ui.adsabs.harvard.edu/abs/2013PhRvL.111w2503B} {111, 232503}

\bibitem[\protect\citeauthoryear{{Bik}, {{\"O}stlin}, {Menacho}, {Adamo},
  {Hayes}, {Herenz}  \& {Melinder}}{{Bik} et~al.}{2018}]{Bik_2018}
{Bik} A.,  {{\"O}stlin} G.,  {Menacho} V.,  {Adamo} A.,  {Hayes} M.,  {Herenz}
  E.~C.,   {Melinder} J.,  2018, \mn@doi [\aap] {10.1051/0004-6361/201833916},
  \href {https://ui.adsabs.harvard.edu/abs/2018A&A...619A.131B} {619, A131}

\bibitem[\protect\citeauthoryear{{Bouchet}, {Jourdain}  \& {Roques}}{{Bouchet}
  et~al.}{2015}]{Bouchet_2015}
{Bouchet} L.,  {Jourdain} E.,   {Roques} J.-P.,  2015, \mn@doi [\apj]
  {10.1088/0004-637X/801/2/142}, \href
  {https://ui.adsabs.harvard.edu/abs/2015ApJ...801..142B} {801, 142}

\bibitem[\protect\citeauthoryear{{Braun}, {Bonaldi}, {Bourke}, {Keane}  \&
  {Wagg}}{{Braun} et~al.}{2019}]{Braun_2019}
{Braun} R.,  {Bonaldi} A.,  {Bourke} T.,  {Keane} E.,   {Wagg} J.,  2019,
  \mn@doi [arXiv e-prints] {10.48550/arXiv.1912.12699}, \href
  {https://ui.adsabs.harvard.edu/abs/2019arXiv191212699B} {p. arXiv:1912.12699}

\bibitem[\protect\citeauthoryear{{Bresolin}, {Kudritzki}  \&
  {Urbaneja}}{{Bresolin} et~al.}{2022}]{Bresolin_2022}
{Bresolin} F.,  {Kudritzki} R.-P.,   {Urbaneja} M.~A.,  2022, \mn@doi [\apj]
  {10.3847/1538-4357/ac9584}, \href
  {https://ui.adsabs.harvard.edu/abs/2022ApJ...940...32B} {940, 32}

\bibitem[\protect\citeauthoryear{{Brinkman}, {Doherty}, {Pols}, {Li},
  {C{\^o}t{\'e}}  \& {Lugaro}}{{Brinkman} et~al.}{2019}]{Brinkman_2019}
{Brinkman} H.~E.,  {Doherty} C.~L.,  {Pols} O.~R.,  {Li} E.~T.,  {C{\^o}t{\'e}}
  B.,   {Lugaro} M.,  2019, \mn@doi [\apj] {10.3847/1538-4357/ab40ae}, \href
  {https://ui.adsabs.harvard.edu/abs/2019ApJ...884...38B} {884, 38}

\bibitem[\protect\citeauthoryear{{Brown} \& {Gnedin}}{{Brown} \&
  {Gnedin}}{2021}]{Brown_2021}
{Brown} G.,  {Gnedin} O.~Y.,  2021, \mn@doi [\mnras] {10.1093/mnras/stab2907},
  \href {https://ui.adsabs.harvard.edu/abs/2021MNRAS.508.5935B} {508, 5935}

\bibitem[\protect\citeauthoryear{{Brown} \& {Leventhal}}{{Brown} \&
  {Leventhal}}{1987}]{Brown_1987}
{Brown} B.~L.,  {Leventhal} M.,  1987, \mn@doi [\apj] {10.1086/165484}, \href
  {https://ui.adsabs.harvard.edu/abs/1987ApJ...319..637B} {319, 637}

\bibitem[\protect\citeauthoryear{{Buzzo} et~al.,}{{Buzzo}
  et~al.}{2021}]{Buzzo_2021}
{Buzzo} M.~L.,  et~al., 2021, \mn@doi [\mnras] {10.1093/mnras/stab426}, \href
  {https://ui.adsabs.harvard.edu/abs/2021MNRAS.503..106B} {503, 106}

\bibitem[\protect\citeauthoryear{{Cabrera-Ziri}, {Bastian}, {Davies}, {Magris},
  {Bruzual}  \& {Schweizer}}{{Cabrera-Ziri} et~al.}{2014}]{Cabrera_Ziri_2014}
{Cabrera-Ziri} I.,  {Bastian} N.,  {Davies} B.,  {Magris} G.,  {Bruzual} G.,
  {Schweizer} F.,  2014, \mn@doi [\mnras] {10.1093/mnras/stu764}, \href
  {https://ui.adsabs.harvard.edu/abs/2014MNRAS.441.2754C} {441, 2754}

\bibitem[\protect\citeauthoryear{{Cabrera-Ziri} et~al.,}{{Cabrera-Ziri}
  et~al.}{2015}]{Cabrera_Ziri_2015}
{Cabrera-Ziri} I.,  et~al., 2015, \mn@doi [\mnras] {10.1093/mnras/stv163},
  \href {https://ui.adsabs.harvard.edu/abs/2015MNRAS.448.2224C} {448, 2224}

\bibitem[\protect\citeauthoryear{{Cadelano}, {Dalessandro}, {Salaris},
  {Bastian}, {Mucciarelli}, {Saracino}, {Martocchia}  \&
  {Cabrera-Ziri}}{{Cadelano} et~al.}{2022}]{Cadelano_2022}
{Cadelano} M.,  {Dalessandro} E.,  {Salaris} M.,  {Bastian} N.,  {Mucciarelli}
  A.,  {Saracino} S.,  {Martocchia} S.,   {Cabrera-Ziri} I.,  2022, \mn@doi
  [\apjl] {10.3847/2041-8213/ac424a}, \href
  {https://ui.adsabs.harvard.edu/abs/2022ApJ...924L...2C} {924, L2}

\bibitem[\protect\citeauthoryear{{Calzetti} et~al.,}{{Calzetti}
  et~al.}{2015}]{Calzetti_2015}
{Calzetti} D.,  et~al., 2015, \mn@doi [\aj] {10.1088/0004-6256/149/2/51}, \href
  {https://ui.adsabs.harvard.edu/abs/2015AJ....149...51C} {149, 51}

\bibitem[\protect\citeauthoryear{{Calzetti} et~al.,}{{Calzetti}
  et~al.}{2023}]{Calzetti_2023}
{Calzetti} D.,  et~al., 2023, \mn@doi [\apj] {10.3847/1538-4357/acbeac}, \href
  {https://ui.adsabs.harvard.edu/abs/2023ApJ...946....1C} {946, 1}

\bibitem[\protect\citeauthoryear{{Canete} et~al.,}{{Canete}
  et~al.}{2023}]{Canete_2023}
{Canete} L.,  et~al., 2023, \mn@doi [\prc] {10.1103/PhysRevC.108.035807}, \href
  {https://ui.adsabs.harvard.edu/abs/2023PhRvC.108c5807C} {108, 035807}

\bibitem[\protect\citeauthoryear{Carretta, Bragaglia, Gratton, Recio-Blanco,
  Lucatello, D’Orazi  \& Cassisi}{Carretta et~al.}{2010}]{Carretta_2010}
Carretta E.,  Bragaglia A.,  Gratton R.~G.,  Recio-Blanco A.,  Lucatello S.,
  D’Orazi V.,   Cassisi S.,  2010, \mn@doi [\aap]
  {10.1051/0004-6361/200913451}, 516, A55

\bibitem[\protect\citeauthoryear{{Chantereau}, {Charbonnel}  \&
  {Meynet}}{{Chantereau} et~al.}{2016}]{Chantereau_2016}
{Chantereau} W.,  {Charbonnel} C.,   {Meynet} G.,  2016, \mn@doi [\aap]
  {10.1051/0004-6361/201628418}, \href
  {https://ui.adsabs.harvard.edu/abs/2016A&A...592A.111C} {592, A111}

\bibitem[\protect\citeauthoryear{Charbonnel}{Charbonnel}{2016}]{Charbonnel_2016}
Charbonnel C.,  2016, \mn@doi [EAS Publications Series] {10.1051/eas/1680006},
  80–81, 177–226

\bibitem[\protect\citeauthoryear{{Chiang} \& {Kong}}{{Chiang} \&
  {Kong}}{2011}]{Chiang_2011}
{Chiang} Y.-K.,  {Kong} A. K.~H.,  2011, \mn@doi [\mnras]
  {10.1111/j.1365-2966.2011.18466.x}, \href
  {https://ui.adsabs.harvard.edu/abs/2011MNRAS.414.1329C} {414, 1329}

\bibitem[\protect\citeauthoryear{{Churazov}, {Sunyaev}, {Sazonov}, {Revnivtsev}
   \& {Varshalovich}}{{Churazov} et~al.}{2005}]{Churazov_2005}
{Churazov} E.,  {Sunyaev} R.,  {Sazonov} S.,  {Revnivtsev} M.,   {Varshalovich}
  D.,  2005, \mn@doi [\mnras] {10.1111/j.1365-2966.2005.08757.x}, \href
  {https://ui.adsabs.harvard.edu/abs/2005MNRAS.357.1377C} {357, 1377}

\bibitem[\protect\citeauthoryear{{Churazov}, {Sazonov}, {Tsygankov}, {Sunyaev}
  \& {Varshalovich}}{{Churazov} et~al.}{2011}]{Churazov_2011}
{Churazov} E.,  {Sazonov} S.,  {Tsygankov} S.,  {Sunyaev} R.,   {Varshalovich}
  D.,  2011, \mn@doi [\mnras] {10.1111/j.1365-2966.2010.17804.x}, \href
  {https://ui.adsabs.harvard.edu/abs/2011MNRAS.411.1727C} {411, 1727}

\bibitem[\protect\citeauthoryear{{Conroy} \& {Spergel}}{{Conroy} \&
  {Spergel}}{2011}]{Conroy_2011}
{Conroy} C.,  {Spergel} D.~N.,  2011, \mn@doi [\apj]
  {10.1088/0004-637X/726/1/36}, \href
  {https://ui.adsabs.harvard.edu/abs/2011ApJ...726...36C} {726, 36}

\bibitem[\protect\citeauthoryear{{Cook} et~al.,}{{Cook}
  et~al.}{2019}]{Cook_2019}
{Cook} D.~O.,  et~al., 2019, \mn@doi [\mnras] {10.1093/mnras/stz331}, \href
  {https://ui.adsabs.harvard.edu/abs/2019MNRAS.484.4897C} {484, 4897}

\bibitem[\protect\citeauthoryear{{Crowther}, {Schnurr}, {Hirschi}, {Yusof},
  {Parker}, {Goodwin}  \& {Kassim}}{{Crowther} et~al.}{2010}]{Crowther_2010}
{Crowther} P.~A.,  {Schnurr} O.,  {Hirschi} R.,  {Yusof} N.,  {Parker} R.~J.,
  {Goodwin} S.~P.,   {Kassim} H.~A.,  2010, \mn@doi [\mnras]
  {10.1111/j.1365-2966.2010.17167.x}, \href
  {https://ui.adsabs.harvard.edu/abs/2010MNRAS.408..731C} {408, 731}

\bibitem[\protect\citeauthoryear{{Crowther} et~al.,}{{Crowther}
  et~al.}{2016}]{Crowther_2016}
{Crowther} P.~A.,  et~al., 2016, \mn@doi [\mnras] {10.1093/mnras/stw273}, \href
  {https://ui.adsabs.harvard.edu/abs/2016MNRAS.458..624C} {458, 624}

\bibitem[\protect\citeauthoryear{{D'Antona}, {Bellazzini}, {Caloi}, {Pecci},
  {Galleti}  \& {Rood}}{{D'Antona} et~al.}{2005}]{D'Antona_2005}
{D'Antona} F.,  {Bellazzini} M.,  {Caloi} V.,  {Pecci} F.~F.,  {Galleti} S.,
  {Rood} R.~T.,  2005, \mn@doi [\apj] {10.1086/431968}, \href
  {https://ui.adsabs.harvard.edu/abs/2005ApJ...631..868D} {631, 868}

\bibitem[\protect\citeauthoryear{{D'Ercole}, {Vesperini}, {D'Antona},
  {McMillan}  \& {Recchi}}{{D'Ercole} et~al.}{2008}]{DErcole_2008}
{D'Ercole} A.,  {Vesperini} E.,  {D'Antona} F.,  {McMillan} S. L.~W.,
  {Recchi} S.,  2008, \mn@doi [\mnras] {10.1111/j.1365-2966.2008.13915.x},
  \href {https://ui.adsabs.harvard.edu/abs/2008MNRAS.391..825D} {391, 825}

\bibitem[\protect\citeauthoryear{{De Angelis} et~al.,}{{De Angelis}
  et~al.}{2017}]{DeAngelis_2017}
{De Angelis} A.,  et~al., 2017, \mn@doi [Experimental Astronomy]
  {10.1007/s10686-017-9533-6}, \href
  {https://ui.adsabs.harvard.edu/abs/2017ExA....44...25D} {44, 25}

\bibitem[\protect\citeauthoryear{{De Cia} et~al.,}{{De Cia}
  et~al.}{2024}]{DeCia_2024}
{De Cia} A.,  et~al., 2024, \mn@doi [arXiv e-prints]
  {10.48550/arXiv.2401.07963}, \href
  {https://ui.adsabs.harvard.edu/abs/2024arXiv240107963D} {p. arXiv:2401.07963}

\bibitem[\protect\citeauthoryear{{Decressin}, {Meynet}, {Charbonnel},
  {Prantzos}  \& {Ekstr{\"o}m}}{{Decressin} et~al.}{2007}]{Decressin_2007}
{Decressin} T.,  {Meynet} G.,  {Charbonnel} C.,  {Prantzos} N.,   {Ekstr{\"o}m}
  S.,  2007, \mn@doi [\aap] {10.1051/0004-6361:20066013}, \href
  {https://ui.adsabs.harvard.edu/abs/2007A&A...464.1029D} {464, 1029}

\bibitem[\protect\citeauthoryear{{Decressin}, {Charbonnel}, {Siess},
  {Palacios}, {Meynet}  \& {Georgy}}{{Decressin} et~al.}{2009}]{Decressin_2009}
{Decressin} T.,  {Charbonnel} C.,  {Siess} L.,  {Palacios} A.,  {Meynet} G.,
  {Georgy} C.,  2009, \mn@doi [\aap] {10.1051/0004-6361/200911822}, \href
  {https://ui.adsabs.harvard.edu/abs/2009A&A...505..727D} {505, 727}

\bibitem[\protect\citeauthoryear{{Denissenkov} \& {Denisenkova}}{{Denissenkov}
  \& {Denisenkova}}{1990}]{Denisenkov_1990}
{Denissenkov} P.~A.,  {Denisenkova} S.~N.,  1990, Soviet Astronomy Letters,
  \href {https://ui.adsabs.harvard.edu/abs/1990SvAL...16..275D} {16, 275}

\bibitem[\protect\citeauthoryear{{Denissenkov} \& {Hartwick}}{{Denissenkov} \&
  {Hartwick}}{2014}]{Denissenkov_2014}
{Denissenkov} P.~A.,  {Hartwick} F.~D.~A.,  2014, \mn@doi [\mnras]
  {10.1093/mnrasl/slt133}, 437, L21

\bibitem[\protect\citeauthoryear{Denissenkov \& Herwig}{Denissenkov \&
  Herwig}{2003}]{Denissenkov_2003}
Denissenkov P.~A.,  Herwig F.,  2003, \mn@doi [\apj] {10.1086/376748}, 590,
  L99–L102

\bibitem[\protect\citeauthoryear{{Denissenkov}, {VandenBerg}, {Hartwick},
  {Herwig}, {Weiss}  \& {Paxton}}{{Denissenkov}
  et~al.}{2015}]{Denissenkov_2015}
{Denissenkov} P.~A.,  {VandenBerg} D.~A.,  {Hartwick} F.~D.~A.,  {Herwig} F.,
  {Weiss} A.,   {Paxton} B.,  2015, \mn@doi [\mnras] {10.1093/mnras/stv211},
  \href {https://ui.adsabs.harvard.edu/abs/2015MNRAS.448.3314D} {448, 3314}

\bibitem[\protect\citeauthoryear{{Deutsch}}{{Deutsch}}{1951}]{Deutsch_1951}
{Deutsch} M.,  1951, \mn@doi [Physical Review] {10.1103/PhysRev.83.866}, \href
  {https://ui.adsabs.harvard.edu/abs/1951PhRv...83..866D} {83, 866}

\bibitem[\protect\citeauthoryear{{Dewdney}, {Hall}, {Schilizzi}  \&
  {Lazio}}{{Dewdney} et~al.}{2009}]{Dewdney_2009}
{Dewdney} P.~E.,  {Hall} P.~J.,  {Schilizzi} R.~T.,   {Lazio} T.~J.~L.~W.,
  2009, \mn@doi [IEEE Proceedings] {10.1109/JPROC.2009.2021005}, \href
  {https://ui.adsabs.harvard.edu/abs/2009IEEEP..97.1482D} {97, 1482}

\bibitem[\protect\citeauthoryear{{Diehl} \& {Timmes}}{{Diehl} \&
  {Timmes}}{1998}]{Diehl_1998}
{Diehl} R.,  {Timmes} F.~X.,  1998, \mn@doi [\pasp] {10.1086/316169}, \href
  {https://ui.adsabs.harvard.edu/abs/1998PASP..110..637D} {110, 637}

\bibitem[\protect\citeauthoryear{{Diehl} et~al.,}{{Diehl}
  et~al.}{2006}]{Diehl_2006}
{Diehl} R.,  et~al., 2006, \mn@doi [\aap] {10.1051/0004-6361:20054301}, \href
  {https://ui.adsabs.harvard.edu/abs/2006A&A...449.1025D} {449, 1025}

\bibitem[\protect\citeauthoryear{{Diehl} et~al.,}{{Diehl}
  et~al.}{2021}]{Diehl_2021}
{Diehl} R.,  et~al., 2021, \mn@doi [\pasa] {10.1017/pasa.2021.48}, \href
  {https://ui.adsabs.harvard.edu/abs/2021PASA...38...62D} {38, e062}

\bibitem[\protect\citeauthoryear{{Doherty}, {Gil-Pons}, {Lau}, {Lattanzio}  \&
  {Siess}}{{Doherty} et~al.}{2014}]{Doherty_2014}
{Doherty} C.~L.,  {Gil-Pons} P.,  {Lau} H. H.~B.,  {Lattanzio} J.~C.,   {Siess}
  L.,  2014, \mn@doi [\mnras] {10.1093/mnras/stt1877}, \href
  {https://ui.adsabs.harvard.edu/abs/2014MNRAS.437..195D} {437, 195}

\bibitem[\protect\citeauthoryear{{Dondoglio}, {Milone}, {Lagioia}, {Marino},
  {Tailo}, {Cordoni}, {Jang}  \& {Carlos}}{{Dondoglio}
  et~al.}{2021}]{Dondoglio_2021}
{Dondoglio} E.,  {Milone} A.~P.,  {Lagioia} E.~P.,  {Marino} A.~F.,  {Tailo}
  M.,  {Cordoni} G.,  {Jang} S.,   {Carlos} M.,  2021, \mn@doi [\apj]
  {10.3847/1538-4357/abc882}, \href
  {https://ui.adsabs.harvard.edu/abs/2021ApJ...906...76D} {906, 76}

\bibitem[\protect\citeauthoryear{{Dopita} et~al.,}{{Dopita}
  et~al.}{2010}]{Dopita_2010}
{Dopita} M.~A.,  et~al., 2010, \mn@doi [\apss] {10.1007/s10509-010-0376-0},
  \href {https://ui.adsabs.harvard.edu/abs/2010Ap&SS.330..123D} {330, 123}

\bibitem[\protect\citeauthoryear{{Doran} et~al.,}{{Doran}
  et~al.}{2013}]{Doran_2013}
{Doran} E.~I.,  et~al., 2013, \mn@doi [\aap] {10.1051/0004-6361/201321824},
  \href {https://ui.adsabs.harvard.edu/abs/2013A&A...558A.134D} {558, A134}

\bibitem[\protect\citeauthoryear{{Ellis} \& {Bland-Hawthorn}}{{Ellis} \&
  {Bland-Hawthorn}}{2009}]{Ellis_2009}
{Ellis} S.~C.,  {Bland-Hawthorn} J.,  2009, \mn@doi [\apj]
  {10.1088/0004-637X/707/1/457}, \href
  {https://ui.adsabs.harvard.edu/abs/2009ApJ...707..457E} {707, 457}

\bibitem[\protect\citeauthoryear{{Engelbracht}, {Rieke}, {Gordon}, {Smith},
  {Werner}, {Moustakas}, {Willmer}  \& {Vanzi}}{{Engelbracht}
  et~al.}{2008}]{Engelbracht_2008}
{Engelbracht} C.~W.,  {Rieke} G.~H.,  {Gordon} K.~D.,  {Smith} J. D.~T.,
  {Werner} M.~W.,  {Moustakas} J.,  {Willmer} C.~N.~A.,   {Vanzi} L.,  2008,
  \mn@doi [\apj] {10.1086/529513}, \href
  {https://ui.adsabs.harvard.edu/abs/2008ApJ...678..804E} {678, 804}

\bibitem[\protect\citeauthoryear{{Fahrion} et~al.,}{{Fahrion}
  et~al.}{2022}]{Fahrion_2022}
{Fahrion} K.,  et~al., 2022, \mn@doi [\aap] {10.1051/0004-6361/202244932},
  \href {https://ui.adsabs.harvard.edu/abs/2022A&A...667A.101F} {667, A101}

\bibitem[\protect\citeauthoryear{{Forestini} \& {Charbonnel}}{{Forestini} \&
  {Charbonnel}}{1997}]{Forestini_1997}
{Forestini} M.,  {Charbonnel} C.,  1997, \mn@doi [\aaps] {10.1051/aas:1997348},
  \href {https://ui.adsabs.harvard.edu/abs/1997A&AS..123..241F} {123, 241}

\bibitem[\protect\citeauthoryear{{F{\"o}rster Schreiber}, {Genzel}, {Lutz},
  {Kunze}  \& {Sternberg}}{{F{\"o}rster Schreiber}
  et~al.}{2001}]{Schreiber_2001}
{F{\"o}rster Schreiber} N.~M.,  {Genzel} R.,  {Lutz} D.,  {Kunze} D.,
  {Sternberg} A.,  2001, \mn@doi [\apj] {10.1086/320546}, \href
  {https://ui.adsabs.harvard.edu/abs/2001ApJ...552..544F} {552, 544}

\bibitem[\protect\citeauthoryear{{Franco} et~al.,}{{Franco}
  et~al.}{2021}]{Franco_2021}
{Franco} M.,  et~al., 2021, \mn@doi [Nature Astronomy]
  {10.1038/s41550-021-01515-9}, \href
  {https://ui.adsabs.harvard.edu/abs/2021NatAs...5.1240F} {5, 1240}

\bibitem[\protect\citeauthoryear{{Freedman} \& {Madore}}{{Freedman} \&
  {Madore}}{1988}]{Freedman_1988}
{Freedman} W.~L.,  {Madore} B.~F.,  1988, \mn@doi [\apjl] {10.1086/185267},
  \href {https://ui.adsabs.harvard.edu/abs/1988ApJ...332L..63F} {332, L63}

\bibitem[\protect\citeauthoryear{{Freedman} et~al.,}{{Freedman}
  et~al.}{1994}]{Freedman_1994}
{Freedman} W.~L.,  et~al., 1994, \mn@doi [\apj] {10.1086/174172}, \href
  {https://ui.adsabs.harvard.edu/abs/1994ApJ...427..628F} {427, 628}

\bibitem[\protect\citeauthoryear{{Fumagalli}, {Krumholz}  \&
  {Hunt}}{{Fumagalli} et~al.}{2010}]{Fumagalli_2010}
{Fumagalli} M.,  {Krumholz} M.~R.,   {Hunt} L.~K.,  2010, \mn@doi [\apj]
  {10.1088/0004-637X/722/1/919}, \href
  {https://ui.adsabs.harvard.edu/abs/2010ApJ...722..919F} {722, 919}

\bibitem[\protect\citeauthoryear{Gieles et~al.,}{Gieles
  et~al.}{2018}]{Gieles_2018}
Gieles M.,  et~al., 2018, \mn@doi [\mnras] {10.1093/mnras/sty1059}, 478,
  2461–2479

\bibitem[\protect\citeauthoryear{Gratton, Carretta  \& Bragaglia}{Gratton
  et~al.}{2012}]{Gratton_2012}
Gratton R.~G.,  Carretta E.,   Bragaglia A.,  2012, \mn@doi [\aapr]
  {10.1007/s00159-012-0050-3}, 20

\bibitem[\protect\citeauthoryear{{Greggio}, {Tosi}, {Clampin}, {De Marchi},
  {Leitherer}, {Nota}  \& {Sirianni}}{{Greggio} et~al.}{1998}]{Greggio_1998}
{Greggio} L.,  {Tosi} M.,  {Clampin} M.,  {De Marchi} G.,  {Leitherer} C.,
  {Nota} A.,   {Sirianni} M.,  1998, \mn@doi [\apj] {10.1086/306100}, \href
  {https://ui.adsabs.harvard.edu/abs/1998ApJ...504..725G} {504, 725}

\bibitem[\protect\citeauthoryear{{Guessoum}, {Jean}  \& {Gillard}}{{Guessoum}
  et~al.}{2005}]{Guessoum_2005}
{Guessoum} N.,  {Jean} P.,   {Gillard} W.,  2005, \mn@doi [\aap]
  {10.1051/0004-6361:20042454}, \href
  {https://ui.adsabs.harvard.edu/abs/2005A&A...436..171G} {436, 171}

\bibitem[\protect\citeauthoryear{{Hartwell}, {Stevens}, {Strickland}, {Heckman}
   \& {Summers}}{{Hartwell} et~al.}{2004}]{Hartwell_2004}
{Hartwell} J.~M.,  {Stevens} I.~R.,  {Strickland} D.~K.,  {Heckman} T.~M.,
  {Summers} L.~K.,  2004, \mn@doi [\mnras] {10.1111/j.1365-2966.2004.07375.x},
  \href {https://ui.adsabs.harvard.edu/abs/2004MNRAS.348..406H} {348, 406}

\bibitem[\protect\citeauthoryear{{Higgins}, {Vink}, {Hirschi}, {Laird}  \&
  {Sabhahit}}{{Higgins} et~al.}{2023}]{Higgins_2023}
{Higgins} E.~R.,  {Vink} J.~S.,  {Hirschi} R.,  {Laird} A.~M.,   {Sabhahit}
  G.~N.,  2023, \mn@doi [arXiv e-prints] {10.48550/arXiv.2308.10941}, \href
  {https://ui.adsabs.harvard.edu/abs/2023arXiv230810941H} {p. arXiv:2308.10941}

\bibitem[\protect\citeauthoryear{{Hillebrandt}, {Thielemann}  \&
  {Langer}}{{Hillebrandt} et~al.}{1987}]{Hillebrandt_1987}
{Hillebrandt} W.,  {Thielemann} F.-K.,   {Langer} N.,  1987, \mn@doi [\apj]
  {10.1086/165669}, \href
  {https://ui.adsabs.harvard.edu/abs/1987ApJ...321..761H} {321, 761}

\bibitem[\protect\citeauthoryear{{Hollyhead}, {Bastian}, {Adamo},
  {Silva-Villa}, {Dale}, {Ryon}  \& {Gazak}}{{Hollyhead}
  et~al.}{2015}]{Hollyhead_2015}
{Hollyhead} K.,  {Bastian} N.,  {Adamo} A.,  {Silva-Villa} E.,  {Dale} J.,
  {Ryon} J.~E.,   {Gazak} Z.,  2015, \mn@doi [\mnras] {10.1093/mnras/stv331},
  \href {https://ui.adsabs.harvard.edu/abs/2015MNRAS.449.1106H} {449, 1106}

\bibitem[\protect\citeauthoryear{{Hunter}, {Shaya}, {Holtzman}, {Light},
  {O'Neil}  \& {Lynds}}{{Hunter} et~al.}{1995}]{Hunter_1995}
{Hunter} D.~A.,  {Shaya} E.~J.,  {Holtzman} J.~A.,  {Light} R.~M.,  {O'Neil}
  Earl~J. J.,   {Lynds} R.,  1995, \mn@doi [\apj] {10.1086/175950}, \href
  {https://ui.adsabs.harvard.edu/abs/1995ApJ...448..179H} {448, 179}

\bibitem[\protect\citeauthoryear{{Hunter}, {O'Connell}, {Gallagher}  \&
  {Smecker-Hane}}{{Hunter} et~al.}{2000}]{Hunter_2000}
{Hunter} D.~A.,  {O'Connell} R.~W.,  {Gallagher} J.~S.,   {Smecker-Hane} T.~A.,
   2000, \mn@doi [\aj] {10.1086/316810}, \href
  {https://ui.adsabs.harvard.edu/abs/2000AJ....120.2383H} {120, 2383}

\bibitem[\protect\citeauthoryear{{Israel}}{{Israel}}{1988}]{Israel_1988}
{Israel} F.~P.,  1988, \aap, \href
  {https://ui.adsabs.harvard.edu/abs/1988A&A...194...24I} {194, 24}

\bibitem[\protect\citeauthoryear{{Jean}, {Kn{\"o}dlseder}, {Gillard},
  {Guessoum}, {Ferri{\`e}re}, {Marcowith}, {Lonjou}  \& {Roques}}{{Jean}
  et~al.}{2006}]{Jean_2006}
{Jean} P.,  {Kn{\"o}dlseder} J.,  {Gillard} W.,  {Guessoum} N.,  {Ferri{\`e}re}
  K.,  {Marcowith} A.,  {Lonjou} V.,   {Roques} J.~P.,  2006, \mn@doi [\aap]
  {10.1051/0004-6361:20053765}, \href
  {https://ui.adsabs.harvard.edu/abs/2006A&A...445..579J} {445, 579}

\bibitem[\protect\citeauthoryear{{Jean}, {Gillard}, {Marcowith}  \&
  {Ferri{\`e}re}}{{Jean} et~al.}{2009}]{Jean_2009}
{Jean} P.,  {Gillard} W.,  {Marcowith} A.,   {Ferri{\`e}re} K.,  2009, \mn@doi
  [\aap] {10.1051/0004-6361/200809830}, \href
  {https://ui.adsabs.harvard.edu/abs/2009A&A...508.1099J} {508, 1099}

\bibitem[\protect\citeauthoryear{{Johnson}, {Hunter}, {Oh}, {Zhang},
  {Elmegreen}, {Brinks}, {Tollerud}  \& {Herrmann}}{{Johnson}
  et~al.}{2012}]{Johnson_2012}
{Johnson} M.,  {Hunter} D.~A.,  {Oh} S.-H.,  {Zhang} H.-X.,  {Elmegreen} B.,
  {Brinks} E.,  {Tollerud} E.,   {Herrmann} K.,  2012, \mn@doi [\aj]
  {10.1088/0004-6256/144/5/152}, \href
  {https://ui.adsabs.harvard.edu/abs/2012AJ....144..152J} {144, 152}

\bibitem[\protect\citeauthoryear{{Kaleida} \& {Scowen}}{{Kaleida} \&
  {Scowen}}{2010}]{Kaleida_2010}
{Kaleida} C.,  {Scowen} P.~A.,  2010, \mn@doi [\aj]
  {10.1088/0004-6256/140/2/379}, \href
  {https://ui.adsabs.harvard.edu/abs/2010AJ....140..379K} {140, 379}

\bibitem[\protect\citeauthoryear{{Kami{\'n}ski} et~al.,}{{Kami{\'n}ski}
  et~al.}{2018}]{Kaminski_2018}
{Kami{\'n}ski} T.,  et~al., 2018, \mn@doi [Nature Astronomy]
  {10.1038/s41550-018-0541-x}, \href
  {https://ui.adsabs.harvard.edu/abs/2018NatAs...2..778K} {2, 778}

\bibitem[\protect\citeauthoryear{{Kami{\'n}ski}, {Steffen}, {Bujarrabal},
  {Tylenda}, {Menten}  \& {Hajduk}}{{Kami{\'n}ski}
  et~al.}{2021}]{Kaminski_2021}
{Kami{\'n}ski} T.,  {Steffen} W.,  {Bujarrabal} V.,  {Tylenda} R.,  {Menten}
  K.~M.,   {Hajduk} M.,  2021, \mn@doi [\aap] {10.1051/0004-6361/202039634},
  \href {https://ui.adsabs.harvard.edu/abs/2021A&A...646A...1K} {646, A1}

\bibitem[\protect\citeauthoryear{{Karachentsev}, {Karachentseva}, {Huchtmeier}
  \& {Makarov}}{{Karachentsev} et~al.}{2004}]{Karachentsev_2004}
{Karachentsev} I.~D.,  {Karachentseva} V.~E.,  {Huchtmeier} W.~K.,   {Makarov}
  D.~I.,  2004, \mn@doi [\aj] {10.1086/382905}, \href
  {https://ui.adsabs.harvard.edu/abs/2004AJ....127.2031K} {127, 2031}

\bibitem[\protect\citeauthoryear{Karakas \& Lattanzio}{Karakas \&
  Lattanzio}{2007}]{Karakas_2007}
Karakas A.,  Lattanzio J.~C.,  2007, \mn@doi [\pasa] {10.1071/AS07021}, 24,
  103–117

\bibitem[\protect\citeauthoryear{{Karakas}, {Fenner}, {Sills}, {Campbell}  \&
  {Lattanzio}}{{Karakas} et~al.}{2006}]{Karakas_2006}
{Karakas} A.~I.,  {Fenner} Y.,  {Sills} A.,  {Campbell} S.~W.,   {Lattanzio}
  J.~C.,  2006, \mn@doi [\apj] {10.1086/508504}, \href
  {https://ui.adsabs.harvard.edu/abs/2006ApJ...652.1240K} {652, 1240}

\bibitem[\protect\citeauthoryear{{Kennicutt} Robert~C. et~al.,}{{Kennicutt}
  et~al.}{2003}]{Kennicutt_2003}
{Kennicutt} Robert~C. J.,  et~al., 2003, \mn@doi [\pasp] {10.1086/376941},
  \href {https://ui.adsabs.harvard.edu/abs/2003PASP..115..928K} {115, 928}

\bibitem[\protect\citeauthoryear{{Kn{\"o}dlseder}}{{Kn{\"o}dlseder}}{1999}]{Knodlseder_1999}
{Kn{\"o}dlseder} J.,  1999, \mn@doi [\apj] {10.1086/306601}, \href
  {https://ui.adsabs.harvard.edu/abs/1999ApJ...510..915K} {510, 915}

\bibitem[\protect\citeauthoryear{{Kn{\"o}dlseder} et~al.,}{{Kn{\"o}dlseder}
  et~al.}{1999}]{Knodlseder_1999comptel}
{Kn{\"o}dlseder} J.,  et~al., 1999, \aap, \href
  {https://ui.adsabs.harvard.edu/abs/1999A&A...344...68K} {344, 68}

\bibitem[\protect\citeauthoryear{{Kobulnicky} \& {Johnson}}{{Kobulnicky} \&
  {Johnson}}{1999}]{Kobulnicky_1999}
{Kobulnicky} H.~A.,  {Johnson} K.~E.,  1999, \mn@doi [\apj] {10.1086/308075},
  \href {https://ui.adsabs.harvard.edu/abs/1999ApJ...527..154K} {527, 154}

\bibitem[\protect\citeauthoryear{{Kobulnicky} \& {Skillman}}{{Kobulnicky} \&
  {Skillman}}{1996}]{Kobulnicky_1996}
{Kobulnicky} H.~A.,  {Skillman} E.~D.,  1996, \mn@doi [\apj] {10.1086/177964},
  \href {https://ui.adsabs.harvard.edu/abs/1996ApJ...471..211K} {471, 211}

\bibitem[\protect\citeauthoryear{{Krause}, {Charbonnel}, {Decressin}, {Meynet}
  \& {Prantzos}}{{Krause} et~al.}{2013}]{Krause_2013}
{Krause} M.,  {Charbonnel} C.,  {Decressin} T.,  {Meynet} G.,   {Prantzos} N.,
  2013, \mn@doi [\aap] {10.1051/0004-6361/201220694}, \href
  {https://ui.adsabs.harvard.edu/abs/2013A&A...552A.121K} {552, A121}

\bibitem[\protect\citeauthoryear{{Krause}, {Charbonnel}, {Bastian}  \&
  {Diehl}}{{Krause} et~al.}{2016}]{Krause_2016}
{Krause} M. G.~H.,  {Charbonnel} C.,  {Bastian} N.,   {Diehl} R.,  2016,
  \mn@doi [\aap] {10.1051/0004-6361/201526685}, \href
  {https://ui.adsabs.harvard.edu/abs/2016A&A...587A..53K} {587, A53}

\bibitem[\protect\citeauthoryear{{Krause} et~al.,}{{Krause}
  et~al.}{2018}]{Krause_2018}
{Krause} M. G.~H.,  et~al., 2018, \mn@doi [\aap] {10.1051/0004-6361/201732416},
  \href {https://ui.adsabs.harvard.edu/abs/2018A&A...619A.120K} {619, A120}

\bibitem[\protect\citeauthoryear{Krause et~al.,}{Krause
  et~al.}{2020}]{Krause_2020}
Krause M.,  et~al., 2020, \mn@doi [\ssr] {10.1007/s11214-020-00689-4}, 216

\bibitem[\protect\citeauthoryear{{Lacki}, {Horiuchi}  \& {Beacom}}{{Lacki}
  et~al.}{2014}]{Lacki_2014}
{Lacki} B.~C.,  {Horiuchi} S.,   {Beacom} J.~F.,  2014, \mn@doi [\apj]
  {10.1088/0004-637X/786/1/40}, \href
  {https://ui.adsabs.harvard.edu/abs/2014ApJ...786...40L} {786, 40}

\bibitem[\protect\citeauthoryear{{Langer}, {Hoffman}  \& {Sneden}}{{Langer}
  et~al.}{1993}]{Langer_1993}
{Langer} G.~E.,  {Hoffman} R.,   {Sneden} C.,  1993, \mn@doi [\pasp]
  {10.1086/133147}, \href
  {https://ui.adsabs.harvard.edu/abs/1993PASP..105..301L} {105, 301}

\bibitem[\protect\citeauthoryear{{Lardo}, {Davies}, {Kudritzki}, {Gazak},
  {Evans}, {Patrick}, {Bergemann}  \& {Plez}}{{Lardo}
  et~al.}{2015}]{Lardo_2015}
{Lardo} C.,  {Davies} B.,  {Kudritzki} R.~P.,  {Gazak} J.~Z.,  {Evans} C.~J.,
  {Patrick} L.~R.,  {Bergemann} M.,   {Plez} B.,  2015, \mn@doi [\apj]
  {10.1088/0004-637X/812/2/160}, \href
  {https://ui.adsabs.harvard.edu/abs/2015ApJ...812..160L} {812, 160}

\bibitem[\protect\citeauthoryear{{Lardo}, {Cabrera-Ziri}, {Davies}  \&
  {Bastian}}{{Lardo} et~al.}{2017}]{Lard0_2017}
{Lardo} C.,  {Cabrera-Ziri} I.,  {Davies} B.,   {Bastian} N.,  2017, \mn@doi
  [\mnras] {10.1093/mnras/stx628}, \href
  {https://ui.adsabs.harvard.edu/abs/2017MNRAS.468.2482L} {468, 2482}

\bibitem[\protect\citeauthoryear{{Leroy}, {Walter}, {Brinks}, {Bigiel}, {de
  Blok}, {Madore}  \& {Thornley}}{{Leroy} et~al.}{2008}]{Leroy_2008}
{Leroy} A.~K.,  {Walter} F.,  {Brinks} E.,  {Bigiel} F.,  {de Blok} W.~J.~G.,
  {Madore} B.,   {Thornley} M.~D.,  2008, \mn@doi [\aj]
  {10.1088/0004-6256/136/6/2782}, \href
  {https://ui.adsabs.harvard.edu/abs/2008AJ....136.2782L} {136, 2782}

\bibitem[\protect\citeauthoryear{{Leroy} et~al.,}{{Leroy}
  et~al.}{2018}]{Leroy_2018}
{Leroy} A.~K.,  et~al., 2018, \mn@doi [\apj] {10.3847/1538-4357/aaecd1}, \href
  {https://ui.adsabs.harvard.edu/abs/2018ApJ...869..126L} {869, 126}

\bibitem[\protect\citeauthoryear{{Levy} et~al.,}{{Levy}
  et~al.}{2021}]{Levy_2021}
{Levy} R.~C.,  et~al., 2021, \mn@doi [\apj] {10.3847/1538-4357/abec84}, \href
  {https://ui.adsabs.harvard.edu/abs/2021ApJ...912....4L} {912, 4}

\bibitem[\protect\citeauthoryear{{Limongi} \& {Chieffi}}{{Limongi} \&
  {Chieffi}}{2006}]{Limongi_2006}
{Limongi} M.,  {Chieffi} A.,  2006, \mn@doi [\apj] {10.1086/505164}, \href
  {https://ui.adsabs.harvard.edu/abs/2006ApJ...647..483L} {647, 483}

\bibitem[\protect\citeauthoryear{{Limongi} \& {Chieffi}}{{Limongi} \&
  {Chieffi}}{2018}]{Limongi_2018}
{Limongi} M.,  {Chieffi} A.,  2018, \mn@doi [\apjs] {10.3847/1538-4365/aacb24},
  \href {https://ui.adsabs.harvard.edu/abs/2018ApJS..237...13L} {237, 13}

\bibitem[\protect\citeauthoryear{{Longmore}}{{Longmore}}{2015}]{Longmore_2015}
{Longmore} S.~N.,  2015, \mn@doi [\mnras] {10.1093/mnrasl/slu203}, \href
  {https://ui.adsabs.harvard.edu/abs/2015MNRAS.448L..62L} {448, L62}

\bibitem[\protect\citeauthoryear{{Ma{\'\i}z-Apell{\'a}niz}}{{Ma{\'\i}z-Apell{\'a}niz}}{2001}]{MaizApellaniz_2001}
{Ma{\'\i}z-Apell{\'a}niz} J.,  2001, \mn@doi [\apj] {10.1086/323775}, \href
  {https://ui.adsabs.harvard.edu/abs/2001ApJ...563..151M} {563, 151}

\bibitem[\protect\citeauthoryear{{Martinet} et~al.,}{{Martinet}
  et~al.}{2022}]{Martinet_2022}
{Martinet} S.,  et~al., 2022, \mn@doi [\aap] {10.1051/0004-6361/202243474},
  \href {https://ui.adsabs.harvard.edu/abs/2022A&A...664A.181M} {664, A181}

\bibitem[\protect\citeauthoryear{{Martins}, {Chantereau}  \&
  {Charbonnel}}{{Martins} et~al.}{2021}]{Martins_2021}
{Martins} F.,  {Chantereau} W.,   {Charbonnel} C.,  2021, \mn@doi [\aap]
  {10.1051/0004-6361/202140800}, \href
  {https://ui.adsabs.harvard.edu/abs/2021A&A...650A.162M} {650, A162}

\bibitem[\protect\citeauthoryear{{Martocchia} et~al.,}{{Martocchia}
  et~al.}{2017}]{Martocchia_2017}
{Martocchia} S.,  et~al., 2017, \mn@doi [\mnras] {10.1093/mnras/stx660}, \href
  {https://ui.adsabs.harvard.edu/abs/2017MNRAS.468.3150M} {468, 3150}

\bibitem[\protect\citeauthoryear{{Martocchia} et~al.,}{{Martocchia}
  et~al.}{2018}]{Martocchia_2018}
{Martocchia} S.,  et~al., 2018, \mn@doi [\mnras] {10.1093/mnras/stx2556}, \href
  {https://ui.adsabs.harvard.edu/abs/2018MNRAS.473.2688M} {473, 2688}

\bibitem[\protect\citeauthoryear{{McCrady}, {Gilbert}  \& {Graham}}{{McCrady}
  et~al.}{2003}]{McCrady_2003}
{McCrady} N.,  {Gilbert} A.~M.,   {Graham} J.~R.,  2003, \mn@doi [\apj]
  {10.1086/377631}, \href
  {https://ui.adsabs.harvard.edu/abs/2003ApJ...596..240M} {596, 240}

\bibitem[\protect\citeauthoryear{{McMillan}}{{McMillan}}{2017}]{McMillan_2017}
{McMillan} P.~J.,  2017, \mn@doi [\mnras] {10.1093/mnras/stw2759}, \href
  {https://ui.adsabs.harvard.edu/abs/2017MNRAS.465...76M} {465, 76}

\bibitem[\protect\citeauthoryear{{McQuinn}, {Skillman}, {Dolphin}, {Berg}  \&
  {Kennicutt}}{{McQuinn} et~al.}{2016}]{McQuinn_2016}
{McQuinn} K. B.~W.,  {Skillman} E.~D.,  {Dolphin} A.~E.,  {Berg} D.,
  {Kennicutt} R.,  2016, \mn@doi [\apj] {10.3847/0004-637X/826/1/21}, \href
  {https://ui.adsabs.harvard.edu/abs/2016ApJ...826...21M} {826, 21}

\bibitem[\protect\citeauthoryear{{Mengel}, {Lehnert}, {Thatte}  \&
  {Genzel}}{{Mengel} et~al.}{2002}]{Mengel_2002}
{Mengel} S.,  {Lehnert} M.~D.,  {Thatte} N.,   {Genzel} R.,  2002, \mn@doi
  [\aap] {10.1051/0004-6361:20011704}, \href
  {https://ui.adsabs.harvard.edu/abs/2002A&A...383..137M} {383, 137}

\bibitem[\protect\citeauthoryear{{Mengel}, {Lehnert}, {Thatte}  \&
  {Genzel}}{{Mengel} et~al.}{2003}]{Mengel_2003}
{Mengel} S.,  {Lehnert} M.~D.,  {Thatte} N.~A.,   {Genzel} R.,  2003, in
  {Guhathakurta} P.,  ed.,  Society of Photo-Optical Instrumentation Engineers
  (SPIE) Conference Series Vol. 4834, Discoveries and Research Prospects from
  6- to 10-Meter-Class Telescopes II. pp 45--56, \mn@doi{10.1117/12.456511}

\bibitem[\protect\citeauthoryear{{Mengel}, {Lehnert}, {Thatte}, {Vacca},
  {Whitmore}  \& {Chandar}}{{Mengel} et~al.}{2008}]{Mengel_2008}
{Mengel} S.,  {Lehnert} M.~D.,  {Thatte} N.~A.,  {Vacca} W.~D.,  {Whitmore} B.,
    {Chandar} R.,  2008, \mn@doi [\aap] {10.1051/0004-6361:200809649}, \href
  {https://ui.adsabs.harvard.edu/abs/2008A&A...489.1091M} {489, 1091}

\bibitem[\protect\citeauthoryear{{Meynet} \& {Arnould}}{{Meynet} \&
  {Arnould}}{2000}]{Meynet_2000}
{Meynet} G.,  {Arnould} M.,  2000, \mn@doi [\aap]
  {10.48550/arXiv.astro-ph/0001170}, \href
  {https://ui.adsabs.harvard.edu/abs/2000A&A...355..176M} {355, 176}

\bibitem[\protect\citeauthoryear{{Meynet}, {Arnould}, {Prantzos}  \&
  {Paulus}}{{Meynet} et~al.}{1997}]{Meynet_1997}
{Meynet} G.,  {Arnould} M.,  {Prantzos} N.,   {Paulus} G.,  1997, \aap, \href
  {https://ui.adsabs.harvard.edu/abs/1997A&A...320..460M} {320, 460}

\bibitem[\protect\citeauthoryear{{Mills} et~al.,}{{Mills}
  et~al.}{2021}]{Mills_2021}
{Mills} E.~A.~C.,  et~al., 2021, \mn@doi [\apj] {10.3847/1538-4357/ac0fe8},
  \href {https://ui.adsabs.harvard.edu/abs/2021ApJ...919..105M} {919, 105}

\bibitem[\protect\citeauthoryear{{Milone} \& {Marino}}{{Milone} \&
  {Marino}}{2022}]{Milone_2022}
{Milone} A.~P.,  {Marino} A.~F.,  2022, \mn@doi [Universe]
  {10.3390/universe8070359}, \href
  {https://ui.adsabs.harvard.edu/abs/2022Univ....8..359M} {8, 359}

\bibitem[\protect\citeauthoryear{{Milone}, {Marino}, {Piotto}, {Bedin},
  {Anderson}, {Aparicio}, {Cassisi}  \& {Rich}}{{Milone}
  et~al.}{2012}]{Milone_2012}
{Milone} A.~P.,  {Marino} A.~F.,  {Piotto} G.,  {Bedin} L.~R.,  {Anderson} J.,
  {Aparicio} A.,  {Cassisi} S.,   {Rich} R.~M.,  2012, \mn@doi [\apj]
  {10.1088/0004-637X/745/1/27}, \href
  {https://ui.adsabs.harvard.edu/abs/2012ApJ...745...27M} {745, 27}

\bibitem[\protect\citeauthoryear{{Milone} et~al.,}{{Milone}
  et~al.}{2013}]{Milone_2013}
{Milone} A.~P.,  et~al., 2013, \mn@doi [\apj] {10.1088/0004-637X/767/2/120},
  \href {https://ui.adsabs.harvard.edu/abs/2013ApJ...767..120M} {767, 120}

\bibitem[\protect\citeauthoryear{{Milone}, {Marino}, {D'Antona}, {Bedin}, {Da
  Costa}, {Jerjen}  \& {Mackey}}{{Milone} et~al.}{2016}]{Milone_2016}
{Milone} A.~P.,  {Marino} A.~F.,  {D'Antona} F.,  {Bedin} L.~R.,  {Da Costa}
  G.~S.,  {Jerjen} H.,   {Mackey} A.~D.,  2016, \mn@doi [\mnras]
  {10.1093/mnras/stw608}, \href
  {https://ui.adsabs.harvard.edu/abs/2016MNRAS.458.4368M} {458, 4368}

\bibitem[\protect\citeauthoryear{{Milone} et~al.,}{{Milone}
  et~al.}{2017}]{Milone_2017}
{Milone} A.~P.,  et~al., 2017, \mn@doi [\mnras] {10.1093/mnras/stw2531}, \href
  {https://ui.adsabs.harvard.edu/abs/2017MNRAS.464.3636M} {464, 3636}

\bibitem[\protect\citeauthoryear{{Milone} et~al.,}{{Milone}
  et~al.}{2018}]{Milone_2018}
{Milone} A.~P.,  et~al., 2018, \mn@doi [\mnras] {10.1093/mnras/sty2573}, 481,
  5098

\bibitem[\protect\citeauthoryear{{Milone} et~al.,}{{Milone}
  et~al.}{2020}]{Milone_2020}
{Milone} A.~P.,  et~al., 2020, \mn@doi [\mnras] {10.1093/mnras/stz2999}, \href
  {https://ui.adsabs.harvard.edu/abs/2020MNRAS.491..515M} {491, 515}

\bibitem[\protect\citeauthoryear{{Mohorovi{\v{c}}i{\'c}}}{{Mohorovi{\v{c}}i{\'c}}}{1934}]{Mohorovicic_1934}
{Mohorovi{\v{c}}i{\'c}} S.,  1934, \mn@doi [Astronomische Nachrichten]
  {10.1002/asna.19342530402}, \href
  {https://ui.adsabs.harvard.edu/abs/1934AN....253...93M} {253, 93}

\bibitem[\protect\citeauthoryear{{Moll}, {Mengel}, {de Grijs}, {Smith}  \&
  {Crowther}}{{Moll} et~al.}{2007}]{Moll_2007}
{Moll} S.~L.,  {Mengel} S.,  {de Grijs} R.,  {Smith} L.~J.,   {Crowther} P.~A.,
   2007, \mn@doi [\mnras] {10.1111/j.1365-2966.2007.12497.x}, \href
  {https://ui.adsabs.harvard.edu/abs/2007MNRAS.382.1877M} {382, 1877}

\bibitem[\protect\citeauthoryear{{Mowlavi}, {Kn{\"o}dlseder}, {Meynet},
  {Dubath}, {Diehl}, {Lichti}, {Schanne}  \& {Winkler}}{{Mowlavi}
  et~al.}{2005}]{Mowlavi_2005}
{Mowlavi} N.,  {Kn{\"o}dlseder} J.,  {Meynet} G.,  {Dubath} P.,  {Diehl} R.,
  {Lichti} G.,  {Schanne} S.,   {Winkler} C.,  2005, \mn@doi [\nphysa]
  {10.1016/j.nuclphysa.2005.05.177}, \href
  {https://ui.adsabs.harvard.edu/abs/2005NuPhA.758..320M} {758, 320}

\bibitem[\protect\citeauthoryear{{Mucciarelli}, {Dalessandro}, {Ferraro},
  {Origlia}  \& {Lanzoni}}{{Mucciarelli} et~al.}{2014}]{Mucciarelli_2014}
{Mucciarelli} A.,  {Dalessandro} E.,  {Ferraro} F.~R.,  {Origlia} L.,
  {Lanzoni} B.,  2014, \mn@doi [\apjl] {10.1088/2041-8205/793/1/L6}, \href
  {https://ui.adsabs.harvard.edu/abs/2014ApJ...793L...6M} {793, L6}

\bibitem[\protect\citeauthoryear{{Norris}}{{Norris}}{2004}]{Norris_2004}
{Norris} J.~E.,  2004, \mn@doi [\apjl] {10.1086/423986}, \href
  {https://ui.adsabs.harvard.edu/abs/2004ApJ...612L..25N} {612, L25}

\bibitem[\protect\citeauthoryear{{Norris}, {Gancarz}, {Rokop}  \&
  {Thomas}}{{Norris} et~al.}{1983}]{Norris_1983}
{Norris} T.~L.,  {Gancarz} A.~J.,  {Rokop} D.~J.,   {Thomas} K.~W.,  1983,
  \mn@doi [Lunar and Planetary Science Conference Proceedings]
  {10.1029/JB088iS01p0B331}, \href
  {https://ui.adsabs.harvard.edu/abs/1983LPSC...14..331N} {88, B331}

\bibitem[\protect\citeauthoryear{{Nowak}, {Krause}  \& {Schaerer}}{{Nowak}
  et~al.}{2022}]{Nowak_2022}
{Nowak} K.,  {Krause} M. G.~H.,   {Schaerer} D.,  2022, \mn@doi [\mnras]
  {10.1093/mnras/stac2547}, \href
  {https://ui.adsabs.harvard.edu/abs/2022MNRAS.516.5507N} {516, 5507}

\bibitem[\protect\citeauthoryear{{Nugis} \& {Lamers}}{{Nugis} \&
  {Lamers}}{2000}]{Nugis_2000}
{Nugis} T.,  {Lamers} H.~J.~G.~L.~M.,  2000, \aap, \href
  {https://ui.adsabs.harvard.edu/abs/2000A&A...360..227N} {360, 227}

\bibitem[\protect\citeauthoryear{{Oberlack} et~al.,}{{Oberlack}
  et~al.}{2000}]{Oberlack_2000}
{Oberlack} U.,  et~al., 2000, \mn@doi [\aap] {10.48550/arXiv.astro-ph/9910555},
  \href {https://ui.adsabs.harvard.edu/abs/2000A&A...353..715O} {353, 715}

\bibitem[\protect\citeauthoryear{{Ore} \& {Powell}}{{Ore} \&
  {Powell}}{1949}]{Ore_1949}
{Ore} A.,  {Powell} J.~L.,  1949, \mn@doi [Physical Review]
  {10.1103/PhysRev.75.1696}, \href
  {https://ui.adsabs.harvard.edu/abs/1949PhRv...75.1696O} {75, 1696}

\bibitem[\protect\citeauthoryear{{{\"O}stlin}, {Amram}, {Bergvall}, {Masegosa},
  {Boulesteix}  \& {M{\'a}rquez}}{{{\"O}stlin} et~al.}{2001}]{Ostlin_2001}
{{\"O}stlin} G.,  {Amram} P.,  {Bergvall} N.,  {Masegosa} J.,  {Boulesteix} J.,
    {M{\'a}rquez} I.,  2001, \mn@doi [\aap] {10.1051/0004-6361:20010832}, \href
  {https://ui.adsabs.harvard.edu/abs/2001A&A...374..800O} {374, 800}

\bibitem[\protect\citeauthoryear{{{\"O}stlin}, {Cumming}  \&
  {Bergvall}}{{{\"O}stlin} et~al.}{2007}]{Ostlin_2007}
{{\"O}stlin} G.,  {Cumming} R.~J.,   {Bergvall} N.,  2007, \mn@doi [\aap]
  {10.1051/0004-6361:20054461}, \href
  {https://ui.adsabs.harvard.edu/abs/2007A&A...461..471O} {461, 471}

\bibitem[\protect\citeauthoryear{{Palacios}, {Meynet}, {Vuissoz},
  {Kn{\"o}dlseder}, {Schaerer}, {Cervi{\~n}o}  \& {Mowlavi}}{{Palacios}
  et~al.}{2005}]{Palacios_2005}
{Palacios} A.,  {Meynet} G.,  {Vuissoz} C.,  {Kn{\"o}dlseder} J.,  {Schaerer}
  D.,  {Cervi{\~n}o} M.,   {Mowlavi} N.,  2005, \mn@doi [\aap]
  {10.1051/0004-6361:20041757}, \href
  {https://ui.adsabs.harvard.edu/abs/2005A&A...429..613P} {429, 613}

\bibitem[\protect\citeauthoryear{{Pavlenko}, {Tennyson}, {Yurchenko},
  {Schmidt}, {Jones}, {Lyubchik}  \& {Su{\'a}rez Mascare{\~n}o}}{{Pavlenko}
  et~al.}{2022}]{Pavlenko_2022}
{Pavlenko} Y.~V.,  {Tennyson} J.,  {Yurchenko} S.~N.,  {Schmidt} M.~R.,
  {Jones} H. R.~A.,  {Lyubchik} Y.,   {Su{\'a}rez Mascare{\~n}o} A.,  2022,
  \mn@doi [\mnras] {10.1093/mnras/stac2588}, \href
  {https://ui.adsabs.harvard.edu/abs/2022MNRAS.516.5655P} {516, 5655}

\bibitem[\protect\citeauthoryear{{Pietrzy{\'n}ski} et~al.,}{{Pietrzy{\'n}ski}
  et~al.}{2013}]{Pietrzynski_2013}
{Pietrzy{\'n}ski} G.,  et~al., 2013, \mn@doi [\nat] {10.1038/nature11878},
  \href {https://ui.adsabs.harvard.edu/abs/2013Natur.495...76P} {495, 76}

\bibitem[\protect\citeauthoryear{{Pineda} et~al.,}{{Pineda}
  et~al.}{2018}]{Pineda_2018}
{Pineda} J.~L.,  et~al., 2018, \mn@doi [\apjl] {10.3847/2041-8213/aaf1ad},
  \href {https://ui.adsabs.harvard.edu/abs/2018ApJ...869L..30P} {869, L30}

\bibitem[\protect\citeauthoryear{{Piotto} et~al.,}{{Piotto}
  et~al.}{2007}]{Piotto_2007}
{Piotto} G.,  et~al., 2007, \mn@doi [\apjl] {10.1086/518503}, \href
  {https://ui.adsabs.harvard.edu/abs/2007ApJ...661L..53P} {661, L53}

\bibitem[\protect\citeauthoryear{{Piotto} et~al.,}{{Piotto}
  et~al.}{2015}]{Piotto_2015}
{Piotto} G.,  et~al., 2015, \mn@doi [\aj] {10.1088/0004-6256/149/3/91}, \href
  {https://ui.adsabs.harvard.edu/abs/2015AJ....149...91P} {149, 91}

\bibitem[\protect\citeauthoryear{{Pleintinger}, {Siegert}, {Diehl}, {Fujimoto},
  {Greiner}, {Krause}  \& {Krumholz}}{{Pleintinger}
  et~al.}{2019}]{Pleintinger_2019}
{Pleintinger} M. M.~M.,  {Siegert} T.,  {Diehl} R.,  {Fujimoto} Y.,  {Greiner}
  J.,  {Krause} M. G.~H.,   {Krumholz} M.~R.,  2019, \mn@doi [\aap]
  {10.1051/0004-6361/201935911}, \href
  {https://ui.adsabs.harvard.edu/abs/2019A&A...632A..73P} {632, A73}

\bibitem[\protect\citeauthoryear{{Pleintinger}, {Diehl}, {Siegert}, {Greiner}
  \& {Krause}}{{Pleintinger} et~al.}{2023}]{Pleintinger_2023}
{Pleintinger} M. M.~M.,  {Diehl} R.,  {Siegert} T.,  {Greiner} J.,   {Krause}
  M. G.~H.,  2023, \mn@doi [\aap] {10.1051/0004-6361/202245069}, \href
  {https://ui.adsabs.harvard.edu/abs/2023A&A...672A..53P} {672, A53}

\bibitem[\protect\citeauthoryear{{Portegies Zwart}, {McMillan}  \&
  {Gieles}}{{Portegies Zwart} et~al.}{2010}]{PortegiesZwart_2010}
{Portegies Zwart} S.~F.,  {McMillan} S. L.~W.,   {Gieles} M.,  2010, \mn@doi
  [\araa] {10.1146/annurev-astro-081309-130834}, \href
  {https://ui.adsabs.harvard.edu/abs/2010ARA&A..48..431P} {48, 431}

\bibitem[\protect\citeauthoryear{{Prantzos} \& {Casse}}{{Prantzos} \&
  {Casse}}{1986}]{Prantzos_1986}
{Prantzos} N.,  {Casse} M.,  1986, \mn@doi [\apj] {10.1086/164419}, \href
  {https://ui.adsabs.harvard.edu/abs/1986ApJ...307..324P} {307, 324}

\bibitem[\protect\citeauthoryear{{Prantzos} \& {Charbonnel}}{{Prantzos} \&
  {Charbonnel}}{2006}]{Prantzos_2006}
{Prantzos} N.,  {Charbonnel} C.,  2006, \mn@doi [\aap]
  {10.1051/0004-6361:20065374}, \href
  {https://ui.adsabs.harvard.edu/abs/2006A&A...458..135P} {458, 135}

\bibitem[\protect\citeauthoryear{{Prantzos} \& {Diehl}}{{Prantzos} \&
  {Diehl}}{1996}]{Prantzos_1996}
{Prantzos} N.,  {Diehl} R.,  1996, \mn@doi [\physrep]
  {10.1016/0370-1573(95)00055-0}, \href
  {https://ui.adsabs.harvard.edu/abs/1996PhR...267....1P} {267, 1}

\bibitem[\protect\citeauthoryear{{Prantzos}, {Charbonnel}  \&
  {Iliadis}}{{Prantzos} et~al.}{2007}]{Prantzos_2007}
{Prantzos} N.,  {Charbonnel} C.,   {Iliadis} C.,  2007, \mn@doi [\aap]
  {10.1051/0004-6361:20077205}, \href
  {https://ui.adsabs.harvard.edu/abs/2007A&A...470..179P} {470, 179}

\bibitem[\protect\citeauthoryear{{Prantzos} et~al.,}{{Prantzos}
  et~al.}{2011}]{Prantzos_2011}
{Prantzos} N.,  et~al., 2011, \mn@doi [Reviews of Modern Physics]
  {10.1103/RevModPhys.83.1001}, \href
  {https://ui.adsabs.harvard.edu/abs/2011RvMP...83.1001P} {83, 1001}

\bibitem[\protect\citeauthoryear{{Prantzos}, {Charbonnel}  \&
  {Iliadis}}{{Prantzos} et~al.}{2017}]{Prantzos_2017}
{Prantzos} N.,  {Charbonnel} C.,   {Iliadis} C.,  2017, \mn@doi [\aap]
  {10.1051/0004-6361/201731528}, \href
  {https://ui.adsabs.harvard.edu/abs/2017A&A...608A..28P} {608, A28}

\bibitem[\protect\citeauthoryear{{Puxley} \& {Skinner}}{{Puxley} \&
  {Skinner}}{1996}]{Puxley_1996}
{Puxley} P.~J.,  {Skinner} G.~K.,  1996, in {Gredel} R.,  ed.,  Astronomical
  Society of the Pacific Conference Series Vol. 102, The Galactic Center.
  p.~439

\bibitem[\protect\citeauthoryear{{Rekola}, {Richer}, {McCall}, {Valtonen},
  {Kotilainen}  \& {Flynn}}{{Rekola} et~al.}{2005}]{Rekola_2005}
{Rekola} R.,  {Richer} M.~G.,  {McCall} M.~L.,  {Valtonen} M.~J.,  {Kotilainen}
  J.~K.,   {Flynn} C.,  2005, \mn@doi [\mnras]
  {10.1111/j.1365-2966.2005.09166.x}, \href
  {https://ui.adsabs.harvard.edu/abs/2005MNRAS.361..330R} {361, 330}

\bibitem[\protect\citeauthoryear{{Renzini}, {Marino}  \& {Milone}}{{Renzini}
  et~al.}{2022}]{Renzini_2022}
{Renzini} A.,  {Marino} A.~F.,   {Milone} A.~P.,  2022, \mn@doi [\mnras]
  {10.1093/mnras/stac973}, \href
  {https://ui.adsabs.harvard.edu/abs/2022MNRAS.513.2111R} {513, 2111}

\bibitem[\protect\citeauthoryear{{Rico-Villas}, {Mart{\'\i}n-Pintado},
  {Gonz{\'a}lez-Alfonso}, {Mart{\'\i}n}  \& {Rivilla}}{{Rico-Villas}
  et~al.}{2020}]{RicoVillas_2020}
{Rico-Villas} F.,  {Mart{\'\i}n-Pintado} J.,  {Gonz{\'a}lez-Alfonso} E.,
  {Mart{\'\i}n} S.,   {Rivilla} V.~M.,  2020, \mn@doi [\mnras]
  {10.1093/mnras/stz3347}, \href
  {https://ui.adsabs.harvard.edu/abs/2020MNRAS.491.4573R} {491, 4573}

\bibitem[\protect\citeauthoryear{{Rizzuti}, {Hirschi}, {Arnett}, {Georgy},
  {Meakin}, {Murphy}, {Rauscher}  \& {Varma}}{{Rizzuti}
  et~al.}{2023}]{Rizzuti_2023}
{Rizzuti} F.,  {Hirschi} R.,  {Arnett} W.~D.,  {Georgy} C.,  {Meakin} C.,
  {Murphy} A.~S.,  {Rauscher} T.,   {Varma} V.,  2023, \mn@doi [\mnras]
  {10.1093/mnras/stad1572}, \href
  {https://ui.adsabs.harvard.edu/abs/2023MNRAS.523.2317R} {523, 2317}

\bibitem[\protect\citeauthoryear{{Sabbi} et~al.,}{{Sabbi}
  et~al.}{2018}]{Sabbi_2018}
{Sabbi} E.,  et~al., 2018, \mn@doi [\apjs] {10.3847/1538-4365/aaa8e5}, \href
  {https://ui.adsabs.harvard.edu/abs/2018ApJS..235...23S} {235, 23}

\bibitem[\protect\citeauthoryear{{Schweizer} et~al.,}{{Schweizer}
  et~al.}{2008}]{Schweizer_2008}
{Schweizer} F.,  et~al., 2008, \mn@doi [\aj] {10.1088/0004-6256/136/4/1482},
  \href {https://ui.adsabs.harvard.edu/abs/2008AJ....136.1482S} {136, 1482}

\bibitem[\protect\citeauthoryear{{Seill{\'e}}, {Buat}, {Haddad}, {Boselli},
  {Boquien}, {Ciesla}, {Roehlly}  \& {Burgarella}}{{Seill{\'e}}
  et~al.}{2022}]{Seille_2022}
{Seill{\'e}} L.~M.,  {Buat} V.,  {Haddad} W.,  {Boselli} A.,  {Boquien} M.,
  {Ciesla} L.,  {Roehlly} Y.,   {Burgarella} D.,  2022, \mn@doi [\aap]
  {10.1051/0004-6361/202243702}, \href
  {https://ui.adsabs.harvard.edu/abs/2022A&A...665A.137S} {665, A137}

\bibitem[\protect\citeauthoryear{{Siegert}}{{Siegert}}{2023}]{Siegert_2023_positronreview}
{Siegert} T.,  2023, \mn@doi [\apss] {10.1007/s10509-023-04184-4}, \href
  {https://ui.adsabs.harvard.edu/abs/2023Ap&SS.368...27S} {368, 27}

\bibitem[\protect\citeauthoryear{{Siegert} \& {Diehl}}{{Siegert} \&
  {Diehl}}{2017}]{Siegert_2017}
{Siegert} T.,  {Diehl} R.,  2017, in {Kubono} S.,  {Kajino} T.,  {Nishimura}
  S.,  {Isobe} T.,  {Nagataki} S.,  {Shima} T.,   {Takeda} Y.,  eds, 14th
  International Symposium on Nuclei in the Cosmos (NIC2016). p. 020305
  (\mn@eprint {arXiv} {1609.08817}), \mn@doi{10.7566/JPSCP.14.020305}

\bibitem[\protect\citeauthoryear{{Siegert}, {Diehl}, {Khachatryan}, {Krause},
  {Guglielmetti}, {Greiner}, {Strong}  \& {Zhang}}{{Siegert}
  et~al.}{2016}]{Siegert_2016}
{Siegert} T.,  {Diehl} R.,  {Khachatryan} G.,  {Krause} M. G.~H.,
  {Guglielmetti} F.,  {Greiner} J.,  {Strong} A.~W.,   {Zhang} X.,  2016,
  \mn@doi [\aap] {10.1051/0004-6361/201527510}, \href
  {https://ui.adsabs.harvard.edu/abs/2016A&A...586A..84S} {586, A84}

\bibitem[\protect\citeauthoryear{{Siegert}, {Pleintinger}, {Diehl}, {Krause},
  {Greiner}  \& {Weinberger}}{{Siegert} et~al.}{2023}]{Siegert_2023}
{Siegert} T.,  {Pleintinger} M. M.~M.,  {Diehl} R.,  {Krause} M. G.~H.,
  {Greiner} J.,   {Weinberger} C.,  2023, \mn@doi [\aap]
  {10.1051/0004-6361/202244457}, \href
  {https://ui.adsabs.harvard.edu/abs/2023A&A...672A..54S} {672, A54}

\bibitem[\protect\citeauthoryear{{Siess}}{{Siess}}{2010}]{Siess_2010}
{Siess} L.,  2010, \mn@doi [\aap] {10.1051/0004-6361/200913556}, \href
  {https://ui.adsabs.harvard.edu/abs/2010A&A...512A..10S} {512, A10}

\bibitem[\protect\citeauthoryear{{Smith}, {Westmoquette}, {Gallagher},
  {O'Connell}, {Rosario}  \& {de Grijs}}{{Smith} et~al.}{2006}]{Smith_2006}
{Smith} L.~J.,  {Westmoquette} M.~S.,  {Gallagher} J.~S.,  {O'Connell} R.~W.,
  {Rosario} D.~J.,   {de Grijs} R.,  2006, \mn@doi [\mnras]
  {10.1111/j.1365-2966.2006.10507.x}, \href
  {https://ui.adsabs.harvard.edu/abs/2006MNRAS.370..513S} {370, 513}

\bibitem[\protect\citeauthoryear{{Staveley-Smith}, {Bond}, {Bekki}  \&
  {Westmeier}}{{Staveley-Smith} et~al.}{2022}]{StaveleySmith_2022}
{Staveley-Smith} L.,  {Bond} E.,  {Bekki} K.,   {Westmeier} T.,  2022, \mn@doi
  [\pasa] {10.1017/pasa.2022.23}, \href
  {https://ui.adsabs.harvard.edu/abs/2022PASA...39...26S} {39, e026}

\bibitem[\protect\citeauthoryear{{Tomsick} et~al.,}{{Tomsick}
  et~al.}{2019}]{Tomsick_2019}
{Tomsick} J.,  et~al., 2019, in Bulletin of the American Astronomical Society.
  p.~98 (\mn@eprint {arXiv} {1908.04334}), \mn@doi{10.48550/arXiv.1908.04334}

\bibitem[\protect\citeauthoryear{{Tomsick} et~al.,}{{Tomsick}
  et~al.}{2023}]{Tomsick_2023}
{Tomsick} J.~A.,  et~al., 2023, \mn@doi [arXiv e-prints]
  {10.48550/arXiv.2308.12362}, \href
  {https://ui.adsabs.harvard.edu/abs/2023arXiv230812362T} {p. arXiv:2308.12362}

\bibitem[\protect\citeauthoryear{{Ventura}, {D'Antona}, {Mazzitelli}  \&
  {Gratton}}{{Ventura} et~al.}{2001}]{Ventura_2001}
{Ventura} P.,  {D'Antona} F.,  {Mazzitelli} I.,   {Gratton} R.,  2001, \mn@doi
  [\apjl] {10.1086/319496}, \href
  {https://ui.adsabs.harvard.edu/abs/2001ApJ...550L..65V} {550, L65}

\bibitem[\protect\citeauthoryear{Ventura, Di~Criscienzo, Carini  \&
  D’Antona}{Ventura et~al.}{2013}]{Ventura_2013}
Ventura P.,  Di~Criscienzo M.,  Carini R.,   D’Antona F.,  2013, \mn@doi
  [\mnras] {10.1093/mnras/stt444}, 431, 3642–3653

\bibitem[\protect\citeauthoryear{{Vink}}{{Vink}}{2018}]{Vink_2018}
{Vink} J.~S.,  2018, \mn@doi [\aap] {10.1051/0004-6361/201832773}, \href
  {https://ui.adsabs.harvard.edu/abs/2018A&A...615A.119V} {615, A119}

\bibitem[\protect\citeauthoryear{{Vink}, {de Koter}  \& {Lamers}}{{Vink}
  et~al.}{2001}]{Vink_2001}
{Vink} J.~S.,  {de Koter} A.,   {Lamers} H.~J.~G.~L.~M.,  2001, \mn@doi [\aap]
  {10.1051/0004-6361:20010127}, \href
  {https://ui.adsabs.harvard.edu/abs/2001A&A...369..574V} {369, 574}

\bibitem[\protect\citeauthoryear{{Vink}, {Muijres}, {Anthonisse}, {de Koter},
  {Gr{\"a}fener}  \& {Langer}}{{Vink} et~al.}{2011}]{Vink_2011}
{Vink} J.~S.,  {Muijres} L.~E.,  {Anthonisse} B.,  {de Koter} A.,
  {Gr{\"a}fener} G.,   {Langer} N.,  2011, \mn@doi [\aap]
  {10.1051/0004-6361/201116614}, \href
  {https://ui.adsabs.harvard.edu/abs/2011A&A...531A.132V} {531, A132}

\bibitem[\protect\citeauthoryear{{Voss}, {Diehl}, {Hartmann}, {Cervi{\~n}o},
  {Vink}, {Meynet}, {Limongi}  \& {Chieffi}}{{Voss} et~al.}{2009}]{Voss_2009}
{Voss} R.,  {Diehl} R.,  {Hartmann} D.~H.,  {Cervi{\~n}o} M.,  {Vink} J.~S.,
  {Meynet} G.,  {Limongi} M.,   {Chieffi} A.,  2009, \mn@doi [\aap]
  {10.1051/0004-6361/200912260}, \href
  {https://ui.adsabs.harvard.edu/abs/2009A&A...504..531V} {504, 531}

\bibitem[\protect\citeauthoryear{{Wallyn}, {Mahoney}, {Durouchoux}  \&
  {Chapuis}}{{Wallyn} et~al.}{1996}]{Wallyn_1996}
{Wallyn} P.,  {Mahoney} W.~A.,  {Durouchoux} P.,   {Chapuis} C.,  1996, \mn@doi
  [\apj] {10.1086/177434}, \href
  {https://ui.adsabs.harvard.edu/abs/1996ApJ...465..473W} {465, 473}

\bibitem[\protect\citeauthoryear{{Wang} et~al.,}{{Wang}
  et~al.}{2009}]{Wang_2009}
{Wang} W.,  et~al., 2009, \mn@doi [\aap] {10.1051/0004-6361/200811175}, \href
  {https://ui.adsabs.harvard.edu/abs/2009A&A...496..713W} {496, 713}

\bibitem[\protect\citeauthoryear{Wang, Liu  \& P{\'e}rez-R{\'\i}os}{Wang
  et~al.}{2024}]{Wang_2024}
Wang W.,  Liu X.,   P{\'e}rez-R{\'\i}os J.,  2024, \mn@doi [Molecules]
  {10.3390/molecules29010222}, 29

\bibitem[\protect\citeauthoryear{{Wasserburg}, {Busso}, {Gallino}  \&
  {Nollett}}{{Wasserburg} et~al.}{2006}]{Wasserburg_2006}
{Wasserburg} G.~J.,  {Busso} M.,  {Gallino} R.,   {Nollett} K.~M.,  2006,
  \mn@doi [\nphysa] {10.1016/j.nuclphysa.2005.07.015}, \href
  {https://ui.adsabs.harvard.edu/abs/2006NuPhA.777....5W} {777, 5}

\bibitem[\protect\citeauthoryear{{Wasserburg}, {Karakas}  \&
  {Lugaro}}{{Wasserburg} et~al.}{2017}]{Wasserburg_2017}
{Wasserburg} G.~J.,  {Karakas} A.~I.,   {Lugaro} M.,  2017, \mn@doi [\apj]
  {10.3847/1538-4357/836/1/126}, \href
  {https://ui.adsabs.harvard.edu/abs/2017ApJ...836..126W} {836, 126}

\bibitem[\protect\citeauthoryear{{Wei}, {Zou}, {Lin}, {Zhou}, {Liu}, {Kong},
  {Ma}  \& {Ma}}{{Wei} et~al.}{2021}]{Wei_2021}
{Wei} P.,  {Zou} H.,  {Lin} L.,  {Zhou} X.,  {Liu} X.,  {Kong} X.,  {Ma} L.,
  {Ma} S.-G.,  2021, \mn@doi [Research in Astronomy and Astrophysics]
  {10.1088/1674-4527/21/1/6}, \href
  {https://ui.adsabs.harvard.edu/abs/2021RAA....21....6W} {21, 006}

\bibitem[\protect\citeauthoryear{{Wik} et~al.,}{{Wik} et~al.}{2014}]{Wik_2014}
{Wik} D.~R.,  et~al., 2014, \mn@doi [\apj] {10.1088/0004-637X/797/2/79}, \href
  {https://ui.adsabs.harvard.edu/abs/2014ApJ...797...79W} {797, 79}

\bibitem[\protect\citeauthoryear{{Winter} \& {Clarke}}{{Winter} \&
  {Clarke}}{2023}]{Winter_2023}
{Winter} A.~J.,  {Clarke} C.~J.,  2023, \mn@doi [\mnras]
  {10.1093/mnras/stad312}, \href
  {https://ui.adsabs.harvard.edu/abs/2023MNRAS.521.1646W} {521, 1646}

\bibitem[\protect\citeauthoryear{{W{\"u}nsch}, {Palou{\v{s}}}, {Tenorio-Tagle}
  \& {Ehlerov{\'a}}}{{W{\"u}nsch} et~al.}{2017}]{Wunsch_2017}
{W{\"u}nsch} R.,  {Palou{\v{s}}} J.,  {Tenorio-Tagle} G.,   {Ehlerov{\'a}} S.,
  2017, \mn@doi [\apj] {10.3847/1538-4357/835/1/60}, \href
  {https://ui.adsabs.harvard.edu/abs/2017ApJ...835...60W} {835, 60}

\bibitem[\protect\citeauthoryear{{Yoast-Hull}, {Everett}, {Gallagher}  \&
  {Zweibel}}{{Yoast-Hull} et~al.}{2013}]{Yoast-Hull_2013}
{Yoast-Hull} T.~M.,  {Everett} J.~E.,  {Gallagher} J.~S. I.,   {Zweibel} E.~G.,
   2013, \mn@doi [\apj] {10.1088/0004-637X/768/1/53}, \href
  {https://ui.adsabs.harvard.edu/abs/2013ApJ...768...53Y} {768, 53}

\bibitem[\protect\citeauthoryear{Yong, Grundahl  \& Norris}{Yong
  et~al.}{2014}]{Yong_2014}
Yong D.,  Grundahl F.,   Norris J.~E.,  2014, \mn@doi [\mnras]
  {10.1093/mnras/stu2334}, 446, 3319–3329

\bibitem[\protect\citeauthoryear{{de Grijs}, {Wilkinson}  \& {Tadhunter}}{{de
  Grijs} et~al.}{2005}]{deGrijs_2005}
{de Grijs} R.,  {Wilkinson} M.~I.,   {Tadhunter} C.~N.,  2005, \mn@doi [\mnras]
  {10.1111/j.1365-2966.2005.09176.x}, \href
  {https://ui.adsabs.harvard.edu/abs/2005MNRAS.361..311D} {361, 311}

\makeatother
\end{thebibliography}





\bsp	
\label{lastpage}
\end{document}